\documentclass[traditabstract, longauth]{aa}  %
\usepackage{natbib}
\bibpunct{(}{)}{;}{a}{}{,} 
\usepackage[utf8]{inputenc}
\usepackage{txfonts}
\usepackage{graphicx}
\usepackage{gensymb} 
\usepackage[dvipsnames]{xcolor}
\usepackage[colorlinks=true, citecolor=blue]{hyperref}
\usepackage{xspace}
\usepackage{siunitx} 
\usepackage{multirow}
\usepackage{hhline}
\usepackage{upgreek} 
\usepackage{pifont} 
\newcommand{\themis}{{\small THEMIS}\xspace}
\newcommand{\mbb}{{\small MBB}\xspace}

\newcommand{\upmicron}{$\upmu$m\xspace}

\begin{document} 

   \title{Spatially-resolved interstellar dust properties in the face-on spiral galaxy M~99 as observed by NIKA2} 
    \author{L.~Pantoni\inst{\ref{Ghent},\ref{CEA}}\fnmsep\thanks{\email{\url{lara.pantoni@ugent.be}}}
              \and  F.~Galliano \inst{\ref{CEA}}
              \and  S.~C.~Madden \inst{\ref{CEA}}
              \and  R.~Adam \inst{\ref{OCA}}
              \and  P.~Ade \inst{\ref{Cardiff}}
              \and  H.~Ajeddig \inst{\ref{CEA}}
              \and  P.~Andr\'e \inst{\ref{CEA}}
              \and  E.~Artis \inst{\ref{LPSC},\ref{Garching}}
              \and  H.~Aussel \inst{\ref{CEA}}
              \and  M.~Baes  \inst{\ref{Ghent}}
              \and  A.~Beelen \inst{\ref{LAM}}
              \and  A.~Beno\^it \inst{\ref{Neel}}
              \and  S.~Berta \inst{\ref{IRAMF}}
              \and  L.~Bing \inst{\ref{LAM}}
              \and  O.~Bourrion \inst{\ref{LPSC}}
              \and  M.~Calvo \inst{\ref{Neel}}
              \and  V.~Casasola \inst{\ref{IRA-INAF}}
              \and  A.~Catalano \inst{\ref{LPSC}}
              \and  I-D.~Chang \inst{\ref{TAIWAN}}
              \and  I.~De~Looze \inst{\ref{Ghent}}
              \and  M.~De~Petris \inst{\ref{Roma}}
              \and  F.-X.~D\'esert \inst{\ref{IPAG}}
              \and  S.~Doyle \inst{\ref{Cardiff}}
              \and  E.~F.~C.~Driessen \inst{\ref{IRAMF}}
              \and  G.~Ejlali \inst{\ref{Tehran}}
              \and  A.~Gomez \inst{\ref{CAB}} 
              \and  J.~Goupy \inst{\ref{Neel}}
              \and  A.~P.~Jones \inst{\ref{IAS}}
              \and  C.~Hanser \inst{\ref{LPSC}}
              \and  A.~Hughes \inst{\ref{IRAP}}
              \and  S.~Katsioli \inst{\ref{Athens_obs},\ref{Athens_univ}}
              \and  F.~K\'eruzor\'e \inst{\ref{Argonne}}
              \and  E.~W.~Koch \inst{\ref{Harvard},\ref{NRAO}}
              \and  C.~Kramer \inst{\ref{IRAMF}}
              \and  B.~Ladjelate \inst{\ref{IRAME}} 
              \and  G.~Lagache \inst{\ref{LAM}}
              \and  S.~Leclercq \inst{\ref{IRAMF}}
              \and  J.-F.~Lestrade \inst{\ref{LUX}}
              \and  J.~F.~Mac\'ias-P\'erez \inst{\ref{LPSC}}
              \and  A.~Maury \inst{\ref{CEA}}
              \and  P.~Mauskopf \inst{\ref{Cardiff},\ref{Arizona}}
              \and  F.~Mayet \inst{\ref{LPSC}}
              \and  A.~Monfardini \inst{\ref{Neel}}
              \and  A.~Moyer-Anin \inst{\ref{LPSC}}
              \and  M.~Mu\~noz-Echeverr\'ia \inst{\ref{IRAP}}
              \and  A.~Nersesian \inst{,\ref{Ghent},\ref{Liege}}
              \and  D.~Paradis \inst{\ref{IRAP}}
              \and  L.~Perotto \inst{\ref{LPSC}}
              \and  G.~Pisano \inst{\ref{Roma}}
              \and  N.~Ponthieu \inst{\ref{IPAG}}
              \and  V.~Rev\'eret \inst{\ref{CEA}}
              \and  A.~J.~Rigby \inst{\ref{Leeds}}
              \and  A.~Ritacco \inst{\ref{ENS}, \ref{INAF}}
              \and  C.~Romero \inst{\ref{Pennsylvanie}}
              \and  H.~Roussel \inst{\ref{IAP}}
              \and  F.~Ruppin \inst{\ref{IP2I}}
              \and  K.~Schuster \inst{\ref{IRAMF}}
              \and  A.~Sievers \inst{\ref{IRAME}}
              \and  M.~W.~L.~Smith \inst{\ref{Cardiff}}
              \and  F.~S.~Tabatabaei \inst{\ref{Tehran}}
              \and  J.~Tedros \inst{\ref{IRAME}}
              \and  C.~Tucker \inst{\ref{Cardiff}}
              \and  N.~Ysard \inst{\ref{IRAP}}
              \and  E.~M.~Xilouris \inst{\ref{Athens_obs}}
              \and  R.~Zylka \inst{\ref{IRAMF}}
              }

   \institute{
    Sterrenkundig Observatorium Universiteit Gent, Krijgslaan 299 S9, B-9000 Gent, Belgium
    \label{Ghent}
    \and
    Universit\'e Paris-Saclay, Universit\'e Paris Cit\'e, CEA, CNRS, AIM, 91191, Gif-sur-Yvette, France
    \label{CEA}
    \and
    Universit\'e C\^ote d'Azur, Observatoire de la C\^ote d'Azur, CNRS, Laboratoire Lagrange, France 
    \label{OCA}
    \and
    School of Physics and Astronomy, Cardiff University, CF24 3AA, UK
    \label{Cardiff}
    \and   
    Universit\'e Grenoble Alpes, CNRS, Grenoble INP, LPSC-IN2P3, 38000 Grenoble, France
    \label{LPSC}
    \and	
    Max Planck Institute for Extraterrestrial Physics, 85748 Garching, Germany
    \label{Garching}
    \and
    Aix Marseille Univ, CNRS, CNES, LAM, Marseille, France
    \label{LAM}
    \and
    Universit\'e Grenoble Alpes, CNRS, Institut N\'eel, France
    \label{Neel}
    \and
    Institut de RadioAstronomie Millim\'etrique (IRAM), Grenoble, France
    \label{IRAMF}
    \and 
    Dipartimento di Fisica, Sapienza Universit\`a di Roma, I-00185 Roma, Italy
    \label{Roma}
    \and
    Univ. Grenoble Alpes, CNRS, IPAG, 38000 Grenoble, France
    \label{IPAG}
    \and
    STAR Institute, Quartier Agora - All\'ee du six Ao\^ut, 19c B-4000 Li\`ege, Belgium
    \label{Liege}
    \and
    INAF - Istituto di Radioastronomia, Via P. Gobetti 101, 40129, Bologna, Italy
    \label{IRA-INAF}
    \and
    Centro de Astrobiolog\'ia (CSIC-INTA), Torrej\'on de Ardoz, 28850 Madrid, Spain
    \label{CAB}
    \and
    Institute for Research in Fundamental Sciences (IPM), School of Astronomy, Tehran, Iran
    \label{Tehran}
    \and
    National Observatory of Athens, IAASARS, GR-15236, Athens, Greece
    \label{Athens_obs}
    \and
    Faculty of Physics, University of Athens, GR-15784 Zografos, Athens, Greece
    \label{Athens_univ}
    \and
    High Energy Physics Division, Argonne National Laboratory, Lemont, IL 60439, USA
    \label{Argonne}
    \and  
    Instituto de Radioastronom\'ia Milim\'etrica (IRAM), Granada, Spain
    \label{IRAME}
    \and
    LUX, Observatoire de Paris, PSL Research Univ., CNRS, Sorbonne Univ., UPMC, 75014 Paris, France  
    \label{LUX}
    \and
    School of Earth \& Space and Department of Physics, Arizona State University, AZ 85287, USA
    \label{Arizona}
    \and
    School of Physics and Astronomy, University of Leeds, Leeds LS2 9JT, UK
    \label{Leeds}
    \and
    INAF-Osservatorio Astronomico di Cagliari, 09047 Selargius, Italy
    \label{INAF}
    \and 
    LPENS, ENS, PSL Research Univ., CNRS, Sorbonne Univ., Universit\'e de Paris, 75005 Paris, France 
    \label{ENS}
    \and  
    Department of Physics and Astronomy, University of Pennsylvania, PA 19104, USA
    \label{Pennsylvanie}
    \and
    Institut d'Astrophysique de Paris, CNRS (UMR7095), 75014 Paris, France
    \label{IAP}
    \and
    University of Lyon, UCB Lyon 1, CNRS/IN2P3, IP2I, 69622 Villeurbanne, France
    \label{IP2I}
    \and
    Institut d'Astrophysique Spatiale (IAS), CNRS, Universit\'e Paris Sud, Orsay, France
    \label{IAS}
    \and
    IRAP, Universit\'e de Toulouse, CNRS, UPS, IRAP, Toulouse Cedex 4, France
    \label{IRAP}
    \and
    centre for Astrophysics, Harvard \& Smithsonian, 60 Garden Street, 02138 Cambridge, MA, USA
    \label{Harvard}
    \and
    National Radio Astronomy Observatory, 800 Bradbury SE, Suite 235, Albuquerque, NM 87106, USA
    \label{NRAO}
    \and Institute of Astronomy and Astrophysics, Academia Sinica, No. 1, Sec. 4, Roosevelt Road, Taipei 106319, Taiwan
    \label{TAIWAN}
    }

   \date{Received ; accepted }
 
  \abstract
  {
  Large dust grains in thermal equilibrium dominate the far-infrared emission of star-forming galaxies and contribute substantially to their millimetre continuum. Constraining dust properties in this regime is challenging due to contamination from free-free and synchrotron emission.
  }
   {
   We investigate spatial variations in the dust spectral index, dust mass, and grain size and composition in the nearby face-on spiral galaxy M~99. To this end, we use new 1.15 and 2~mm continuum observations obtained with NIKA2 on the IRAM~30~m telescope as part of the IMEGIN Guaranteed Time Large Programme, combined with ancillary data spanning ultraviolet to radio wavelengths.
   }
   {
   We decompose the infrared-to-radio spectral energy distribution of M~99 into dust, free-free, and synchrotron components using the hierarchical Bayesian SED-fitting code \texttt{HerBIE}. Dust emission is modelled using both a modified blackbody (\mbb) with a variable millimetre spectral index $\beta$ and the \themis dust model with a fixed $\beta$. Our spatially resolved analysis is performed on $\sim$1.75 kpc ($25^{\prime\prime}$) scales, encompassing the galaxy centre, spiral arms, and inter-arm regions.
   }
   {
   From the \mbb modelling, we find significant spatial variations in $\beta$, ranging from $\sim1.6-1.7$ in diffuse regions to $\sim2.3-2.5$ in denser, star-forming environments. These variations likely reflect dust grain evolution driven by coagulation and changes in the silicate-to-carbonaceous grain abundance. Dust masses inferred with variable $\beta$ are up to a factor of $\sim4$ higher than those derived assuming a fixed $\beta$ (1.6 on average). Variable-$\beta$ models recover expected correlations with dust-to-stellar and dust-to-gas ratios, whereas fixed-$\beta$ models systematically bias these quantities. The small grain fraction increases from $\sim$10\% in the centre to $\sim$15\% in the diffuse disc and is anti-correlated with the interstellar radiation field intensity, while gas-phase metallicity plays only a minor role within the central 8 kpc. The synchrotron spectral index varies from $\sim0.6-0.7$ in star-forming regions to $\sim$1.2 in the diffuse medium, consistent with cosmic-ray electron ageing.
   }
   {}
    

    \titlerunning{A NIKA2 perspective on M~99 spatially-resolved dust properties.}
    \authorrunning{L. Pantoni}

   \maketitle

\section{Introduction}\label{introduction}

Interstellar dust grains are ubiquitous in galaxies and are generally well mixed with the gas in the interstellar medium \citep[ISM;][]{Galliano2018ARA&A..56..673G}. Although they contribute only about 1\% of the total ISM mass \citep{Whittet2022dge..book.....W}, dust grains play a significant role in shaping the spectral energy distribution 
(SED) of galaxies. They attenuate stellar light in the ultraviolet (UV) to optical range, through absorption and scattering, and re-emit the absorbed energy thermally in the infrared \citep[IR;][]{Whittet2022dge..book.....W}.

Beyond studies of dust in the Milky Way (MW), observations of galaxies in the nearby Universe (within $\sim$100 Mpc) provide access to a wide diversity of environments, albeit typically at limited spatial resolution ($\sim$~kpc). Edge-on systems, for example, enable investigations of extraplanar dust at large vertical distances from the disc \citep[e.g.,][]{Holwerda2012A&A...541L...5H, Yoon2021MNRAS.502..969Y, Chastenet2026}, while face-on galaxies offer a favourable geometry for probing dust properties in the galaxy centre, spiral arms, and inter-arm regions, as well as for characterising radial dust distributions \citep[e.g.,][]{Casasola2017A&A...605A..18C, Tailor2025A&A...701A..74T}.
Spatially resolved studies of nearby galaxies hosting active galactic nuclei \citep[see][for a review]{Li2007ASPC..373..561L}, dwarf galaxies \citep[e.g.,][]{RemyRuyer2013A&A...557A..95R}, and local starbursts \citep[e.g.,][]{Contini2003MNRAS.342..299C} further extend these investigations to extreme physical conditions. Collectively, observations of nearby galaxies constitute a crucial intermediate step toward understanding dust evolution across cosmic time and, in particular, in high-redshift systems \citep{Galliano2018ARA&A..56..673G}.

Several key open questions in dust studies concern the physical properties of dust grains at millimetre (mm) wavelengths. Robust constraints on the dust millimetre opacity $\kappa$ and its slope $\beta$ (i.e., the dust spectral index) are essential for accurately deriving dust masses and for mapping dust-to-stellar, dust-to-gas, and dust-to-metal ratios \citep[e.g.,][]{Lamperti2019MNRAS.489.4389L}. These quantities provide direct insight into the chemical evolution of galaxies and the reservoirs available for dust production \citep[e.g.,][]{Casasola2022A&A...668A.130C, Park2024MNRAS.535..729P}.

Since 2017, the New IRAM KID Array~2 (NIKA2) camera installed on the IRAM~30~m telescope at Pico Veleta (Spain) has enabled continuum observations of galaxies at 1.15 and 2~mm, with angular resolutions of $12^{\prime\prime}$ and $18^{\prime\prime}$, respectively \citep{Adam2018A&A...609A.115A, Calvo2016JLTP..184..816C, Perotto2020A&A...637A..71P}. Observations of nearby galaxies with NIKA2 provide a unique opportunity to spatially resolve dust emission in the mm regime and to directly probe variations in dust properties across different environments.

In this context, the IRAM~30~m Guaranteed Time Large Programme Interpreting the millimetre Emission of Galaxies with IRAM–NIKA2 (IMEGIN; PI: S. Madden) devoted approximately 200 hours to mapping the mm continuum of 22 nearby galaxies spanning a wide range of stellar masses, morphologies, and metallicities. The sample was selected to lie within 30 Mpc and to benefit from high-quality infrared-to-(sub)mm imaging for resolved SED modelling, complemented by matched-resolution UV, CO, and HI data.
Notable results from the IMEGIN programme include the characterisation of extraplanar dust in the halo of NGC~891 \citep[][]{Katsioli2023A&A...679A...7K}, studies of dust emission in the starburst regions of NGC~2146 and in the peculiar dwarf galaxy NGC~2976 \citep[][]{Ejlali2025A&A...693A..88E}, as well as an analysis of dust properties in the barred spiral and AGN-host galaxy NGC~3627 \citep{Katsioli2026MNRAS.tmp...78K}.

As with other ground-based facilities, such as SCUBA-2 \citep[e.g.,][]{Smith2021ApJS..257...52S, Pattle2023MNRAS.522.2339P}, NIKA2 imaging of extended sources is subject to large-scale filtering. The severity of this filtering depends on a combination of atmospheric and instrumental noise, the observing strategy, and data-reduction choices (e.g., source masking), and must be carefully quantified and accounted for in scientific analyses (see Appendix~\ref{App:large-scale-filtering} and Ejlali et al., in prep.).

This paper, part of the IMEGIN publication series, presents a spatially resolved analysis of the IR-to-radio emission of the nearby spiral galaxy M~99 (NGC~4254) based on SED fitting, with a particular emphasis on dust properties at mm wavelengths. M~99 is the first galaxy in the IMEGIN sample for which NIKA2 maps affected by large-scale filtering are analysed and published.

M~99 has a stellar mass of $M_\star \sim 4.2 \times 10^{10}$ M$_\odot$ \citep{Chemin2016A&A...588A..48C}, a star formation rate (SFR) $\sim$ 3.1 M$_\odot$ yr$^{-1}$ \citep{Leroy2021ApJS..257...43L}, and an average gas-phase metallicity of $12 + \log(\mathrm{O/H}) \sim 8.6$ \citep{Kreckel2019ApJ...887...80K}, making it an excellent analogue to the MW. Its nearly face-on orientation, with an inclination of only 20 deg \citep{Clark2018A&A...609A..37C}, offers a clear view of the galaxy centre, spiral arms, and disc. 
This favourable geometry is one of the primary motivations for selecting M~99 as a case study in the IMEGIN programme.
Located at a distance of 14.40~Mpc \citep{Poznanski2009ApJ...694.1067P}, M~99 has an optical isophotal diameter of $D_{25} \sim 5^{\prime}$ (corresponding to $\sim20$ kpc; \citealt{deVaucouleurs:1991rc3..book.....D, Clark2018A&A...609A..37C}). The galaxy exhibits a mild asymmetry in its morphology, characterised by a prominent spiral arm extending nearly 15 kpc perpendicular to the major axis in the optical band. This feature has been interpreted as a relic of a past tidal interaction with a massive companion, likely within the last Gyr \citep{Soria2006MNRAS.372.1531S, Duc2008ApJ...673..787D, Chemin2016A&A...588A..48C}. The main geometric parameters of M~99 are summarised in Table~\ref{tab:NGC4254prop}.
   \begin{table} 
   \caption[]{List of the main geometric parameters for the face-on spiral galaxy M~99.}
   \label{tab:NGC4254prop}
   \centering
        \begin{tabular}{lcc}
        \hline\hline
        \noalign{\smallskip}
        \textbf{Quantity} & \textbf{Value} & \textbf{Ref}\\
        \noalign{\smallskip}
        \hline
        \noalign{\smallskip}
        RA (J2000) & 184.7059746 deg& 1\\
        DEC (J2000) & 14.4231206 deg& 1\\
        Semi-major axis (a) &  $170^{\prime\prime}$ & 1\\
        Axial ratio (b/a) & 0.755 & 1\\
        Inclination & 20.1 deg & 2\\
        Position angle & $-28$ deg& 1\\
        Distance & 14.40 Mpc & 3\\
        $25^{\prime\prime}$ &  1.75 kpc & \\
        \noalign{\smallskip}
        \hline
        \noalign{\smallskip}
    \end{tabular}
\tablebib{ 
(1) This work (see Appendix \ref{App:photometry}); 
(2) DustPedia \citep{Clark2018A&A...609A..37C};
(3) SNII optical \citep[SN 1986I;][]{Poznanski2009ApJ...694.1067P}.
}
\end{table}

As part of the DustPedia \citep{Davies2017PASP..129d4102D,Clark2018A&A...609A..37C} and KINGFISH \citep{Kennicutt2011PASP..123.1347K} samples, the integrated dust properties of M~99 have been already explored up to $\sim500$~\upmicron (sampled by Herschel/SPIRE). Previous studies estimate a total dust mass of $M_{\rm dust} \sim (2-5) \times 10^7$ M$_\odot$ \citep{Nersesian2019A&A...624A..80N, Aniano2020ApJ...889..150A}. In this work, we extend the analysis to the mm and radio regimes. 

\begin{figure*}[!h]
\centering
\includegraphics[width=.95\columnwidth]{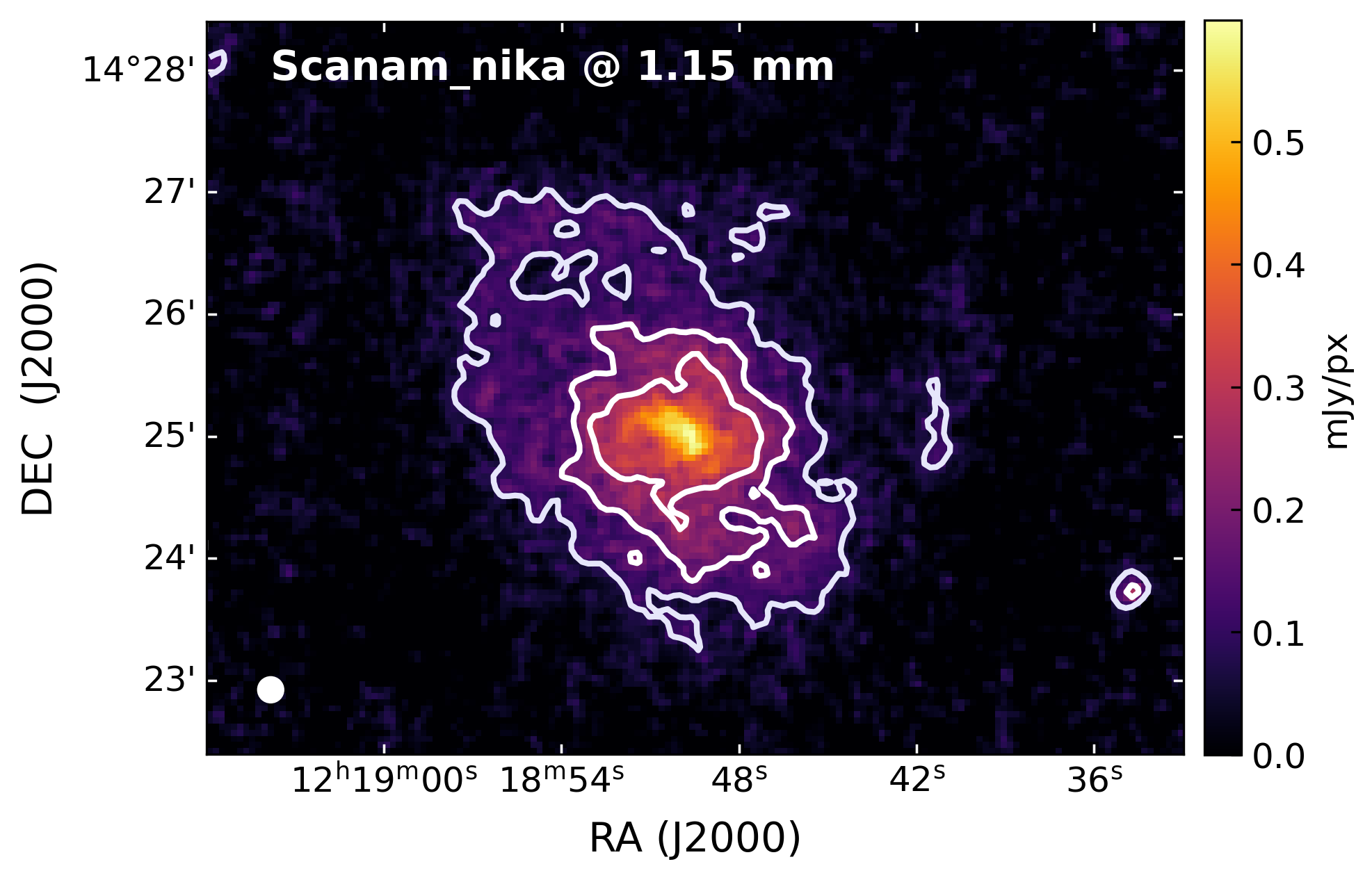}
\includegraphics[width=.95\columnwidth]{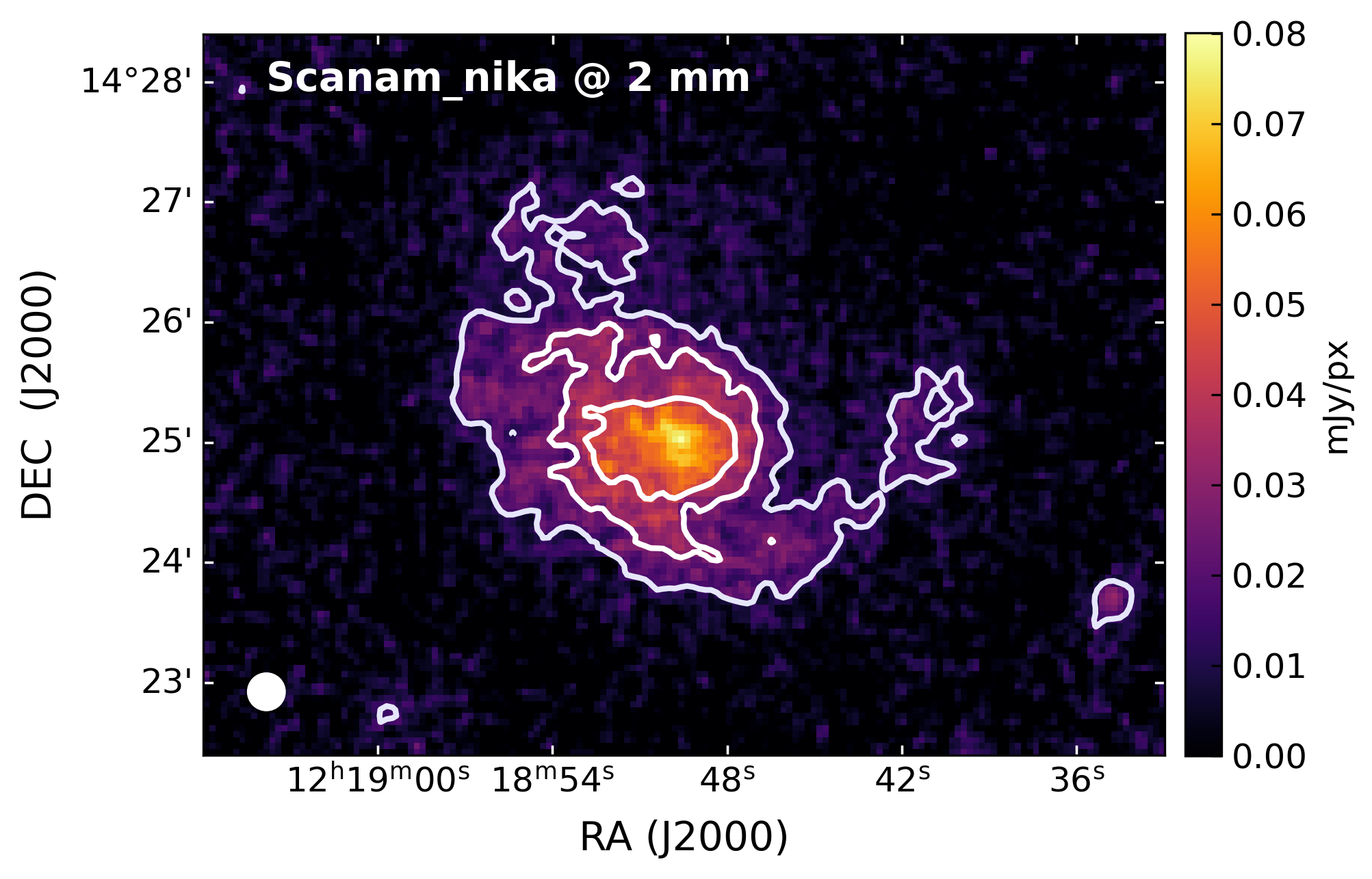}
\caption{NIKA2 maps of M~99 at 1.15 mm (left) and 2 mm (right) obtained through the Scanam\_nika data reduction pipeline in mJy/px (cutouts of $8^\prime\times6^\prime$; pixel size of $3^{\prime\prime}$). The mean RMS is 0.034 mJy/px at 1.15 mm and 0.0055 mJy/px at 2 mm. White solid lines show the $[2.5,5,7.5]\times\sigma$ contours. The FWHM is $\sim12^{\prime\prime}$ at 1.15 mm and $\sim18^{\prime\prime}$ at 2 mm, corresponding to $\sim0.8$ and $\sim1.25$ kpc (cf. white circles, bottom-left).}
\label{fig-NIKA2_SCANAM}       
\end{figure*}
The structure of the paper is as follows. In Sect.~\ref{sect:data}, we present the NIKA2 observations of M~99, describe the ancillary datasets, and outline the procedures used to homogenise the multi-wavelength data. Sect.~\ref{Sect:method} details our SED fitting methodology. The results, both on integrated and spatially-resolved scales, are presented and discussed in Sect.~\ref{Sect:results}. Finally, we summarise our main conclusions in Sect.~\ref{Sect:conclusions}.

\section{Data}\label{sect:data}
In this Sect. we present the NIKA2 maps of M~99, along with the multi-wavelength ancillary data we use in this work.

\subsection{NIKA2 observations at 1.15 mm and 2 mm}\label{sect:NIKA2-maps}
 
We observed M~99 with the IRAM 30~m/NIKA2 camera over multiple observing sessions conducted between February 2018 and January 2023. NIKA2 provides simultaneous continuum observations at 1.15 and 2 mm. The corresponding angular resolutions of $12^{\prime\prime}$ and $18^{\prime\prime}$ (FWHM) translate into linear scales of approximately 0.8 kpc and 1.25 kpc, respectively, at the distance of M~99. We acquired 159 scans, for a total of 24 hours of on-source integration time.
Each sub-scan was conducted at speeds ranging from $40^{\prime\prime}$s$^{-1}$ to $94^{\prime\prime}$s$^{-1}$, with typical time separations of 10~s between sub-scans. Scanning directions alternated between position angles of 30 deg and 120 deg. For reference, each NIKA2 observing scan lasts approximately 420~s. Telescope pointing was updated hourly, and focus was adjusted every three hours. During observations, the elevation of the source ranged from 30 deg to 67 deg, while the atmospheric optical depth at 225 GHz ($\tau_{225\,\mathrm{GHz}}$) varied between 0.07 and 0.45, with a median value of 0.25.

The data were reduced using two independent pipelines: \texttt{PIIC} \citep[Pointing and Imaging in Continuum;][]{Zylka2013:ascl.soft03011Z, BertaZylka} and \texttt{Scanam\_nika} v14.0. The latter is adapted from the \texttt{Scanamorphos} algorithm originally developed for Herschel on-the-fly imaging \citep{Roussel2013PASP..125.1126R} and subsequently optimised for NIKA2 observations \citep[see Appendix~A of][]{Ejlali2025A&A...693A..88E}.
Comparisons with space-based measurements (e.g., Planck) indicate that the \texttt{PIIC} maps suffer from substantial loss of extended emission, with up to $\sim$40\% of the total flux filtered out at 1.15~mm (Ejlali et al., in prep.). In contrast, the \texttt{Scanam\_nika} maps are consistent with Planck photometry once differences in filter transmission and central wavelength are taken into account.
We further assess the reliability of the \texttt{Scanam\_nika} products through dedicated simulations, in which the Herschel/SPIRE 250~$\upmu$m map, convolved to the NIKA2 2~mm resolution, is processed identically to the NIKA2 data (see Appendix \ref{App:large-scale-filtering} and Appendix A3 of \citealt{Ejlali2025A&A...693A..88E}). These tests show that residual filtering of large-scale emission in the \texttt{Scanam\_nika} maps varies on a pixel-by-pixel basis, reaching levels of up to $\sim$15\%. We incorporate this effect into the flux density uncertainties, as described in Appendix~\ref{App:large-scale-filtering}.
Ejlali et al. (in prep.) will present a comprehensive overview of the data-reduction methods (\texttt{PIIC} and \texttt{Scanam\_nika}) applied to the IMEGIN sample, and will describe the feathering of NIKA2 maps with data from cosmic microwave background (CMB) experiments (ACT or Planck). This technique generally enables the recovery of extended emission that is partially filtered out in some NIKA2 observations \citep[see][]{Smith2021ApJS..257...52S}. In this work, we adopt the \texttt{Scanam\_nika} maps as our fiducial mm data products.

Fig.~\ref{fig-NIKA2_SCANAM} presents the final NIKA2 maps of M~99 in units of mJy px$^{-1}$, with a pixel size of $3^{\prime\prime}$. To convert the flux densities to mJy beam$^{-1}$, beam areas of $(188 \pm 11)$ arcsec$^2$ at 1.15 mm and $(381 \pm 11)$ arcsec$^2$ at 2 mm should be used. The 1.15 mm map is constructed from 74 scans (out of 159), corresponding to $\sim9.6$ hours of usable on-source time; the remaining scans were excluded due to poor weather conditions or data corruption. The 2 mm map includes 151 scans ($\sim16.3$ hours), with only 8 scans discarded. The selected scans span atmospheric opacities of $0.22-0.54$ at 1.15 mm and $0.14-0.69$ at 2 mm. The peak flux density in the 1.15 mm map is $\sim0.6$ mJy px$^{-1}$ and $\sim0.08$ mJy px$^{-1}$ at 2 mm. We measure the median sky brightness and root mean square (RMS) noise outside the source mask, ensuring inclusion of residual correlated noise.
The sky level is then subtracted from the map as the final step of the \texttt{Scanam\_nika} processing pipeline. The RMS is $\sim0.034$ mJy px$^{-1}$ at 1.15 mm $\sim0.006$ mJy px$^{-1}$ at 2~mm.
These Scanam\_nika maps will be publicly available through the IRAM Science Archive\footnote{\url{https://iram-institute.org/science-portal/data-archive/}} by June 2026.

\subsection{Ancillary data}\label{Sect:ancillary}

   \begin{table*} 
     \caption[]{Multi-wavelength continuum maps used in this work.}
     \label{tab:ancillary_SEDfit}
     \centering
         \begin{tabular}{lccccccc}
            \hline\hline
            \noalign{\smallskip}
            \textbf{Telescope/Instrument} & $\mathbf{\lambda_{obs}}$ & $\mathbf{\theta_{res}^{FWHM}}$ & \textbf{Pixel size} & \textbf{Calibration} &\textbf{SED fitting} & \textbf{Calibrators} & \textbf{Ref.} \\  
             & [\upmicron] & [$^{\prime\prime}$] & [$^{\prime\prime}$] & \textbf{uncertainty} [\%] & & &\\
            \noalign{\smallskip}
            \hline
            \noalign{\smallskip}
            GALEX / FUV & 0.153 & 4 & 3 & 4.5\tablefootmark{a} &  & \checkmark & 1\\ 
            GALEX / NUV & 0.227 & 5.6 & 3 & 2.7\tablefootmark{a} & & \checkmark & 1\\
            Spitzer / IRAC & 3.6 & 1.66 & 0.6 & 10.2\tablefootmark{a} & \checkmark & \checkmark & 1\\ 
            Spitzer / IRAC & 4.5 & 1.72 & 0.6 & 10.2\tablefootmark{a} & \checkmark & & 1\\ 
            Spitzer / IRAC & 5.2  & 1.88 & 0.6 & 10.2\tablefootmark{a} & \checkmark & & 1 \\ 
            Spitzer / IRAC & 8  & 1.98 & 0.6 & 10.2\tablefootmark{a} & \checkmark & & 1 \\ 
            Spitzer / MIPS & 25 & 6 & 1.5 & 4\tablefootmark{a}& \checkmark & \checkmark & 1 \\ 
            Herschel / PACS  & 70 & 9 & 2 &5.4\tablefootmark{a} & \checkmark & & 1\\ 
            Herschel / PACS & 100 & 10 & 3 & 5.4\tablefootmark{a}& \checkmark & & 1 \\ 
            Herschel / PACS & 160 & 13 & 4 & 5.4\tablefootmark{a}& \checkmark & & 1 \\ 
            Herschel / SPIRE & 250 & 18 & 6 & 5.9\tablefootmark{a}& \checkmark & & 1 \\ 
            Herschel / SPIRE & 350 & 25 & 8 & 5.9\tablefootmark{a}& \checkmark & & 1 \\ 
            Herschel / SPIRE & 500 & 36 & 12 & 5.9\tablefootmark{a} & \checkmark & & 1 \\ 
            IRAM 30~m / NIKA2 & 1150 & 12 & 3 & 7.7\tablefootmark{b} & \checkmark & & 2 \\
            Planck / HFI4 & 1380 & 300 & 103 & 1\tablefootmark{a} & \checkmark & & 2 \\
            IRAM 30~m / NIKA2 & 2000 & 18 & 3 & 5.8\tablefootmark{b} & \checkmark & & 2 \\
            X-VLA + Effelsberg & $3\times10^4$ & 15 & 1 & 5\tablefootmark{c} & \checkmark & & 3\\ 
            C-VLA + Effelsberg & $6\times10^4$ & 15 & 1 & 5\tablefootmark{c} &\checkmark & & 3\\ 
            L-VLA & $20\times10^4$ & 4.5 & 0.7 & 3\tablefootmark{d} &\checkmark & & 4\\ 
            \noalign{\smallskip}
            \hline
         \end{tabular} 
    \tablefoot{In the order, we list (increasing wavelengths): the telescope and instrument name; the central wavelength; the angular resolution (FWHM); the pixel size; and the calibration uncertainty. Finally, we indicate whether the map was used for SED fitting and/or as a calibrator. \\
\tablefoottext{a}{\citet{Galliano2021A&A...649A..18G} and references therein.}
\tablefoottext{b}{\citet[][their Appendix A1]{Ejlali2025A&A...693A..88E} for uncertainty on the beam solid angle (i.e., 5.8\% at 1.15 mm and 2.9\% at 2 mm), \citet{Perotto2020A&A...637A..71P} and reference therein for uncertainty on Uranus model (i.e., 5\%).}
\tablefoottext{c}{\cite{Tabatabaei2017ApJ...836..185T, Chemin2016A&A...588A..48C}.}
\tablefoottext{d}{\cite{Perley2015AAS...22531106P} and the official VLA NRAO website: \url{https://science.nrao.edu/facilities/vla/}.}}
    \tablebib{$(1)$ DustPedia archive; $(2)$ this work; $(3)$ \cite{Chy2007A&A...474..415C}; $(4)$ Chiang, Koch, and the PHANGS collaboration (in prep.).
    }
   \end{table*}
   \begin{table}
   \caption[]{Spectral line intensity maps used in this work.}
   \centering
         \begin{tabular}{lcccc}
            \hline\hline
            \noalign{\smallskip}
             \textbf{Transition} & \textbf{Telescope} & $\mathbf{\lambda_{rest}}$ & $\mathbf{\theta_{res}^{FWHM}}$ & \textbf{Pixel size} \\ 
             & & [mm] & [$^{\prime\prime}$] & [$^{\prime\prime}$] \\
            \noalign{\smallskip}
            \hline 
            \noalign{\smallskip}
            CO(2--1)\tablefootmark{a} & IRAM$-$30~m & 1.3 & 13.4 & 2 \\ 
            CO(1--0)\tablefootmark{b} & IRAM$-$30~m & 2.6 & 25.6 & 4\\
            HI\tablefootmark{c} & VLA & 210 & 30 & 5 \\ 
            \noalign{\smallskip}
            \hline
         \end{tabular}
     \tablefoot{We use the listed transitions to compute the gas surface density of M~99 and correct the mm continuum for CO contamination. \\
     \tablefoottext{a}{HERACLES programme \citep{Leroy2009}.}
     \tablefoottext{b}{EMPIRE programme \citep{Jimenez-Donaire2019ApJ...880..127J}.}
     \tablefoottext{c}{VIVA VLA Atlas \citep{Chung2009AJ....138.1741C}.}}
    \label{tab:ancillary_spectrallines}
   \end{table}
   
M~99 is part of the DustPedia sample \citep{Davies2017PASP..129d4102D,Clark2018A&A...609A..37C}, which comprises 875 nearby galaxies with homogeneous multi-wavelength continuum coverage from the far-ultraviolet (FUV) to the mm regime.
From DustPedia\footnote{\url{http://DustPedia.astro.noa.gr/}} we use: (i) the UV maps from GALEX \citep{Martin2005ApJ...619L...1M} to calibrate the spatially-resolved SFR surface density (not used in the SED modelling); 
(ii) the IR maps from Spitzer \citep{Werner2004} and Herschel \citep{Pilbratt2010A&A...518L...1P} for SED modelling,
and Spitzer/MIPS maps to calibrate both the stellar mass and SFR, as detailed in Appendix \ref{App:ancillary_maps};
(iii) the Planck 1.38 mm map to model the integrated mm emission of M~99 (Sect. \ref{Sect:globalSED}) and validate the total flux recovered by NIKA2 (Appendix \ref{App:mm-maps}). 

Spitzer/IRAC images are calibrated for point sources and require surface brightness corrections when applied to extended objects, such as nearby galaxies. Following Sect. 8.2 of the IRAC Instrument Handbook \citep[v4.0;][]{IRAChandbook}, we apply correction factors of 0.91, 0.94, 0.66, and 0.74 to the 3.6, 4.5, 5.8, and 8.0~\upmicron maps, respectively. These corrections, accurate to within 10\%, are included in the systematic uncertainties reported in Table \ref{tab:ancillary_SEDfit}.

We supplement the DustPedia dataset with new and archival radio continuum maps from the Karl G. Jansky Very Large Array \citep[VLA;][]{Perley2011ApJ...739L...1P}
and the Effelsberg 100~m Radio Telescope \citep{Wielebinski2011JAHH...14....3W}. In Table \ref{tab:ancillary_SEDfit} we provide a complete list of the continuum maps used in this work. 
We also incorporate full HI and CO maps (Table \ref{tab:ancillary_spectrallines}), along with $\sim 1900$ gas-phase metallicity measurements from the literature \citep[][]{DeVis2019A&A...623A...5D, Kreckel2019ApJ...887...80K}.

This comprehensive multi-wavelength data set enables the construction of spatially-resolved maps of stellar mass, SFR, and atomic and molecular gas surface densities using widely adopted empirical calibrations. The maps identified as ``calibrators'' in Table~\ref{tab:ancillary_SEDfit} are used as inputs for these derivations. In addition, we construct a map of the gas-phase metallicity of M~99, extending out to $2^{\prime}$ ($\sim$8.4 kpc) from the galactic centre. Further details on the construction of these ancillary products are provided in Appendix~\ref{App:ancillary_maps}.

\subsection{Data processing and homogenisation}

Spatially resolved SED modelling requires all input maps to be homogenised in angular resolution, pixel scale, orientation, and units. In addition, the images must be corrected for potential contamination from astrophysical sources (e.g., bright foreground stars), large-scale foreground or background emission, instrumental artefacts or residual gradients.

To meet these requirements, we developed and applied the Homogenisation for IMEGIN Photometry post-processing pipeline (\texttt{HIP})\footnote{Publicly available on GitLab: \url{https://gitlab.com/imegin/hip}}. \texttt{HIP} performs several key steps: (i) identification and masking of bright foreground stars in the NIR images; (ii) modelling and subtraction of sky emission and instrumental gradients, particularly relevant for the GALEX, Spitzer, and Herschel datasets; (iii) estimation and removal of CO(2--1) line contribution from the NIKA2 1.15~mm and Planck/HFI4 continuum bands; (iv) measurement of integrated photometry; (v) convolution to a common angular resolution followed by reprojection to a uniform pixel scale and orientation. Uncertainties are propagated through these steps using either a bootstrap Monte Carlo (MC) approach or a faster gradient-based propagation method, validated against MC runs with $N_{\rm MC}\sim 10^{3}-10^{4}$ iterations. In this work, we use $N_{\rm MC}=10^{3}$ to generate uncertainty maps and $N_{\rm MC}=100$ to estimate uncertainties on integrated fluxes.

Using \texttt{HIP}, we convolved the multi-wavelength maps of M~99 to the poorest angular resolution among the images included in each analysis, balancing adequate FIR sampling with the high resolution of NIKA2. For the global analysis (Sect.~\ref{Sect:globalSED}), we used all maps listed in Table~\ref{tab:ancillary_SEDfit}. For the analysis of the three morphological components of M~99 (disc, spiral arms, and galaxy centre; Sect.~\ref{sect:centre-arms-disc}), we convolved the maps to $18^{\prime\prime}$ ($\sim 1.25$~kpc), excluding the SPIRE 350~\upmicron, SPIRE 500~\upmicron, and Planck/HFI4 bands. For the pixel-by-pixel analysis (Sect.~\ref{sect:spatially-res-anal}), we convolved the maps to $25^{\prime\prime}$ ($\sim 1.75$~kpc), corresponding to the SPIRE 350~\upmicron resolution, and resampled them onto an $8^{\prime\prime}$ ($\sim 0.56$~kpc) pixel grid, excluding the SPIRE 500~\upmicron and Planck/HFI4 data.

A detailed description of \texttt{HIP}, including the specific prescriptions adopted in this work, is provided in Appendix~\ref{App:hip}.

\section{SED fitting}\label{Sect:method}

We derived the dust properties of M~99 through SED fitting with \texttt{HerBIE}, a hierarchical Bayesian dust SED fitting code \citep{Galliano2018MNRAS.476.1445G,Galliano2021A&A...649A..18G}.

\subsection{\texttt{HerBIE}: hierarchical Bayesian SED fitting}\label{sect:HerBIE}

\texttt{HerBIE} uses a hierarchical Bayesian framework to fit physical dust models, including realistic optical properties, stochastic heating, and a mixture of radiation environments. Unlike least-squares or non-hierarchical Bayesian methods, the hierarchical approach suppresses noise-driven correlations and scatter, yielding more accurate parameter estimates and uncertainties \citep{Shetty2009ApJ...696..676S, Kelly2012ApJ...752...55K, Lamperti2019MNRAS.489.4389L}.

As for any Bayesian approach, \texttt{HerBIE} computes posterior probability density functions (PDFs) as the product of the likelihood and a prior. When the data poorly constrain a parameter, the prior dominates. In the hierarchical Bayesian application to our spatially resolved M~99 data, the priors are estimated from the ensemble of all pixels via a set of shared hyperparameters. These hyperparameters describe the distribution of physical parameters (e.g., dust mass, temperature) across all pixels in the maps and influence the posteriors of each individual pixel \citep[][Eq. 19]{Galliano2018MNRAS.476.1445G}. Posterior distributions are sampled using a Gibbs-within-Metropolis-Hastings Markov chain Monte Carlo (MCMC) algorithm \citep{Geman4767596}. Finally, \texttt{HerBIE} explicitly accounts for both statistical and systematic uncertainties in the input maps \citep[see Sect.~3 of][]{Galliano2018MNRAS.476.1445G}.

\subsection{Physical components of the \texttt{HerBIE} SED model}\label{sect:HerBIEmodels}

We model the NIR-to-radio SED with \texttt{HerBIE} using three emission components: (i) dust emission in the MIR-to-mm regime; (ii) stellar continuum in the NIR; (iii) radio continuum, including both free-free and synchrotron radiation.

Dust emission is described by the non-uniformly illuminated dust mixture model \citep[\texttt{powerU};][Eq. 11]{Galliano2018MNRAS.476.1445G}. It assumes the dust mass is illuminated with a range of radiation field intensities, $U$, following a power law distribution \citep[e.g.,][]{Dale2001ApJ...549..215D}. The dust emission is thus modelled with six free parameters: $U_{\rm min}$, the minimum radiation field intensity; $\Delta U$, the range of radiation field intensities; $\alpha_{\rm dust}$, the spectral index describing the power-law distribution of radiation field intensities; $M_{\rm dust}$, the total dust mass; $q$, the small-grain mass fraction; and $q^{+}$, the fraction of charged small grains.

The NIR stellar continuum is modelled (using the \texttt{starBB} module) as a blackbody with $T_\star=5\times10^4$ K, ensuring the spectrum lies in the Rayleigh–Jeans regime. In \texttt{HerBIE} this component is described by a single free parameter, the bolometric luminosity, $L_\star$.

The radio continuum, comprising both free–free and synchrotron emission, is modelled  (using the \texttt{radio} module) as two power laws \citep[][Eq. 13]{Galliano2018MNRAS.476.1445G} with three free parameters: $L_{\rm radio}$, the luminosity at $\lambda=1$ cm; $f_{\rm FF}$, the fraction of free-free emission at $\lambda=1$ cm; and $\alpha_{\rm sync}$, the synchrotron spectral index, assumed to be constant across the modelled range of frequencies.

In total, our SED fits involve ten free parameters. In the following, we refer to the mass-weighted starlight intensity heating the dust grains ${\langle}{U}{\rangle}$, as the average interstellar radiation field (ISRF). This quantity is a function of three underlying parameters:  $U_{\rm min}$, $\Delta U$ and $\alpha_{\rm dust}$, for which the adopted priors are $0.01<U_{\rm min}<1000$; $0<\Delta U<10^6$; $1<\alpha_{\rm dust}<2.5$. Other prior ranges are summarised in Table~\ref{tab:global_sed_paramvalue}.

An equivalent equilibrium temperature for large grains, $T_{\rm eq}$, can be estimated from \citet{Boulanger1996AA...312..256B}:
\begin{equation}\label{eq:equiv_temp}
    {\langle}{U}{\rangle}\sim \Bigg(\frac{T_{\rm eq}}{18 \,{\rm K}}\Bigg)^{4+\beta}
\end{equation}
where $\langle U \rangle$ is normalised to the average ISRF in the solar neighbourhood, i.e. $\langle U \rangle_\odot = 2.2\times10^{-5}$ W m$^{-2}$ \citep{Mathis1983A&A...128..212M}, and $\beta$ is the dust spectral index.

\subsection{\themis: a laboratory-based dust model}\label{Sect:themis}

We adopt the physically motivated \themis dust model \citep[][]{Jones2013A&A...558A..62J, Jones2017A&A...602A..46J, Kohler2014A&A...565L...9K} as the basis for the \texttt{powerU} component in \texttt{HerBIE}. 

\themis dust properties are based on laboratory measurements of the optical properties of interstellar dust analogues. \themis dust mixture is comprised of:
(i) small (radius $a < 150$ nm), partially hydrogenated amorphous carbon grains, the smallest of which ($a \lesssim 10$ nm) exhibit spectroscopic properties similar to polycyclic aromatic hydrocarbons (PAHs) and are responsible for the aromatic and aliphatic features observed in the MIR; (ii) large dehydrogenated amorphous carbon and silicate grains dominating the FIR continuum. Carbonaceous grains with $a > 20$ nm are coated with an amorphous carbon mantle, while silicate grains have both a carbon mantle and iron inclusions.

In \texttt{HerBIE}, the small, partially hydrogenated, amorphous carbon grains (AC) are modelled using three size-based populations: (i) very small (VSAC, $a < 0.7$ nm) grains are responsible for the shortest-wavelength MIR aromatic features; (ii) small (SAC, $0.7 < a < 1.5$ nm) grains dominate the longer-wavelength MIR features; and (iii) medium/large (MLAC, $a > 1.5$ nm) grains produce the featureless MIR continuum and contribute to the FIR peak and long-wavelength continuum.

We define the mass fraction of aromatic feature carriers $q_{\rm AF}$ as $q_{\rm AF} = q_{\rm VSAC} + q_{\rm SAC}$ 
where $q$ denotes the
mass fraction. This definition is analogous to the PAH mass fraction ($q_{\rm PAH}$) introduced by \citet{Draine2007ApJ...657..810D}\footnote{With $q_{\rm PAH}=0.45\times q_{\rm AF}$ \citep{Galliano2021A&A...649A..18G}.}.

The default \themis composition has 69\% silicates, 14\% MLAC, and 17\% VSAC+SAC by mass. In our \texttt{HerBIE} fits, we adopt the default \themis grain properties but allow $q_{\rm AF}$ to vary. The intrinsic dust millimetre spectral index (i.e., determined by the chemical composition of large grains) is thus constant, with $\beta=1.79$ \citep{Bianchi2019A&A...631A.102B} for a given interstellar field intensity U, but mixing multiple U components can flatten the effective (i.e., observed) slope. This effect is illustrated, for example, in Fig. III.13 of \citet[][]{Galliano2022HabT.........1G} and in \citet{Galliano2018MNRAS.476.1445G}.

\subsection{Single-temperature Modified Blackbody}\label{Sect:MBB}

In addition to the \themis dust model, we employ a single-T modified blackbody (\mbb) model to describe the FIR SED of M~99, which is referred to as \texttt{MBB1} in the \texttt{HerBIE} framework \citep[][Eq. 2]{Galliano2018MNRAS.476.1445G}. The \mbb model describes the thermal emission from dust grains heated by the ISRF. The free parameters are three: the dust mass, $M_{\rm dust}$; the dust temperature, $T_{\rm dust}$; and the dust spectral index, $\beta$. 

This simplified model assumes that the FIR dust emission is dominated by a single temperature component, which is a reasonable approximation for wavelengths longer than 100~\upmicron. A key advantage of using the \mbb approach is that it allows us to explore spatial or environmental variations in $\beta$, which is otherwise fixed in the \themis model.

\begin{figure*}[!h]
\centering
\includegraphics[width=1\columnwidth]{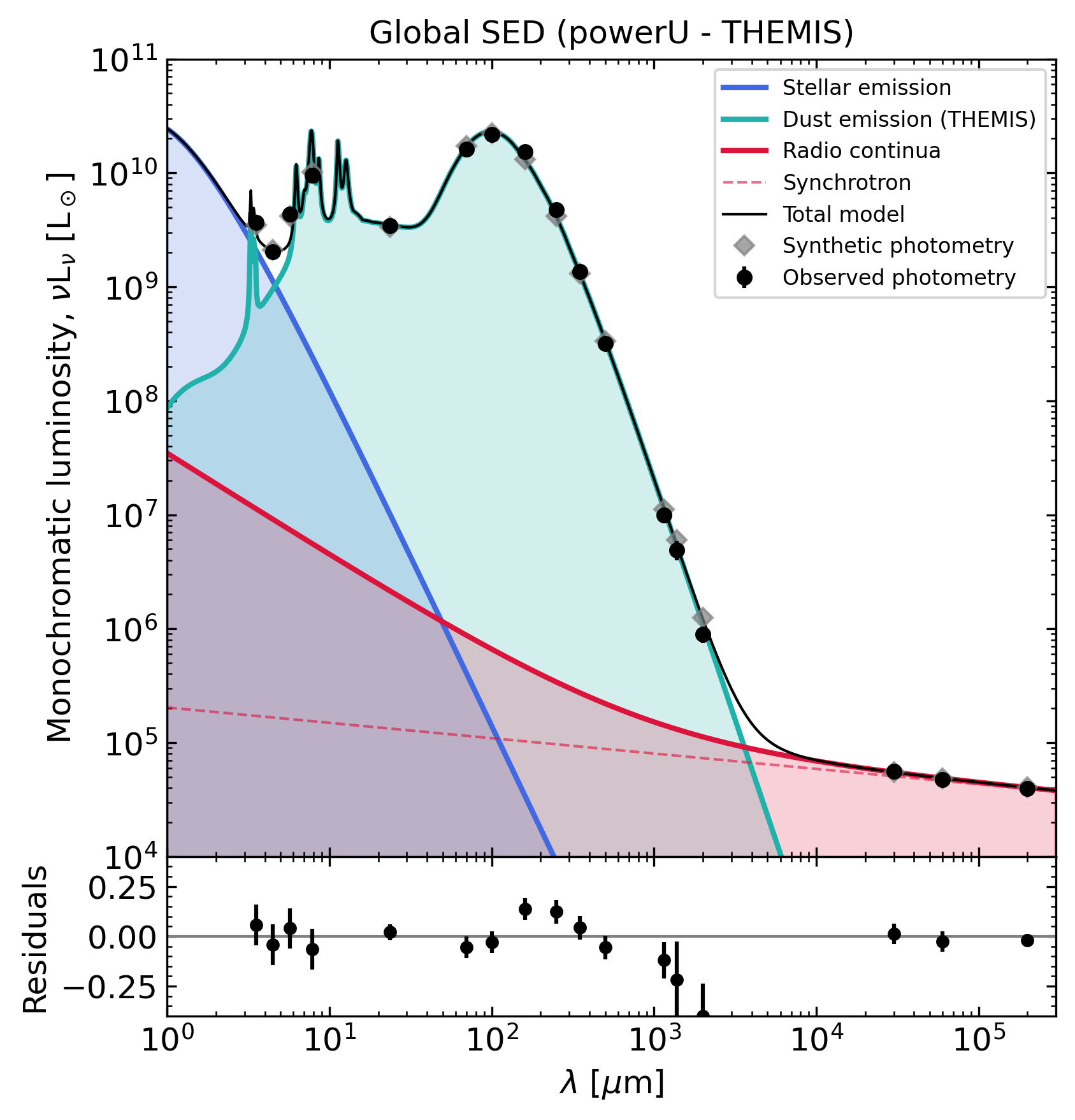}
\includegraphics[width=.99\columnwidth]{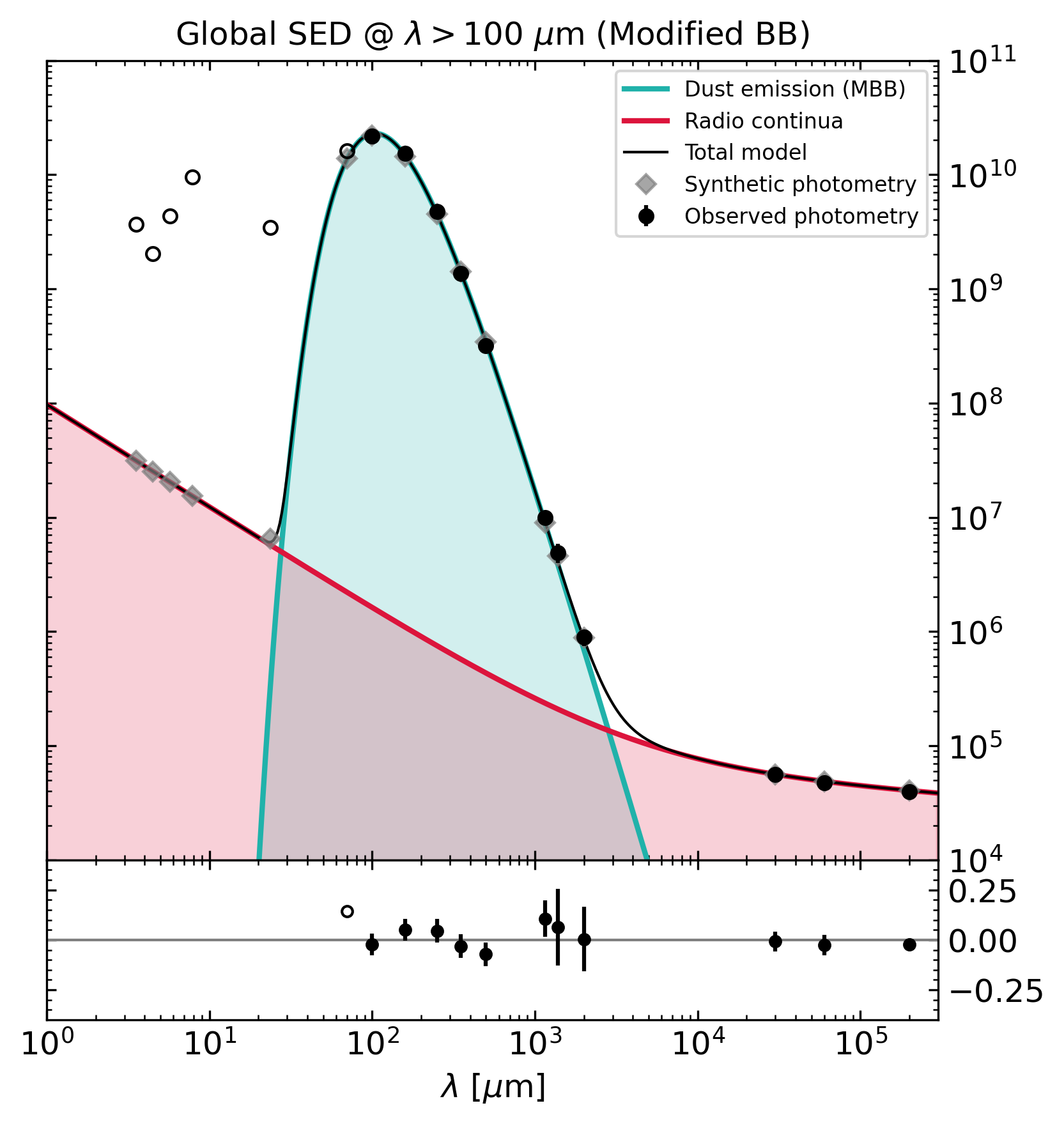}
\caption{ 
Integrated NIR-to-radio SED modelling of M~99 with \texttt{HerBIE}. Filled black circles show observed photometry used in the fit; grey diamonds indicate corresponding synthetic photometry. \textit{Left panel}: full SED fit (black line), including stellar emission (\texttt{starBB}, blue), dust emission using \texttt{powerU} with \themis (green), and radio continuum (\texttt{radio}, red). \textit{Right panel}: fit limited to $\lambda \ge 100$~\upmicron using a single-T \mbb (\texttt{MBB1}, green), excluding stellar emission. Empty circles denote data not included in the fit. Residuals are shown below each panel.}
\label{fig-GLOBAL-SED}
\end{figure*}

\begin{table}
\caption[]{Multi-band integrated photometry of M~99 computed with \texttt{HIP}.}
\label{tab:integrated_photometry} 
    \begin{tabular}{lcccc}
    \hline\hline
    \noalign{\smallskip}
    \textbf{Instrument} & \textbf{Luminosity} & \textbf{Flux}  & \textbf{RMS} \\
    & [$10^{7}$ $L_\odot$] & [Jy] & [mJy] \\
    \noalign{\smallskip}
    \hline
    \noalign{\smallskip}
    IRAC 3.6~\upmicron & 370 & $0.68\pm0.07$ & 2\\ 
    IRAC 4.5~\upmicron & 200 & $0.47\pm0.05$ & 1.4\\ 
    IRAC 5.2~\upmicron & 435 & $1.3\pm0.1$ & 4\\ 
    IRAC 8~\upmicron & 950 & $3.9\pm0.4$ & 3\\ 
    MIPS 24~\upmicron & 340 & $4.2\pm0.2$ & 6\\ 
    PACS 70~\upmicron & 1625 & $59\pm3$ & 140\\ 
    PACS 100~\upmicron & 2160 & $112\pm6$ & 150 \\ 
    PACS 160~\upmicron & 1530 & $126\pm7$ & 220\\ 
    SPIRE 250~\upmicron & 475 & $61\pm4$ & 204 \\ 
    SPIRE 350~\upmicron & 137 & $25\pm2$ & 100 \\ 
    SPIRE 500~\upmicron & 32 & $8.3\pm0.5$ & 55\\ 
    NIKA2 1.15 mm & 0.99 & $0.59\pm0.05$ & 30\\
    HFI4 1.38 mm & 0.49 & $0.35\pm0.07$ & 68 \\
    NIKA2 2 mm & 0.089 & $0.09\pm0.01$ & 9 \\ 
    X-VLA+Eff. 3 cm & 0.0056 & $0.087\pm0.004$ & 0.04\\ 
    C-VLA+Eff. 6 cm & 0.0048 & $0.148\pm0.007$ & 0.3\\ 
    L-VLA 20 cm & 0.004 & $0.41\pm0.01$ & 0.02\\ 
    \noalign{\smallskip}
    \hline
    \end{tabular} 
\tablefoot{NIKA2 1.15 and Planck/HFI4 integrated fluxes are CO-subtracted (cf. Appendix \ref{Sect:CO_sub}). The NIKA2 RMS includes the scatter driven by large-scale filtering (cf. Appendix \ref{App:Scanam_nika_maps}). Prior to aperture photometry, the sky emission was subtracted from all images. Monochromatic luminosities (i.e., $\nu\,$L$_\nu$) listed in the second column were computed at $d=14.4$ Mpc (Table \ref{tab:NGC4254prop}).}
\end{table}

\begin{table} 
\caption[]{M~99 integrated properties used in this work.}
\label{tab:global_properties}
\centering
         \begin{tabular}{lccc}
            \hline\hline
            \noalign{\smallskip}
             \textbf{Quantity (symbol)} & \textbf{Adopted value} & \textbf{Units} & \textbf{Ref.} \\
              \noalign{\smallskip}
            \hline
            \noalign{\smallskip}
            $M_\star$ & $(4\pm1)\times10^{10}$ & M$_\odot$ & 1\\ 
            SFR & $3^{+10}_{-2}$ & M$_\odot$ yr$^{-1}$ & 2\\ 
            $M_{\rm H_2}$ & $(6\pm2)\times10^{9}$ & M$_\odot$ & 3\\ 
            $M_{\rm HI}$ & $(7\pm1)\times10^{9}$ & M$_\odot$ & 3\\ 
            $12+\log({\rm O/H})$ & $8.55\pm0.04$ & & 4\\ 
            \noalign{\smallskip}
            \hline
         \end{tabular} 
     \tablefoot{We list: the stellar mass, $M_\star$; the SFR; the molecular gas mass, $M_{\rm H_2}$; the atomic gas mass, $M_{\rm HI}$; the oxygen abundance, $12+\log({\rm O/H})$.}
     \tablebib{(1) this work, Appendix \ref{Sect:stellarsurfdensity}; (2) this work, Appendix \ref{Sect:SFRsurfdensity}; (3) this work, Appendix \ref{Sect:gassurfdensity}; (4) \cite{Kreckel2019ApJ...887...80K}, using the empirical S-calibration (Scal) by \cite{PilyuginGrebel2016MNRAS.457.3678P}; (5) this work, using the \themis model.}
\end{table}

\begin{table*} 
\caption[]{Best-fit parameter values obtained with \texttt{HerBIE} (non-hierarchical Bayesian framework) and derived quantities.}
\label{tab:global_sed_paramvalue}
\centering
    \begin{tabular}{lccccc}
    \hline\hline
    \noalign{\smallskip}
    \textbf{Parameter} & \multicolumn{2}{c}{\textbf{\themis}} & \multicolumn{2}{c}{\textbf{\mbb}} & \textbf{Units} \\
     & Value & Range & Value & Range & \\
    \noalign{\smallskip}
    \hline
    \noalign{\smallskip}
    $M_{\rm dust}$ & $(3.2\pm0.2)\times10^7$ & [unlimited] & $(3.8\pm0.7)\times10^7$ & [unlimited] & M$_\odot$ \\
    $T_{\rm dust}$ & $\sim24.2$ & [derived] & $23.7\pm0.9$ & [$2.7-1000$] & K \\
    $\beta$ & 1.79 & [fixed] & $1.85\pm0.09$ & [$0-5$] &  \\
    $q_{\rm AF}$& $15\pm1$ & [$10^{-3}-90$] & $-$ &  & \%\\ 
    $L_\star$ & $(3.8\pm0.7)\times10^{10}$ & [unlimited] & $-$ &  & L$_\odot$\\
    ${\langle}{U}{\rangle}$ & $5.5\pm0.9$ & [derived] & $-$ & & ${\langle}{U}{\rangle}_\odot$\\
    $L_{\rm radio}$ & $(7.0\pm0.6)\times10^4$ & [unlimited] & $(8\pm1)\times10^4$ & [unlimited] & L$_\odot$ \\
    $\alpha_{\rm sync}$ & $0.87\pm0.04$ & [$0.5-1.2$] & $0.92\pm0.09$ & [$0.5-1.2$] & \\
    $f_{\rm FF}$ & $15\pm12$ & [$0-100$] & $30\pm20$ & [$0-100$] &\%\\ 
    DGR & $1/400 \sim 0.0025$ & [derived] & $1/345 \sim 0.0029$& [derived] & \\
    DSR & $1/1290 \sim0.0008$ & [derived] & $1/1000 \sim0.001$ & [derived] & \\
    \noalign{\smallskip}
    \hline
    \end{tabular}
\tablefoot{In the order, we list: the dust mass, $M_{\rm dust}$; the dust temperature, $T_{\rm dust}$ (for \themis, derived using Eq.~\eqref{eq:equiv_temp}); the dust spectral index, $\beta$; the fraction of small dust grains (i.e., aromatic feature carriers), $q_{\rm AF}$; the NIR stellar luminosity, $L_\star$; the average ISRF, ${\langle}{U}{\rangle}$ (in units of ${\langle}{U}{\rangle}_\odot = 2.2\times10^{-5}$ W m$^{-2}$); the radio luminosity at $\lambda=1$ cm, $L_{\rm radio}$; the radio spectral index, $\alpha_{\rm sync}$; the fraction of free-free emission at $\lambda=1$ cm, $f_{\rm FF}$; the dust-to-total gas ratio, DGR; the dust-to-stellar mass ratio, DSR. Values in square brackets denote the adopted priors.}
\end{table*}

\section{Results and discussion}
\label{Sect:results}

In this Sect., we present our results and discuss their implications.
As the hierarchical Bayesian framework is primarily designed for spatially-resolved analyses, we perform the integrated SED fitting using \texttt{HerBIE} in its non-hierarchical mode.

\subsection{A global view of the dust in M~99}\label{Sect:globalSED}

We measured the integrated multi-wavelength photometry of M~99 (Table~\ref{tab:NGC4254prop}) using HIP (Appendix~\ref{App:photometry}), within the elliptical aperture shown in Fig. \ref{fig-SPIRE250_photometry}, based on the geometry from \citet[][]{Clark2018A&A...609A..37C}. Prior to extraction, we subtracted the sky emission (Appendix \ref{Sect:skysub}) and corrected the NIKA2~1.15~mm and Planck/HFI4 maps for CO(2--1) contamination (Appendix~\ref{Sect:CO_sub}).
Additional integrated properties of M~99 are summarised in Table~\ref{tab:global_properties}. Our photometric measurements agree within uncertainties with DustPedia \citep{Clark2018A&A...609A..37C}, KINGFISH \citep{Aniano2020ApJ...889..150A, Chang2020ApJ...900...53C, Chastenet2025ApJS..276....2C}, and the 3 and 6 cm measurements of  \citet{Tabatabaei2017ApJ...836..185T}.
Fig.~\ref{fig-GLOBAL-SED} shows the integrated SED of M~99 fitted with the \themis and \mbb models.
\themis reproduces the data well ($\chi^2_{\rm red} = 1.35$), though it slightly underestimates the steepness of the FIR slope. The \mbb fit ($\lambda\gtrsim100$~\upmicron), instead, gives a better match in the FIR-mm ($\chi^2_{\rm red} = 0.43$). The \mbb fit yields $M_{\rm dust} = (3.8 \pm 0.7) \times 10^7$ M$_\odot$, dust temperature $T_{\rm dust} = 23.7 \pm 0.9$ K, and spectral index $\beta = 1.85 \pm 0.09$ (Table \ref{tab:global_sed_paramvalue}). The \mbb $\beta$ is marginally steeper than the default \themis value ($\beta = 1.79$), although the two are consistent within the uncertainties. From the \themis fit, we derive $M_{\rm dust}=(3.2\pm0.2)\times10^7$ M$_\odot$ and, using Eq.~\eqref{eq:equiv_temp}, an equivalent temperature $T_{\rm eq}\sim24.2$ K, which is consistent with the \mbb outcome. 
The dust-to-stellar and dust-to-gas ratios inferred from our dust mass estimates (Table \ref{tab:global_sed_paramvalue}) are typical of local star-forming spirals \citep[e.g.,][]{Boselli2010PASP..122..261B, DeVis2017MNRAS.464.4680D, Aniano2020ApJ...889..150A, Casasola2020A&A...633A.100C, Casasola2022A&A...668A.130C}.

About 15\% of the \themis dust mass is in small, partially hydrogenated carbonaceous grains (VSAC + SAC). \themis typically yields higher mass fractions of these grains than other models such as \citet{Draine2007ApJ...657..810D} or \citet{Compiegne2011A&A...525A.103C} because the carriers are less emissive \citep{Galliano2021A&A...649A..18G,Galliano2022HabT.........1G}. For comparison, using the \citet{Draine2007ApJ...657..810D} model, \citet{Aniano2020ApJ...889..150A} derived, for M~99, $q_{\rm PAH} = (4.1 \pm 0.8)$\% and $M_{\rm dust}$ = $(5.2 \pm 0.4) \times 10^{7}$ M$_\odot$, while \citet{Chastenet2025ApJS..276....2C} found $q_{\rm PAH} = (4.4 \pm 0.2)$\%  and $M_{\rm dust}$ = $(7.5 \pm 0.6) \times 10^{7}$ M$_\odot$. The latter estimate is consistent with our dust masses after applying the recommended $1/3$ scaling \citep[e.g.,][]{Chastenet2021ApJ...912..103C}. We note that \citet{Draine2007ApJ...657..810D} model has previously been shown to produce a higher dust mass than other physical ISM dust models by a factor of $\sim$ 2 \citep[e.g.,][]{Galliano2011A&A...536A..88G,Dalcanton2015ApJ...814....3D, Galliano2018MNRAS.476.1445G}.

Earlier studies of the integrated SED of M~99 by the DustPedia collaboration adopted \themis-like grains: \citet{Nersesian2019A&A...624A..80N} obtained $M_{\rm dust} = (2.2 \pm 0.1) \times 10^{7}$ M$_\odot$ with CIGALE, and $(2.6 \pm 0.2) \times 10^{7}$ M$_\odot$ from an \mbb fit; \citet{Galliano2021A&A...649A..18G} found $(2.1 \pm 0.2) \times 10^{7}$ M$_\odot$ using \texttt{HerBIE}. These dust masses are slightly lower than ours, likely due to methodological differences, although the dust temperatures ($24-25$ K) agree well with our estimates.

The new NIKA2 and radio observations enable a clear separation of the dust, free-free, and synchrotron components in the integrated SED of M~99. Dust emission dominates at 1.15 mm ($\sim$99\%), but its contribution decreases to 91\% at 2 mm, 21\% at 5 mm, and less than 2\% at 1 cm.
Adopting the \themis best-fit free-free fraction at 1 cm ($f_{\mathrm{ff}} = 0.15$; Table~\ref{tab:global_sed_paramvalue}), free-free emission contributes 52\% of the radio continuum at 1.15 mm, declining to 40\% at 2 mm and 23\% at 5 mm, while synchrotron emission dominates at $\lambda>1$ cm. The radio parameters derived from the \mbb fit ($\alpha_{\mathrm{sync}}^{\mathrm{MBB}} = 0.92 \pm 0.09$ and $f_{\mathrm{ff}}^{\mathrm{MBB}} = 35 \pm 23$\%) are consistent within uncertainties with those obtained using the \themis model. However, the \mbb fit allows for higher free-free fractions (up to $\sim$50–60\%), in line with typical values reported for normal star-forming galaxies \citep[$f_{\mathrm{ff}} \sim 50$\%;][]{Condon1992ARA&A..30..575C}.

\subsection{Dust properties in M~99 centre, spiral arms, and disc}\label{sect:centre-arms-disc}

\begin{figure}[!h]
\centering
\includegraphics[width=.98\columnwidth]{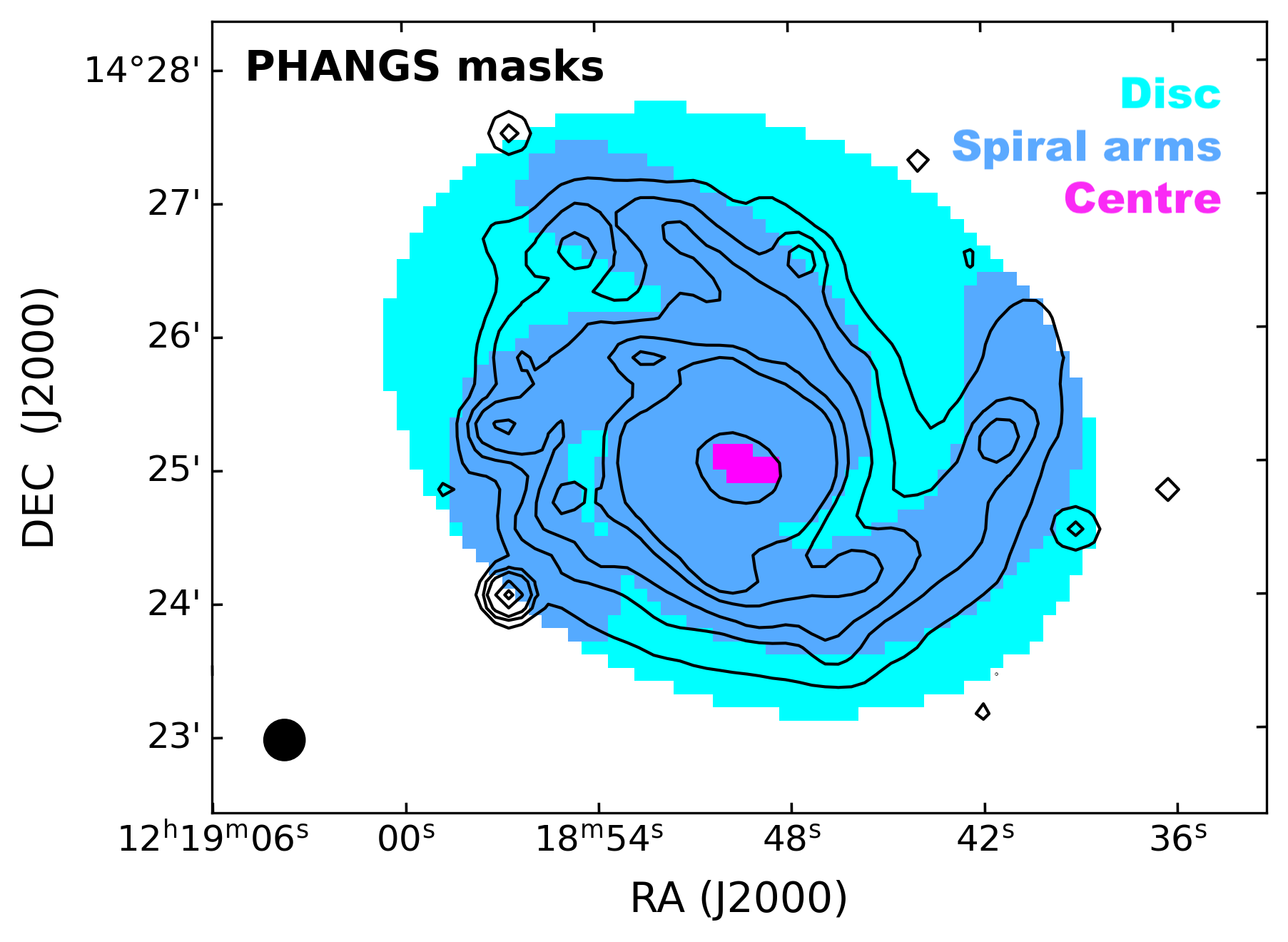}
\caption{PHANGS environmental masks by \citet{Querejeta2021AA...656A.133Q} degraded to match SPIRE~250~\upmicron angular resolution (i.e., 18$^{\prime\prime}\sim1.25$ kpc) and pixel size (i.e., 6$^{\prime\prime}\sim0.4$ kpc). We show the disc in cyan, the spiral arms in blue and the centre in magenta. For reference, we overlay the IRAC 3.6~\upmicron contours (black solid lines).}
\label{fig-PHANGS-masks}
\end{figure}

The nearly face-on orientation of M~99 allows us to examine dust properties in the galaxy centre, spiral arms, and disc. To isolate these components, we adopt the morphological masks from \citet[][PHANGS]{Querejeta2021AA...656A.133Q}, which are based on the Spitzer/IRAC 3.6~\upmicron map (see Fig.~\ref{fig-PHANGS-masks}).
\begin{table*}
\caption[]{Multi-band photometry of spiral arms, disc and centre of M~99 computed with \texttt{HIP}.}
\label{tab:env_photometry}
\centering
    \begin{tabular}{lcccc}
    \hline\hline
    \noalign{\smallskip}
    \textbf{Instrument} & \textbf{Spiral arms} & \textbf{Disc} & \textbf{Centre} & \textbf{Total}\\ 
    & Flux [Jy] & Flux [Jy] & Flux [Jy] & Flux [Jy]\\
    \noalign{\smallskip}
    \hline
    \noalign{\smallskip}
    IRAC 3.6~\upmicron & $0.50\pm0.05$ (0.04) & $0.12\pm0.01$ (0.04) & $0.058\pm0.006$ (0.002) & $0.68\pm0.07$\\ 
    IRAC 4.5~\upmicron & $0.35\pm0.04$ (0.05) & $0.082\pm0.008$ (0.06) &$0.039\pm0.004$ (0.003) & $0.5\pm0.1$\\ 
    IRAC 5.2~\upmicron & $1.0\pm0.1$ (0.2)& $0.24\pm0.02$ (0.2) & $0.076\pm0.008$ (0.015) & $1.3\pm0.1$ \\ 
    IRAC 8~\upmicron & $3.1\pm0.3$ (0.3)& $0.72\pm0.07$ (0.4)& $0.21\pm0.02$ (0.04) & $4.0\pm0.4$\\ 
    MIPS 24~\upmicron & $3.4\pm0.1$ (0.7) &$0.67\pm0.03$ (0.6)& $0.28\pm0.01$ (0.2) & $4.3\pm0.1$\\ 
    PACS 70~\upmicron & $48\pm3$ (74) & $8.4\pm0.5$ (43)& $5.1\pm0.3$ (8)& $61\pm4$\\ 
    PACS 100~\upmicron & $92\pm5$ (54) & $17\pm1$ (62)& $8.8\pm0.5$ (8) & $117\pm6$\\ 
    PACS 160~\upmicron & $103\pm6$ (48)& $23\pm1$ (43)& $8.5\pm0.5$ (8) & $126\pm7$\\ 
    SPIRE~250~\upmicron & $46\pm3$ (35) & $12.6\pm0.7$ (26) &$3.4\pm0.2$ (10) & $62\pm4$\\ 
    NIKA2 1.15 mm &$0.41\pm0.04$ (20)& $0.17\pm0.02$ (9) &$0.018\pm0.001$ (0.2) & $0.60\pm0.06$\\
    NIKA2 2 mm & $0.068\pm0.006$ (5)& $0.025\pm0.004$ (4)& $0.0031\pm0.0002$ (0.07)& $0.095\pm0.010$\\
    X-VLA+Eff. 3 cm & $0.066\pm0.003$ (0.02) & $0.0188\pm0.0009$ (0.02) & $0.0034\pm0.0002$ (0.0008) & $0.088\pm0.004$\\ 
    C-VLA+Eff. 6 cm & $0.107\pm0.005$ (0.03)& $0.036\pm0.002$ (0.03) & $0.0051\pm0.0003$ (0.001) & $0.148\pm0.007$\\ 
    L-VLA 20 cm & $0.285\pm0.009$ (0.04) & $0.127\pm0.004$ (0.04) & $0.0113\pm0.0003$ (0.002) & $0.42\pm0.01$\\ 
    \noalign{\smallskip}
    \hline
    \end{tabular} 
\tablefoot{Integrated fluxes (in Jy) are measured on the images degraded to $18^{\prime\prime}$ resolution.
The associated error includes the calibration uncertainty. In parentheses we report the statistical uncertainty (i.e., RMS) in mJy. NIKA2 RMS values include the contribution from the scatter driven by large-scale filtering (cf. Appendix \ref{App:Scanam_nika_maps}). 
The last column reports the sum of the three components.}
\end{table*}
\begin{figure*}[!h]
\centering
\includegraphics[width=1\columnwidth]{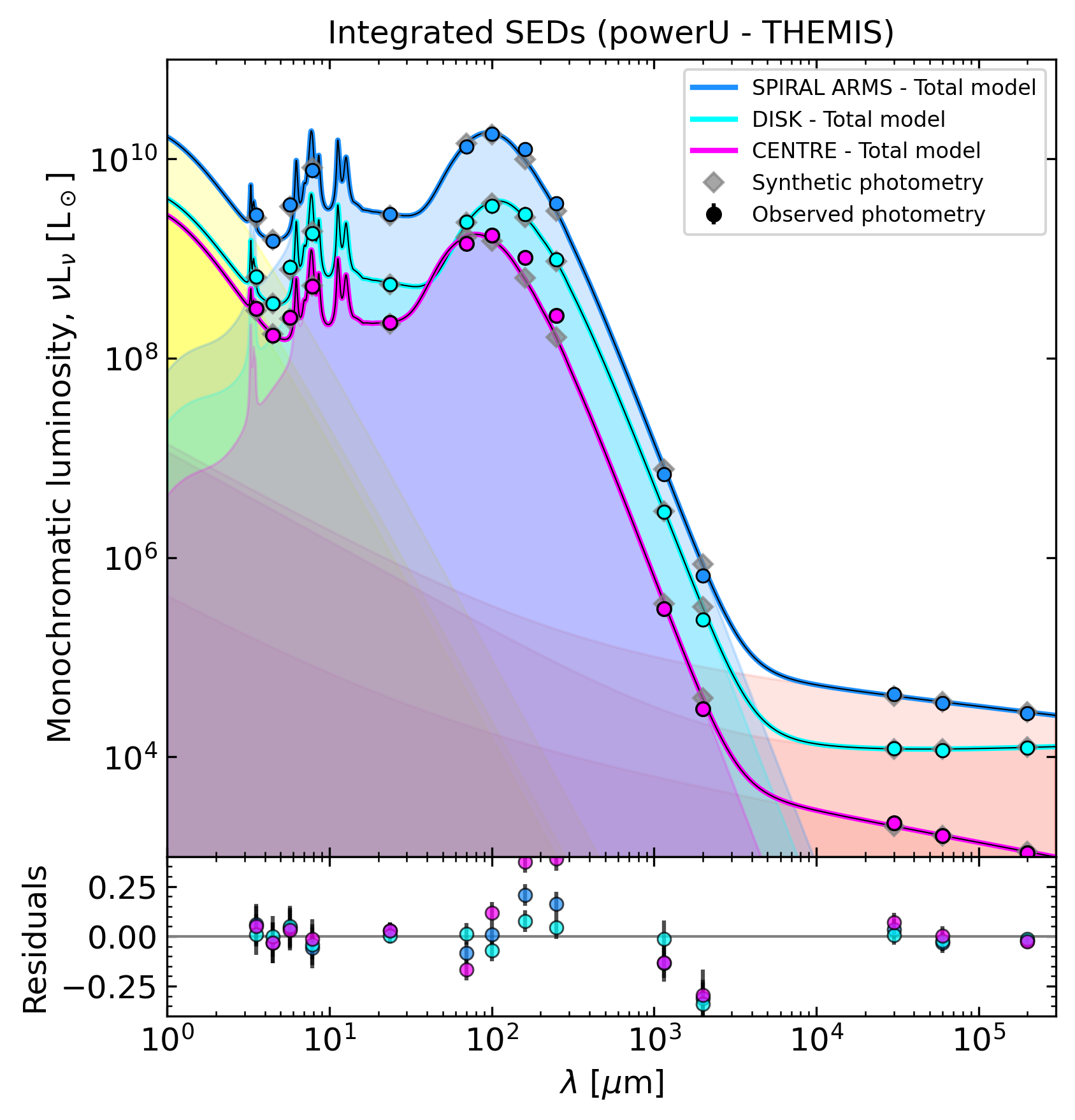}
\includegraphics[width=.99\columnwidth]{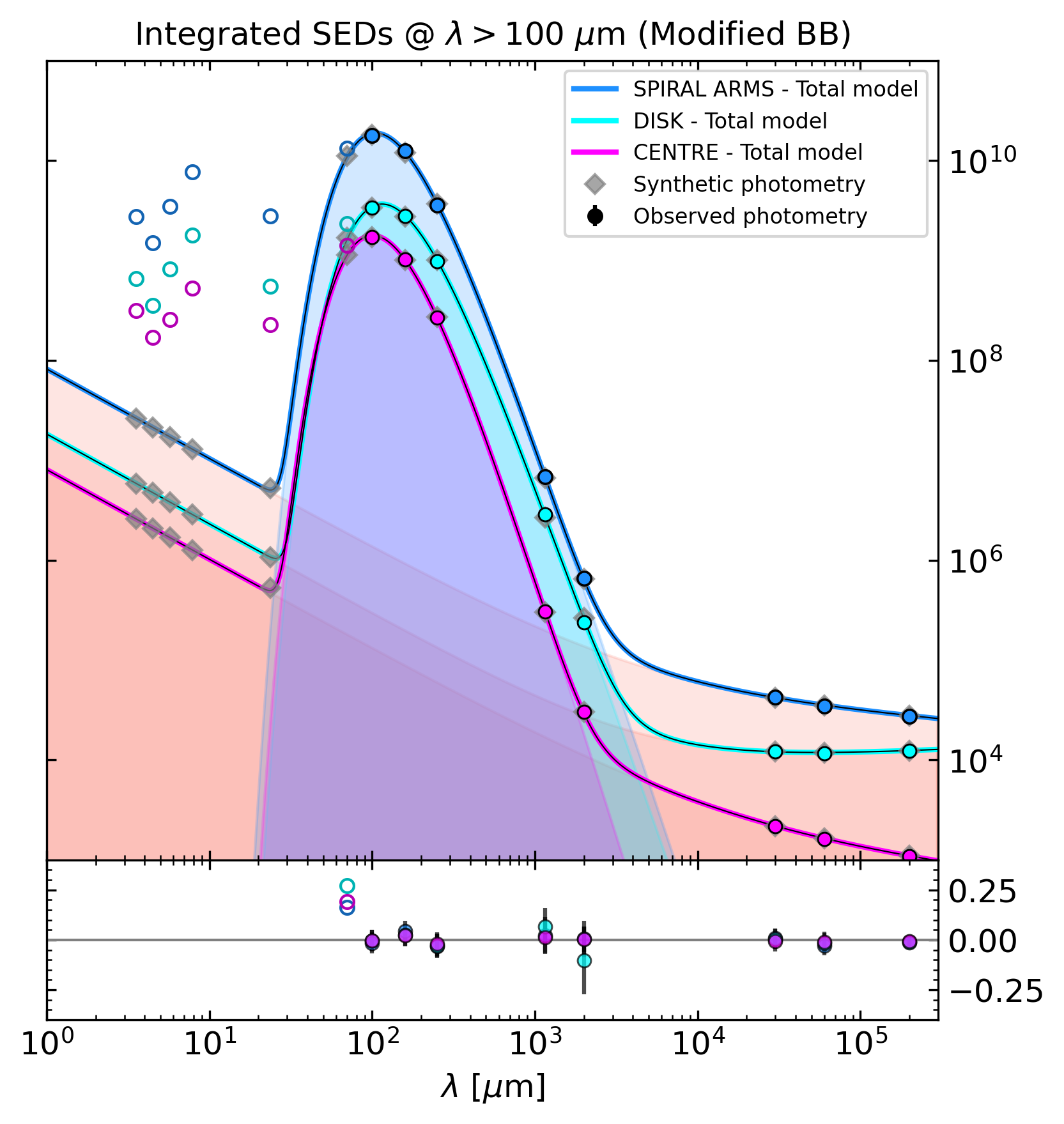}
\caption{
NIR-to-radio SED modelling of M~99's spiral arms (blue), disc (cyan), and centre (magenta) with \texttt{HerBIE}. Filled circles show observed photometry; grey diamonds indicate synthetic photometry. \textit{Left panel}: full SED fits (solid lines) including stellar emission (\texttt{starBB}, yellow areas), dust emission using \texttt{powerU} with \themis (blue, cyan, magenta areas), and radio continuum (\texttt{radio}, red areas). \textit{Right panel}: fits restricted to $\lambda \ge 100$~\upmicron, modelling dust with single-T \mbb (\texttt{MBB1}, coloured areas) and excluding stellar emission. Empty circles were not included in the fit. Monochromatic luminosities ($\nu\,$L$_\nu$) assume a distance of 14.4 Mpc (see Table~\ref{tab:NGC4254prop}). Residuals are shown below each panel.}
\label{fig-ENVs-SED}
\end{figure*}

We integrate the flux densities of each component using \texttt{HIP} (Appendix~\ref{App:hip}) and fit the resulting integrated SEDs with \texttt{HerBIE} \citep{Galliano2018MNRAS.476.1445G}, as detailed in Sect. \ref{Sect:method}. The multi-wavelength fluxes for the centre, spiral arms, and disc of M~99 are reported in Table~\ref{tab:env_photometry}. Their sum is consistent with the integrated photometry in Table~\ref{tab:integrated_photometry}.

Figure \ref{fig-ENVs-SED} shows the integrated SEDs and their decomposition into stellar (\themis only), dust, and radio components. Table~\ref{tab:envs_sed_paramvalue} lists the best-fit parameters, and Table~\ref{tab:env_properties} summarises the derived ISM properties. We note that these values average over large ISM volumes (a few to $\sim$10 kpc) and encompass a mix of ISM environments, from molecular clouds and HII regions to diffuse media.

\begin{table}
    \caption[]{Best-fit parameter values obtained with \texttt{HerBIE} (non-hierarchical Bayesian framework) for the three morphological components : disc, spiral arms and galaxy centre.}
\centering
    \begin{tabular}{llccc}
    \hline\hline
    \noalign{\smallskip}
    \textbf{Parameter} &  & \textbf{\themis} & \textbf{\mbb} & \textbf{Units}\\
    \noalign{\smallskip}
    \hline
    \noalign{\smallskip}
    & Disc & $9.7\pm0.8$ & $10\pm2$ & \\ 
    $M_{\rm dust}$ & Arms & $20\pm1$ & $34\pm6$ & $10^6$ M$_\odot$\\ 
    & Centre & $0.70\pm0.04$ & $3.0\pm0.7$  & \\ 
    \noalign{\smallskip}
    \hline
    \noalign{\smallskip} 
    & Disc & $\sim22$ & $22.0\pm0.9$ & \\
    $T_{\rm dust}$ & Arms & $\sim25$ & $23.2\pm0.9$ & K\\    
    & Centre & $\sim30$ & $23\pm1$ & \\
    \noalign{\smallskip}
    \hline
    \noalign{\smallskip}                       
    & Disc & 1.79 & $1.72\pm0.09$ & \\
    $\beta$ & Arms & 1.79 & $1.92\pm0.08$ & \\
    & Centre& 1.79 & $2.27\pm0.09$ & \\
    \noalign{\smallskip}
    \hline
    \noalign{\smallskip}
    & Disc & $16\pm1$ & $-$ & \\ 
    $q_{\rm AF}$ & Arms & $15\pm1$ & $-$ & \%\\ 
    & Centre & $10\pm1$ & $-$ & \\ 
    \noalign{\smallskip}
    \hline
    \noalign{\smallskip}                       
    & Disc & $6\pm1$ & $-$ & \\
    $L_\star$ & Arms & $26\pm5$ & $-$ & $10^{9}$ L$_\odot$\\
    & Centre& $4.0\pm0.5$ & $-$ & \\
    \noalign{\smallskip}
    \hline
    \noalign{\smallskip} 
    & Disc & $3.1\pm0.3$ & $-$ & \\
    ${\langle}{U}{\rangle}$ & Arms & $6.8\pm0.7$ & $-$ & ${\langle}{U}{\rangle}_\odot$\\           
    & Centre & $18\pm2$ & $-$ & \\
    \noalign{\smallskip}
    \hline
    \noalign{\smallskip} 
    & Disc & $13\pm2$ & $14\pm2$ & \\
    $L_{\rm radio}$ & Arms & $52\pm4$ & $61\pm8$ & $10^3$ L$_\odot$ \\
    & Centre & $2.9\pm0.2$ & $3.8\pm0.4$ & \\
    \noalign{\smallskip}
    \hline
    \noalign{\smallskip}
    & Disc & $1.06\pm0.05$ & $1.09\pm0.07$ & \\
    $\alpha_{\rm sync}$ & Arms & $0.82\pm0.03$ & $0.87\pm0.08$ & \\
    & Centre& $0.70\pm0.03$ & $0.8\pm0.1$ & \\
    \noalign{\smallskip}
    \hline
    \noalign{\smallskip} 
    & Disc & $24\pm16$ & $30\pm20$ & \\ 
    $f_{\rm FF}$ & Arms & $9\pm8$ & $35\pm20$ & \%\\ 
    & Centre & $5\pm4$ & $50\pm20$ & \\ 
    \noalign{\smallskip}
    \hline
    \end{tabular} 
\tablefoot{The listed quantities and their associated priors are identical to those in Table \ref{tab:global_sed_paramvalue}}.
\label{tab:envs_sed_paramvalue}
\end{table}

\begin{table}
\caption[]{Integrated ISM properties in the spiral arms, disc and centre of M~99.}
\label{tab:env_properties}
\centering
         \begin{tabular}{llcc}
            \hline\hline
            \noalign{\smallskip}
             \textbf{Quantity} &  & \textbf{Value} & \textbf{Units} \\
            \noalign{\smallskip}
            \hline
            \noalign{\smallskip}
            & Disc & 30 &  \\
            $\Sigma_\star$  & Arms & 150 & M$_\odot\,{\rm pc}^{-2}$ \\ 
            & Centre & 1400 &  \\ 
            \noalign{\smallskip}
            \hline
            \noalign{\smallskip}
             & Disc & 0.002 & \\ 
            $\Sigma_{\rm SFR}$ & Arms & 0.013 & M$_\odot\,{\rm yr}^{-1}\,{\rm kpc}^{-2}$\\ 
            & Centre & 0.08 & \\
            \noalign{\smallskip}
            \hline
            \noalign{\smallskip}
             & Disc & 5 & \\ 
            $\Sigma_{\rm mol}$& Arms & 30 & M$_\odot\,{\rm pc}^{-2}$ \\ 
            &Centre & 170 & \\ 
            \noalign{\smallskip}
            \hline
            \noalign{\smallskip}
             & Disc & $76\pm6$ &  \\
            $\Sigma_{\rm dust}$ & Arms & $114\pm6$ & $\times10^{-3}$ M$_\odot\,{\rm pc}^{-2}$ \\ 
            & Centre & $360\pm20$ &  \\ 
            \noalign{\smallskip} 
            \hline
            \noalign{\smallskip}
             & Disc & 3 & \\
            $\tau_{\rm depl}$& Arms & 2 & Gyr \\ 
            &Centre & 2 & \\
            \noalign{\smallskip}
            \hline
            \noalign{\smallskip}
             & Disc & 0.07 & \\
            sSFR & Arms & 0.09 & ${\rm Gyr}^{-1}$ \\
            &Centre & 0.06 & \\
            \noalign{\smallskip}
            \hline
         \end{tabular}
     \tablefoot{In the order we list: the stellar mass surface density, $\Sigma_\star$; the SFR density, $\Sigma_{\rm SFR}$; the molecular gas surface density, $\Sigma_{\rm mol}$; the dust mass surface density, $\Sigma_{\rm dust}$; the depletion timescale, $\tau_{\rm depl}$; the specific SFR, sSFR. When not given, the uncertainty is dominated by calibration uncertainty, reported in Appendix \ref{App:ancillary_maps}. $\Sigma_{\rm mol}$ and $\tau_{\rm depl}$ are obtained assuming that the CO(1--0) emission outside the EMPIRE masks is zero (cf. Fig. \ref{fig-CO}).} 
\end{table}

The \themis model reproduces the disc emission well ($\chi^2_{\rm red} = 0.6$; Fig.~\ref{fig-ENVs-SED}, left panel), as expected given that it is calibrated on the diffuse ISM of the Milky Way. The \mbb model also provides an excellent fit ($\chi^2_{\rm red} = 0.2$; Fig.~\ref{fig-ENVs-SED}, right panel) and yields $\beta = 1.72 \pm 0.09$, consistent with the \themis results.
In contrast, the performance of \themis degrades in the spiral arms ($\chi^2_{\rm red} = 2.8$) and even more markedly in the central region ($\chi^2_{\rm red} = 9.4$). The corresponding \mbb fits achieve significantly lower reduced $\chi^2$ values (0.15 in the arms and 0.04 in the centre) and reveal a systematic increase in $\beta$ from the disc ($\sim1.7$) to the spiral arms ($\sim1.9$) and the galaxy centre ($\sim2.3$), as summarised in Table~\ref{tab:envs_sed_paramvalue}.
This steepening of $\beta$ is consistent with predictions from dust models and laboratory measurements \citep[e.g.,][]{Kohler2015A&A...579A..15K, Ysard2018A&A...617A.124Y, Demyk2022A&A...666A.192D}, as well as with observed variations of $\beta$ in nearby galaxies at $\lambda = 250-500$~\upmicron \citep[e.g.,][]{Bianchi2022A&A...664A.187B}. 
The \themis~2.0 dust model \citep[][]{Ysard2024A&A...684A..34Y}, which incorporates composition-dependent variations in grain optical properties and allows for $\beta$ values in the range $[1.4,1.9]$ under diffuse ISM conditions, may improve the \themis fit and yield more realistic values of $\beta$. However, at the time of writing, \themis~2.0 has not yet been implemented in \texttt{HerBIE}. Variations in $\beta$ are discussed in more detail in Sect.~\ref{Sect:results_beta}.

Dust temperatures inferred from the \mbb fits in the three regions are consistent within uncertainties (Table \ref{tab:envs_sed_paramvalue}), whereas \themis shows a stronger gradient: $T_{\rm dust} \sim30$ K in the centre, $\sim25$ K in the arms, and $\sim22$ K in the disc. This trend could partly arise from the shorter-wavelength data included in the \themis fit or from $\beta-T_{\rm dust}$ degeneracy. However, similar radial decreases of dust temperature have been reported for other face-on spirals \citep{Tailor2025A&A...701A..74T, Marsh2017MNRAS.471.2730M, Smith2016MNRAS.462..331S}. Assuming the centre and inner arms dominate the emission at small galactocentric radii, and the disc emission dominates at large radii,  our results are consistent with these findings.

\begin{figure*}[!ht]
\centering
\includegraphics[width=.95\columnwidth]{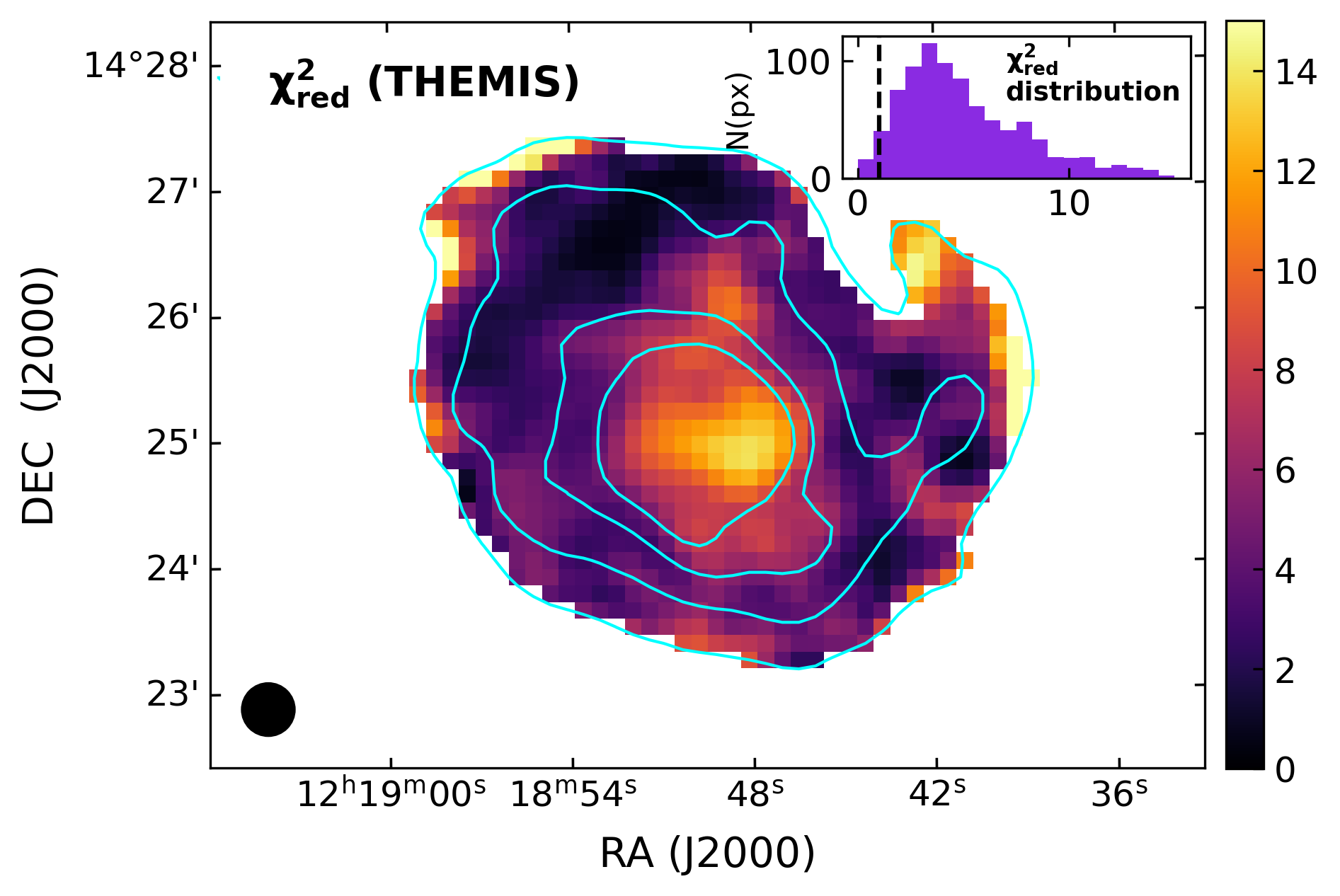}
\includegraphics[width=.94\columnwidth]{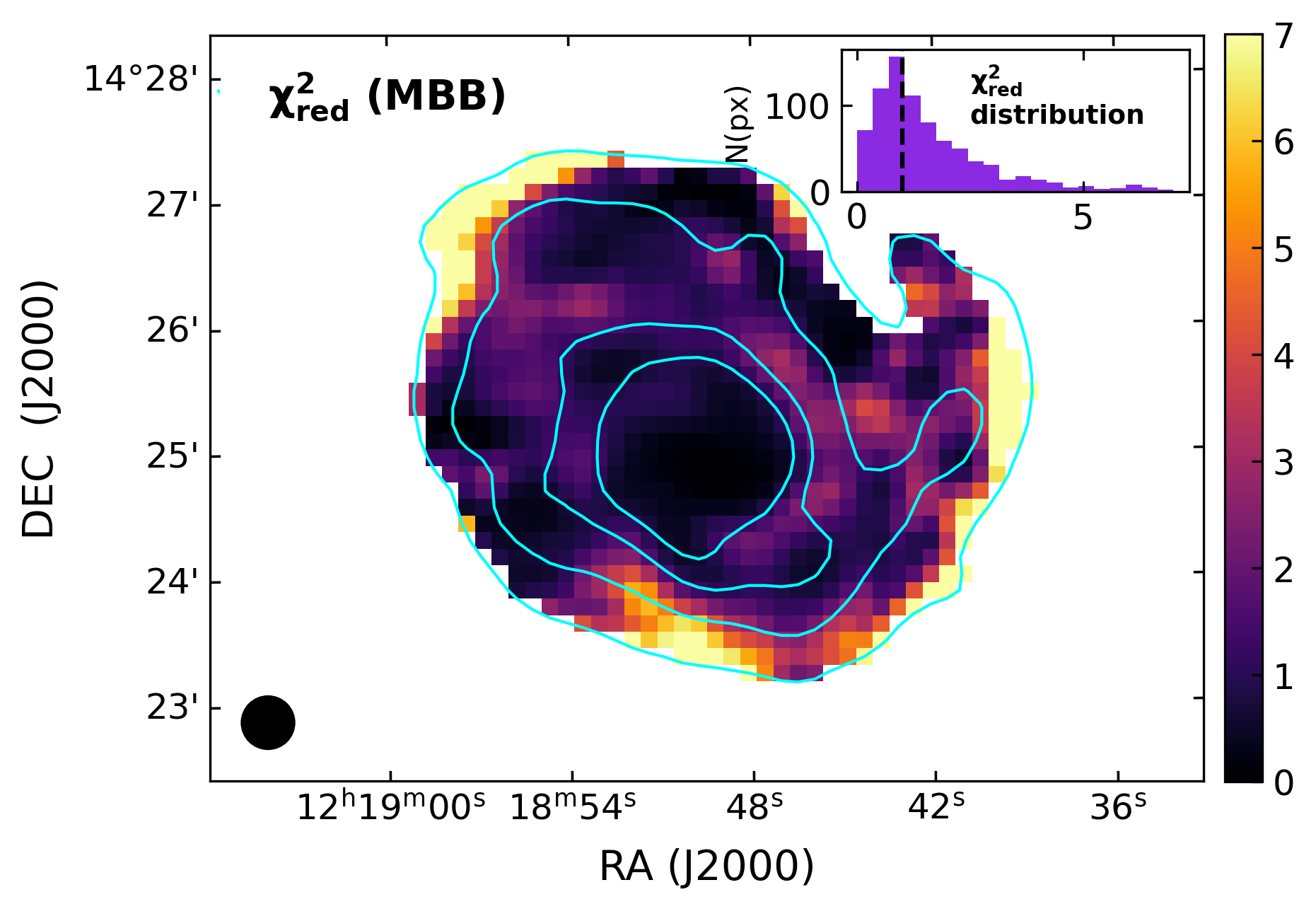}
\caption{Pixel-by-pixel $\chi^2_{\rm red}$ maps at $25^{\prime\prime}$ resolution; bottom left corner). \textit{Left panel}: $\chi^2_{\rm red}$ from the NIR-to-radio SED fitting with \themis. \textit{Right panel}: $\chi^2_{\rm red}$ from the FIR-to-radio ($\lambda \geq 100$~\upmicron) SED fitting with a single-T \mbb. Insets show the $\chi^2_{\rm red}$ distribution: the black dashed vertical lines indicate a $\chi^2_{\rm red}=1$. Cyan solid lines represent the SPIRE 350~\upmicron contours at $[5, 15, 35, 55]\times\sigma$.}
\label{fig-chi2}
\end{figure*}

The fraction of small grains ($q_{\rm AF}$) anti-correlates with the mean radiation field (${\langle}{U}{\rangle}$, Table \ref{tab:envs_sed_paramvalue}): $q_{\rm AF}$ is $\sim 16$\% in the disc (${\langle}{U}{\rangle} \sim 3 \,{\langle}{U}{\rangle}_{\odot}$), 15\% in the spiral arms (${\langle}{U}{\rangle} \sim 7 \,{\langle}{U}{\rangle}_{\odot}$), and $\sim 10$\% in the centre (${\langle}{U}{\rangle} \sim 18 \,{\langle}{U}{\rangle}_{\odot}$), consistent with destruction of small grains exposed to intense and hard radiation fields \citep[e.g.,][]{Boulanger1988ApJ...332..328B, Puget1989ARA&A..27..161P, Contursi2000A&A...362..310C, Aniano2020ApJ...889..150A}.

The synchrotron spectral index ($\alpha_{\rm sync}$) steepens from $\sim0.7$ in the centre and $\sim0.8$ in the spiral arms to $\sim1.1$ in the disc (\themis; Table~\ref{tab:envs_sed_paramvalue}). This result is consistent with results for M~51 \citep{Gajovic2024A&A...689A..68G, Fletcher2011MNRAS.412.2396F} and other nearby star-forming galaxies \citep{Tabatabaei2013A&A...552A..19T, Basu2015MNRAS.449.3879B}, and agrees within uncertainties with our \mbb-based estimates (Table~\ref{tab:envs_sed_paramvalue}). Variations in $\alpha_{\rm sync}$ are discussed further in Sect. \ref{Sect:alpha_sync}. 

The free–free fraction at 1 cm ($f_{\mathrm{ff}}$) remains poorly constrained but is estimated to be $<10-40\%$, and negligible at longer wavelengths. The \mbb fit allows for higher free-free fractions, up to $\sim50-70$\% (Table \ref{tab:envs_sed_paramvalue}). Our fractions appear systematically lower than those reported from radio SED studies of isolated star-forming regions \citep[$\sim50-80\%$ on scales $\lesssim1$ kpc;][]{Linden2020ApJS..248...25L, Dignan2025ApJ...988..216D}, likely reflecting differences in spatial resolution and/or observational sensitivity.

As shown in Table \ref{tab:env_properties}, the disc has the lowest dust ($\Sigma_{\rm dust}$) and molecular gas ($\Sigma_{\rm mol}$) surface densities. Star formation is therefore concentrated in the arms and centre. The SFR surface density ($\Sigma_{\rm SFR}$) in the disc is about 20\% of that in the arms and only 2.5\% of that in the centre. The arms and centre also have shorter molecular gas depletion times ($\tau_{\rm depl}$), implying higher star formation efficiencies. The specific SFR (sSFR $= \Sigma_{\rm SFR}/\Sigma_{*}$) reaches its highest values in the spiral arms and its lowest values in the centre, highlighting the arms as the main sites of recent star formation. We note that $\Sigma_{\rm SFR}$ in the centre may be overestimated due to known biases in the FUV and 24~\upmicron tracers (Appendix~\ref{App:ancillary_maps}).

\subsection{The spatially-resolved properties of dust in M~99}
\label{sect:spatially-res-anal}

We use the full capabilities of \texttt{HerBIE}’s hierarchical Bayesian framework \citep{Galliano2018MNRAS.476.1445G} to model the dust SED of M~99 on a pixel-by-pixel basis. External parameters (gas mass, stellar mass, SFR, gas-phase metallicity) were determined from our ancillary data (Appendix \ref{App:ancillary_maps}) and included as priors to improve the recovery of physical correlations between dust properties and the external parameters.

As for the integrated analysis, we fit the SED in each pixel using both the \themis and \mbb models (Sect.~\ref{Sect:method}), with the \mbb fit restricted to wavelengths $\lambda \ge 100$~\upmicron. We initialised the hierarchical Bayesian run with the best-fit parameters obtained from a fit via classic $\chi^2$ minimisation. We ran \texttt{HerBIE} with $10^{5}$ iterations and a burn-in of $10^{4}$ iterations. We used the autocorrelation functions (ACFs) to assess the convergence of the MCMC chains, which for most parameters occurs after a few $10^4$ iterations. 
We cross-checked our best-fit models by comparing the 850~$\upmu$m and 1.38~mm flux densities with predictions from a neural network trained on IR and sub-mm data \citep{Paradis2024A&A...691A.241P}, as described in Appendix~\ref{App:val_Deb}. Fig.~\ref{fig-chi2} shows the $\chi^2_{\rm red}$ maps of our fits.
In the following analysis, we exclude pixels with low signal-to-noise (S/N $<3$) in the NIKA2 maps (see Appendix \ref{App:masked_pixels}). 
\begin{figure}[!h]
\centering
\includegraphics[width=1\columnwidth]{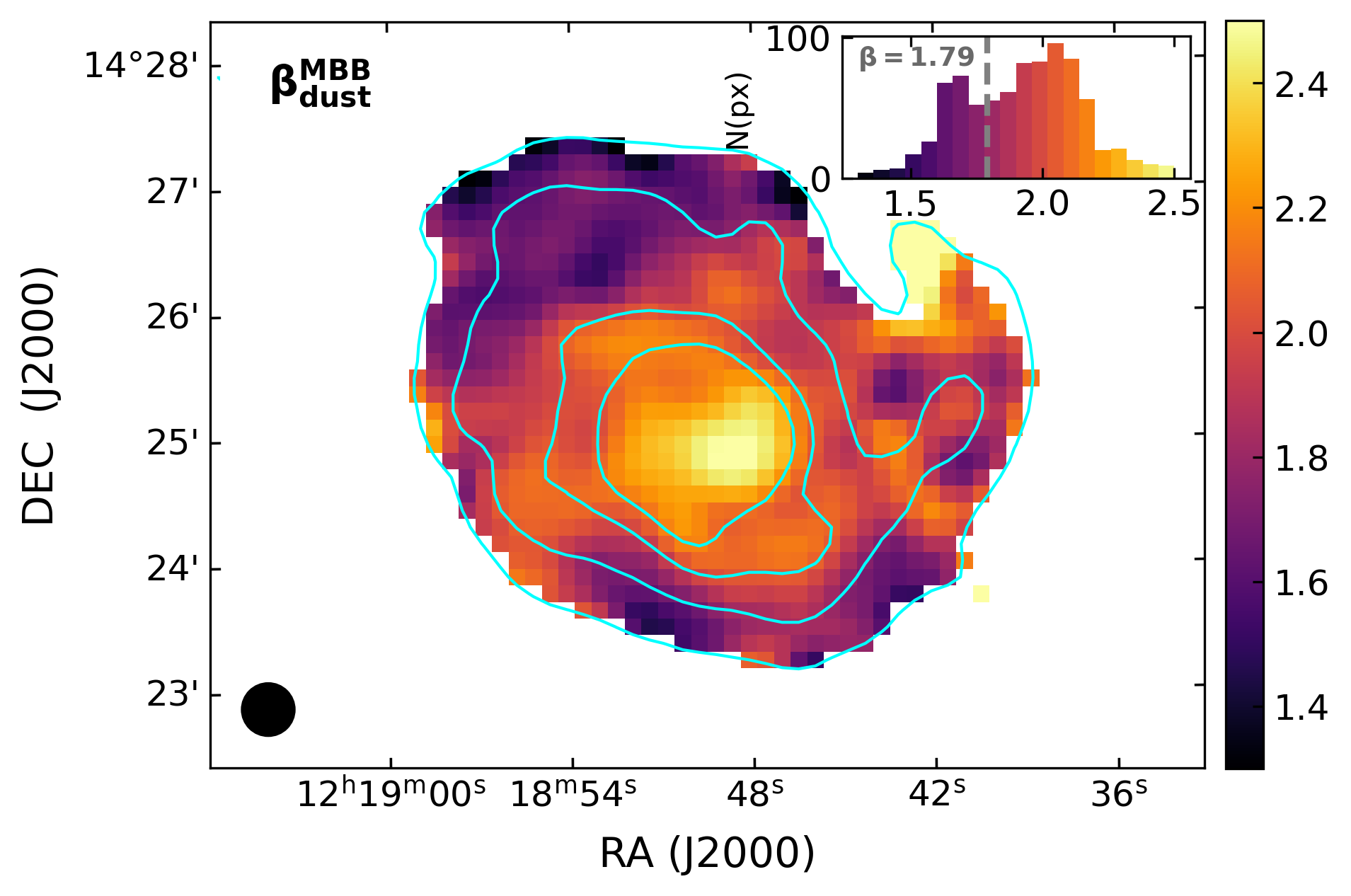}
\caption{Dust spectral index ($\beta$) map derived from spatially-resolved SED fitting at $25^{\prime\prime}$ resolution for $\lambda \geq 100$~\upmicron, using a single-T \mbb model. The inset shows the $\beta$ distribution; the grey dashed line marks the \themis value ($\beta = 1.79$). SPIRE 350~\upmicron contours at $[5, 15, 35, 55] \times \sigma$ are overlapped in cyan.}
\label{fig-beta}
\end{figure}

\subsubsection{Spatially-resolved variations of the dust spectral index}\label{Sect:results_beta}

\themis fits show systematically elevated $\chi^2_{\rm red}$ values, peaking at $\sim4$ and exceeding 10 in the central regions (left panel of Fig.\ref{fig-chi2}), mainly due to residuals at FIR wavelengths. The \mbb fits produce $\chi^2_{\rm red} \sim 1$, with higher values confined to low S/N in the galaxy outskirts (Fig.\ref{fig-chi2}, right panel). This suggests that the fixed $\beta$ value assumed by \themis ($\beta = 1.79$) may be inadequate, especially in the central regions.
\begin{figure*}[!h]
\centering
\includegraphics[width=.98\columnwidth]{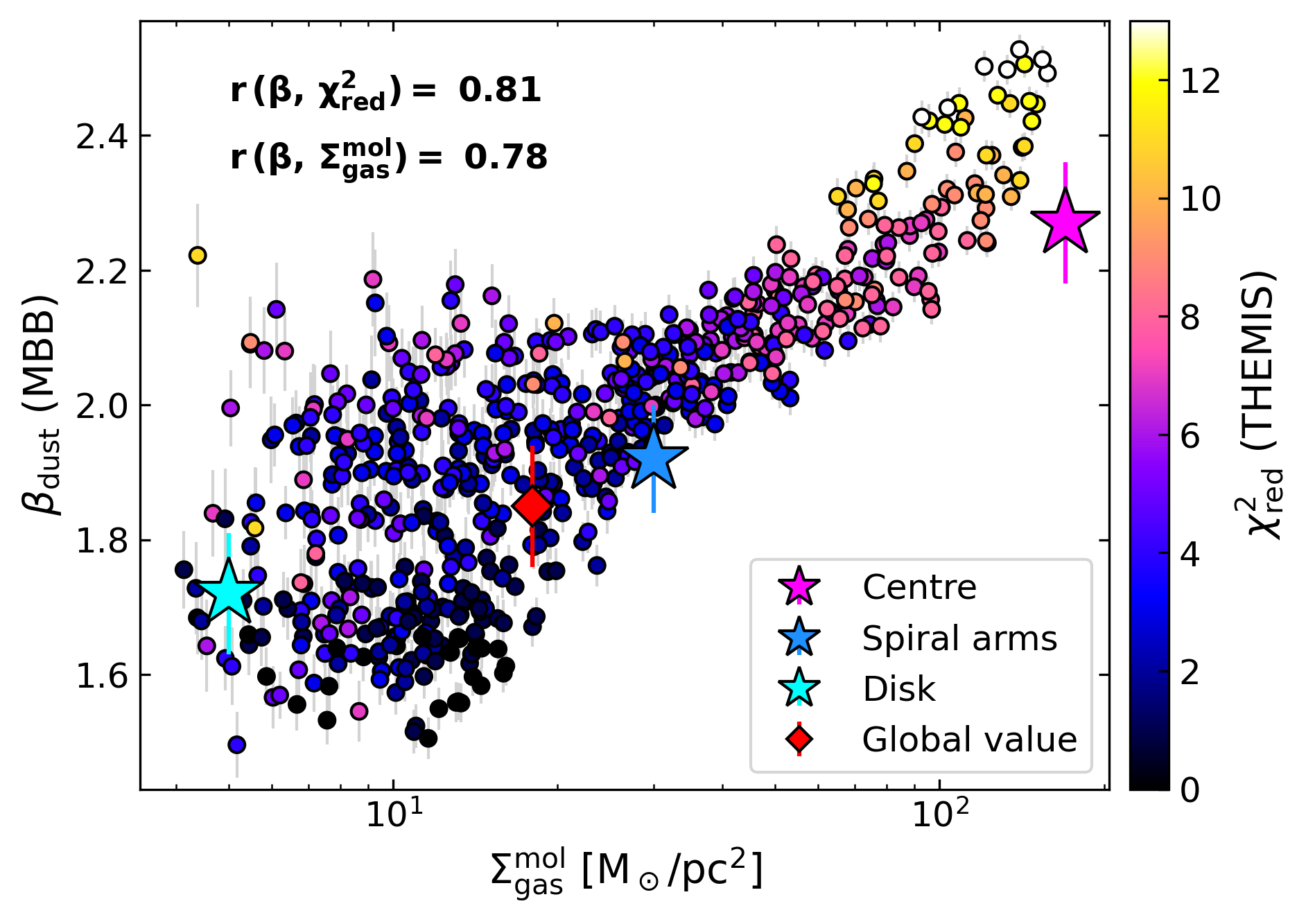}
\includegraphics[width=1\columnwidth]{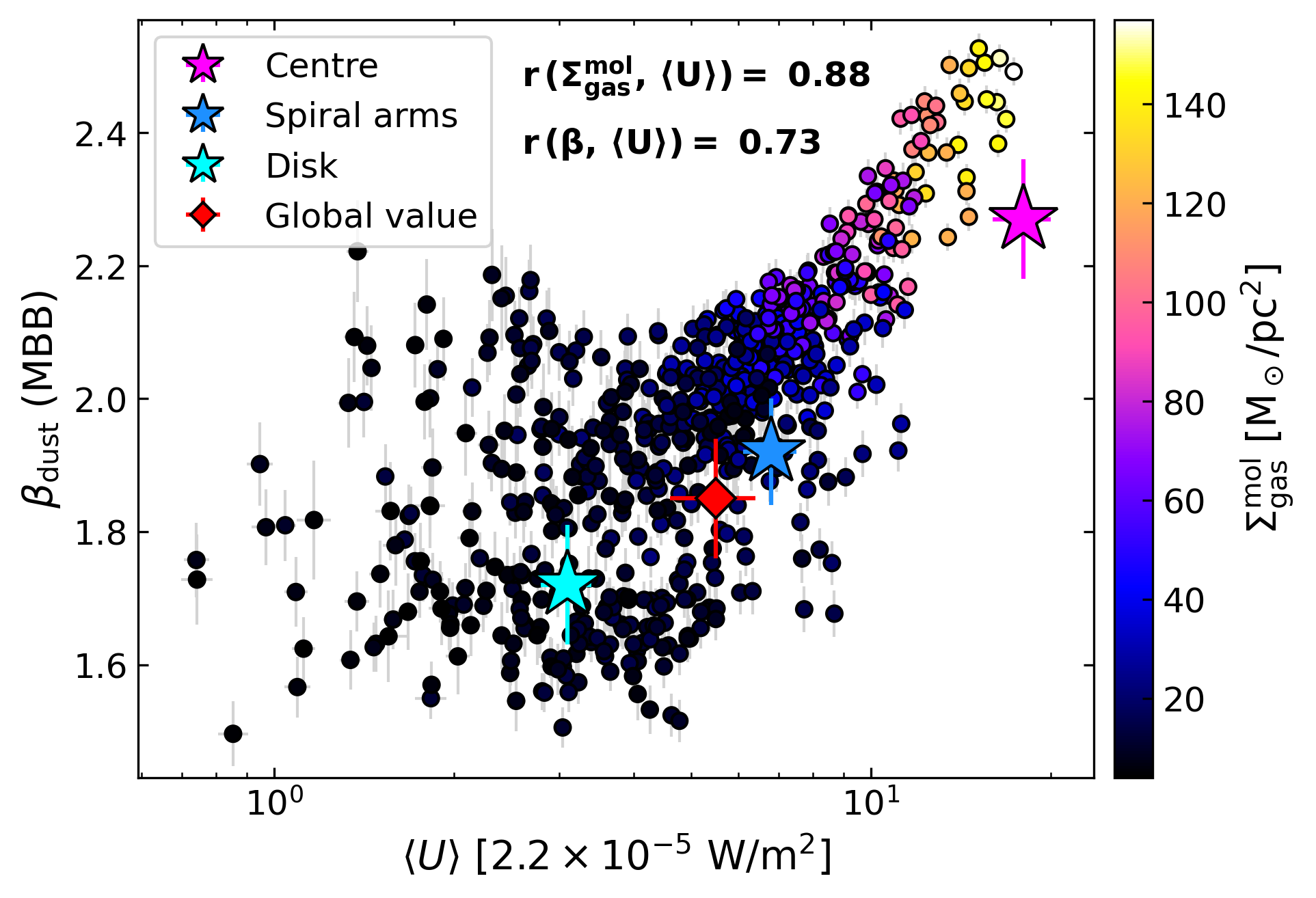}
\caption{Dust spectral index $\beta$ from the pixel-by-pixel single-$T$ \mbb fit vs. molecular gas surface density, $\Sigma_{\rm gas}^{\rm mol}$ (left panel), and average ISRF, $\langle U \rangle$ (right panel). Each point represents an $8^{\prime\prime} \sim 560$ pc pixel, colour-coded by \themis $\chi^2_{\rm red}$ (left) or $\Sigma_{\rm gas}^{\rm mol}$ (right). Analysis is limited to pixels within the CO(1--0) mask (Fig.~\ref{fig-CO}); angular resolution is $25^{\prime\prime}$. Pearson correlation coefficients ($r$) are shown at the top. Filled stars mark the centre (magenta), spiral arms (blue), and disc (cyan) regions (Sect.~\ref{sect:centre-arms-disc}); the red diamond indicates integrated values over the whole galaxy (Sect.~\ref{Sect:globalSED}).}
\label{fig-beta_chi2_U_scatter}
\end{figure*}

The spatial distribution of $\beta$ from our \mbb fit (Fig.~\ref{fig-beta}) mirrors the structure in the \themis $\chi^2_{\rm red}$ map: higher $\beta$ values coincide with poor \themis fits. The $\beta$ distribution is bimodal\footnote{A Hartigan's dip test with $N=100$ gives a dip statistic of 0.12 and a p-value of 0.045. 
A dip value $\gtrsim 0.03-0.05$ with p-value $< 0.05$ is typically considered strong evidence against unimodality \citep{Hartigan1985}.}, with peaks near $\beta \sim 1.7$ in the outer disc and $\beta \sim 2.1$ in the centre, suggesting a mild radial gradient similar to that reported for M33 \citep[e.g.,][]{Tabatabaei2014A&A...561A..95T}.
Although the typical uncertainties on $\beta$ ($0.05-0.15$) make this trend only marginally significant, the correlation between $\beta$ and the \themis $\chi^2_{\rm red}$ is strong ($r \sim 0.8$)\footnote{All reported correlations in this work are statistically significant, with p-values $\ll10^{-3}$. If provided, the uncertainty on the Pearson correlation coefficient $r$ is estimated using the MCMC method within the hierarchical Bayesian framework.}.

The $\beta$ values determined from our \mbb fit are also tightly correlated with the molecular gas surface density ($\Sigma_{\rm gas}^{\rm mol}$; $r = 0.78\pm0.01$): steeper FIR slopes appear in denser regions, echoing our integrated results for the centre, arms, and disc (filled stars overlaid on Fig. \ref{fig-beta_chi2_U_scatter}; see Sect. \ref{sect:centre-arms-disc}). These spatial variations likely trace intrinsic dust property changes rather than temperature mixing, which would produce the opposite trend \citep[i.e., a lower $\beta$ in dense regions; e.g.,][]{Galliano2018ARA&A..56..673G, Galliano2022HabT.........1G}.
Moreover, at wavelengths around $\lambda \sim 1$ mm and for $T_{\rm dust}>10$ K, we are in the Rayleigh-Jeans regime, where the impact of temperature mixing on the derived $\beta$ is expected to be minimal. We consider this result unlikely to be driven by data-processing artefacts, since the regions with $\beta \gtrsim 2$ correspond to the densest areas, where the S/N ratio is high and missing flux is negligible (see Appendix~\ref{App:large-scale-filtering}).
In cold, dense regions ($A_V \gtrsim 1$), grains may coagulate into larger aggregates and acquire aliphatic-rich amorphous carbon mantles, as well as ice coatings, processes that lead to steeper values of $\beta$ \citep{Ossenkopf1994A&A...291..943O, Ormel2011A&A...532A..43O, Kohler2011A&A...528A..96K, Ysard2016A&A...588A..44Y, Carpine2025A&A...698A.200C}.

The dust spectral index further correlates with the mean ISRF intensity, $\langle U \rangle$ ($r = 0.7\pm0.2$; right panel of Fig.~\ref{fig-beta_chi2_U_scatter}). Enhanced $\langle U \rangle$ promotes photo-processing and mantle erosion, which shift the grain population toward silicate-rich compositions with steeper emissivity\footnote{Silicate grains with no or thinner carbonaceous mantles have steeper FIR emissivity if exposed to the same ISRF \citep{Kohler2014A&A...565L...9K}.} ($\beta \gtrsim 2$), while low-$\langle U \rangle$  regions retain less-processed carbonaceous mantles and show flatter $\beta$ ($\sim 1.6 - 2$). These effects have been extensively discussed in dust evolution models \citep{Jones2013A&A...558A..62J, Kohler2015A&A...579A..15K, Ysard2016A&A...588A..44Y} and are supported by observational studies linking radiation field intensity to variations in dust emissivity and grain structure \citep[e.g.,][]{Compiegne2011A&A...525A.103C, Galliano2018ARA&A..56..673G}. 
The broad dispersion in $\beta$ observed at low $\langle U \rangle$ may also arise from a distinct population of large carbonaceous grains, whose optical properties are inherently broader.
At higher $\langle U \rangle$, this population becomes more homogenised or undergoes significant photo-destruction, increasing the dominance of silicates to the observed $\beta$.

The radiation field intensity $\langle U \rangle$ is strongly correlated with $\Sigma^{\rm mol}_{\rm gas}$ ($r = 0.877\pm0.003$), suggesting that dense regions also host the strongest radiation fields\footnote{Spatial scales of $\sim$1.75 kpc, corresponding to our working resolution, encompass both dense molecular clouds and HII regions.}. Replacing $\Sigma^{\rm mol}_{\rm gas}$ with total gas surface density yields similar correlations with $\beta$ ($r \sim 0.7$) and with $\langle U \rangle$ ($r \sim 0.9$), as expected from the tight coupling between atomic and molecular gas ($r \sim 0.99$). Overall, the $\beta-\Sigma^{\rm mol}_{\rm gas}-\langle U \rangle$ correlations indicate that steepening of the FIR slope in M~99 is driven by dust processing in dense and strongly irradiated environments.

\subsubsection{Spatial variations of large and small dust grains }\label{Sect:qAF_variations} 

\begin{figure}[!h]
\centering
\includegraphics[width=.99\columnwidth]{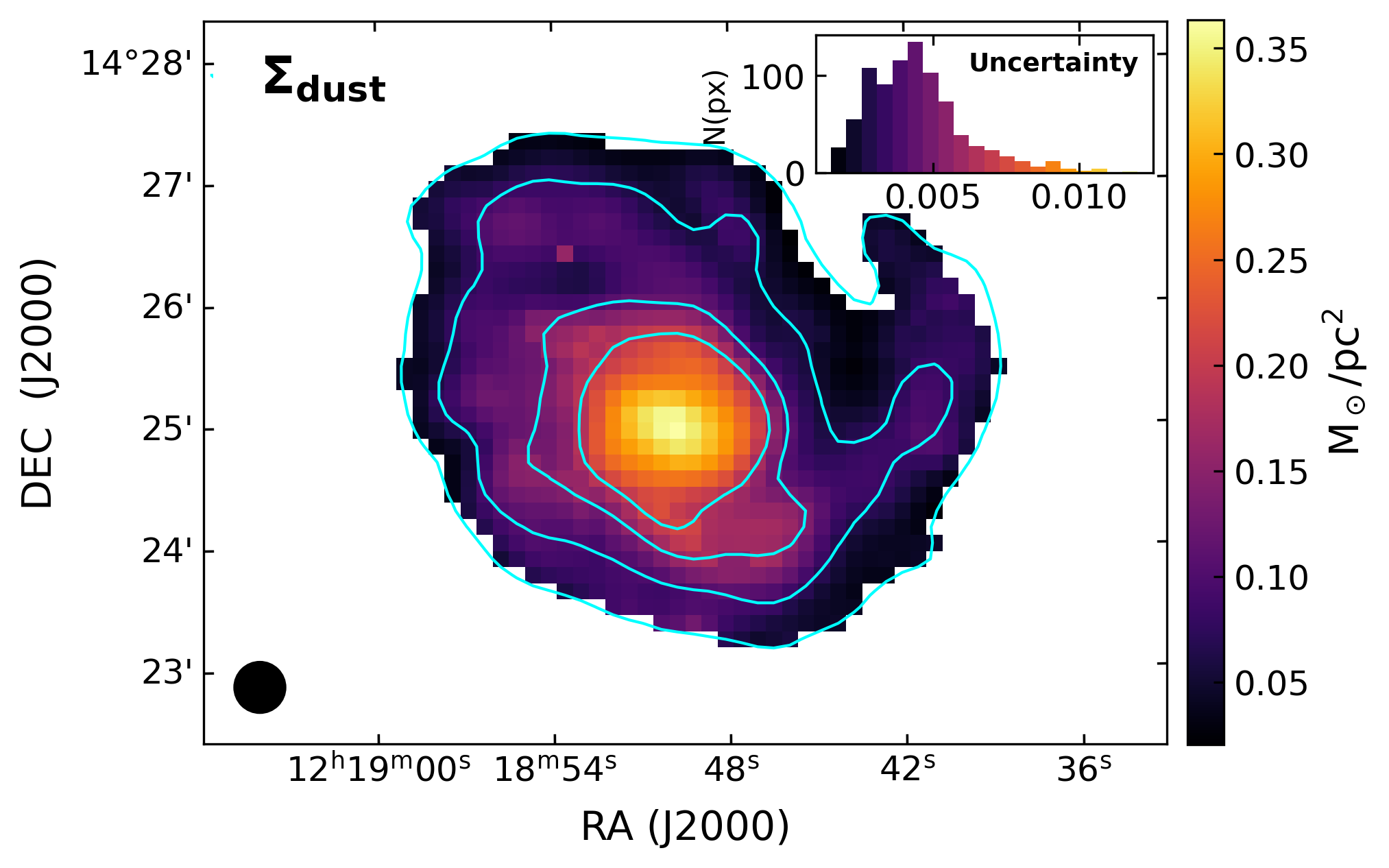}\\
\includegraphics[width=.96\columnwidth]{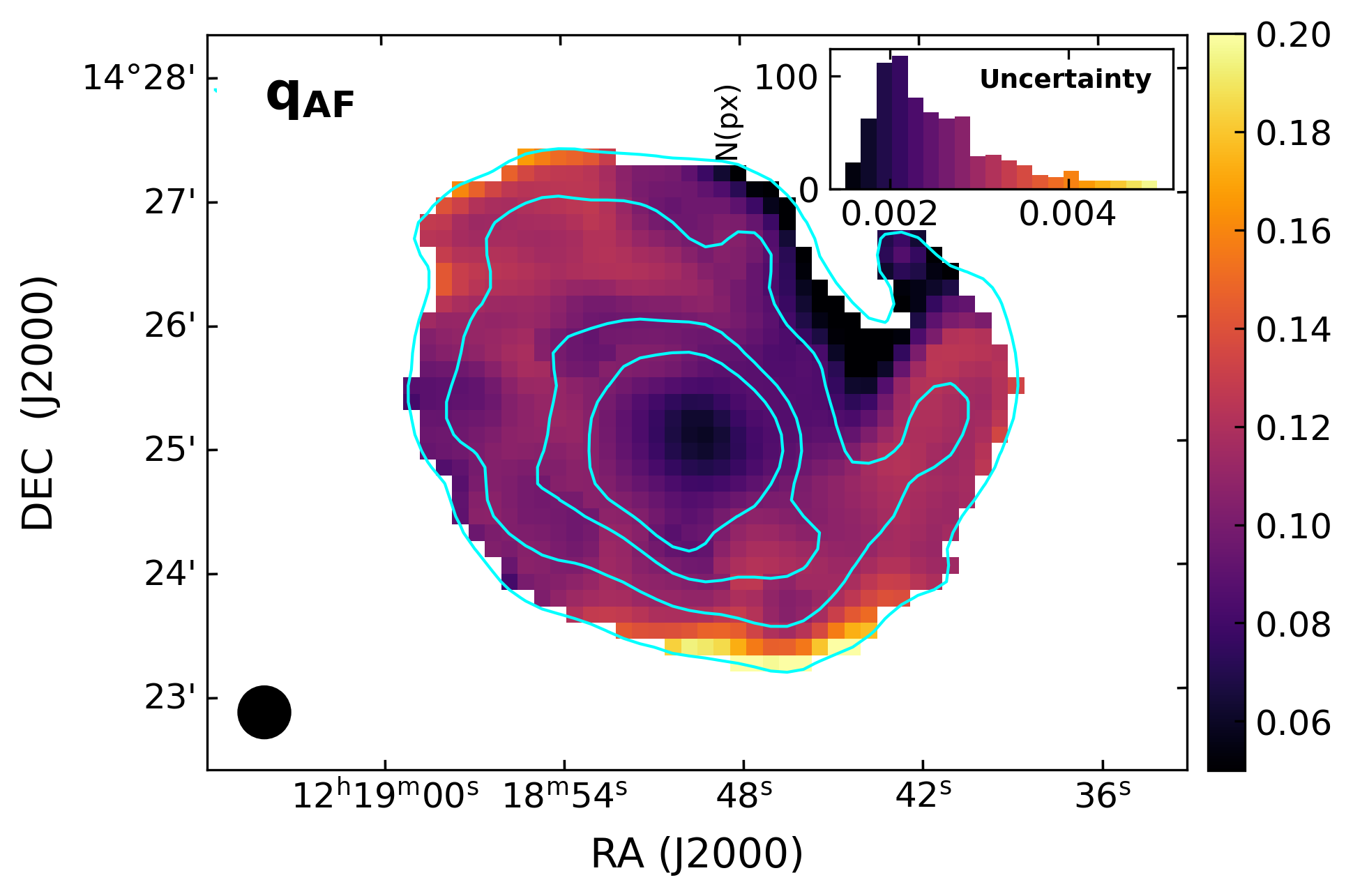}\\
\includegraphics[width=.97\columnwidth]{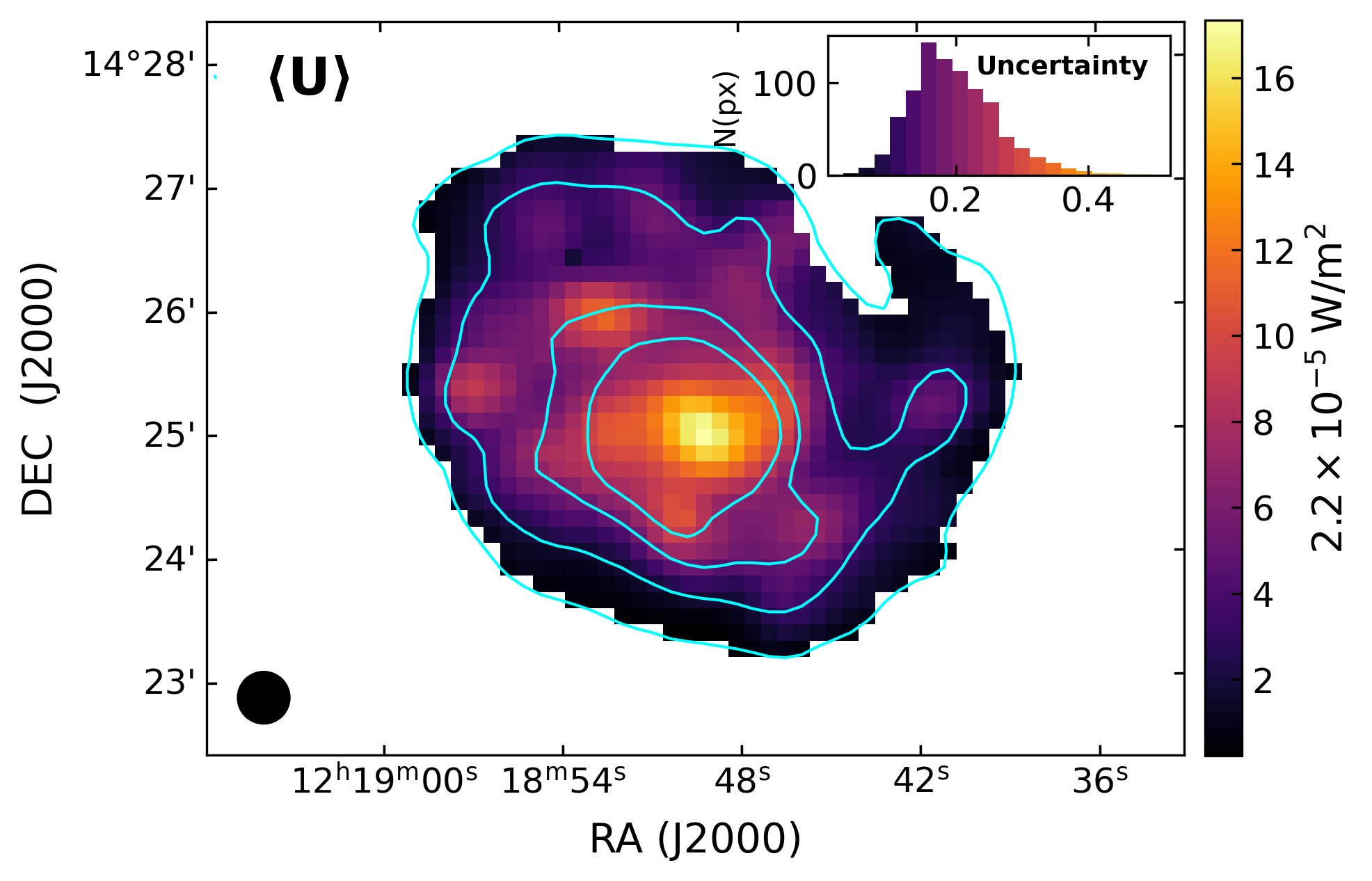}
\caption{Pixel-by-pixel maps at $25^{\prime\prime}$ resolution of dust mass surface density, $\Sigma_{\rm dust}$ (top), small grain fraction, $q_{\rm AF}$ (middle), and average interstellar radiation field, $\langle U \rangle$ (bottom), from the \themis fit. SPIRE 350~\upmicron contours at $[5, 15, 35, 55] \times \sigma$ are overlaid in cyan. Insets show the corresponding uncertainty distributions, colour-coded as in the maps.}
\label{fig-dust_maps}
\end{figure}

It is well established that large grains dominate the FIR emission and the dust mass of galaxies. The \themis dust mass surface density ($\Sigma_{\rm dust}$) of M~99 (Fig.~\ref{fig-dust_maps}, top panel), with a statistical uncertainty of only $\sim 0.5$\%, shows that the bulk of the dust is concentrated in the central regions and along the spiral arms. Compared to the \themis estimates, the \mbb-derived $\Sigma_{\rm dust}$ is higher by factors of about 4 in the centre, 2 in the spiral arms, and 1.1 in the diffuse disc (median $\sim 1.5$), in agreement with our integrated analyses (Table \ref{tab:envs_sed_paramvalue}).

\begin{figure*}[!h]
\centering
\includegraphics[width=1.01\columnwidth]{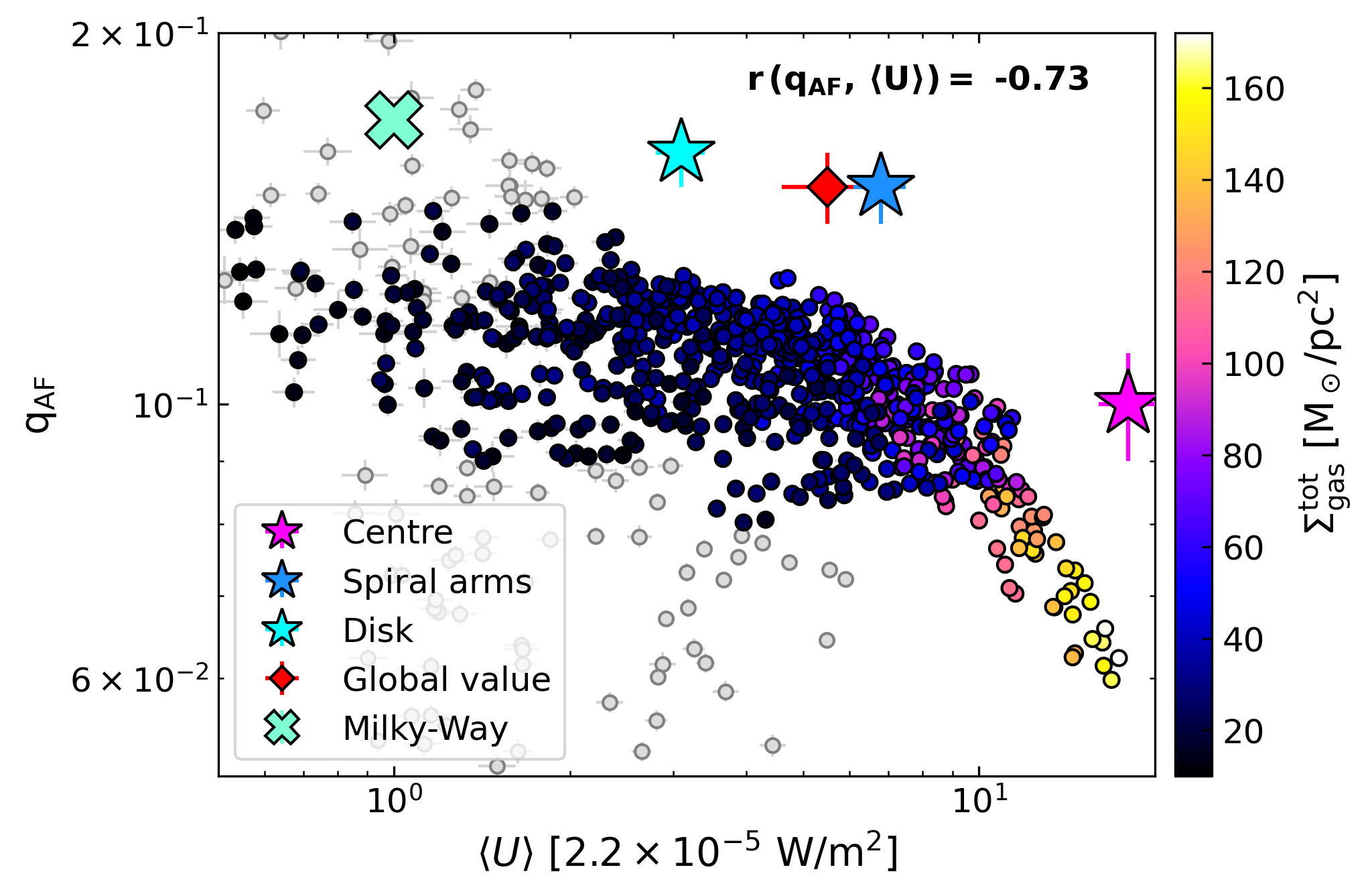}
\includegraphics[width=.99\columnwidth]{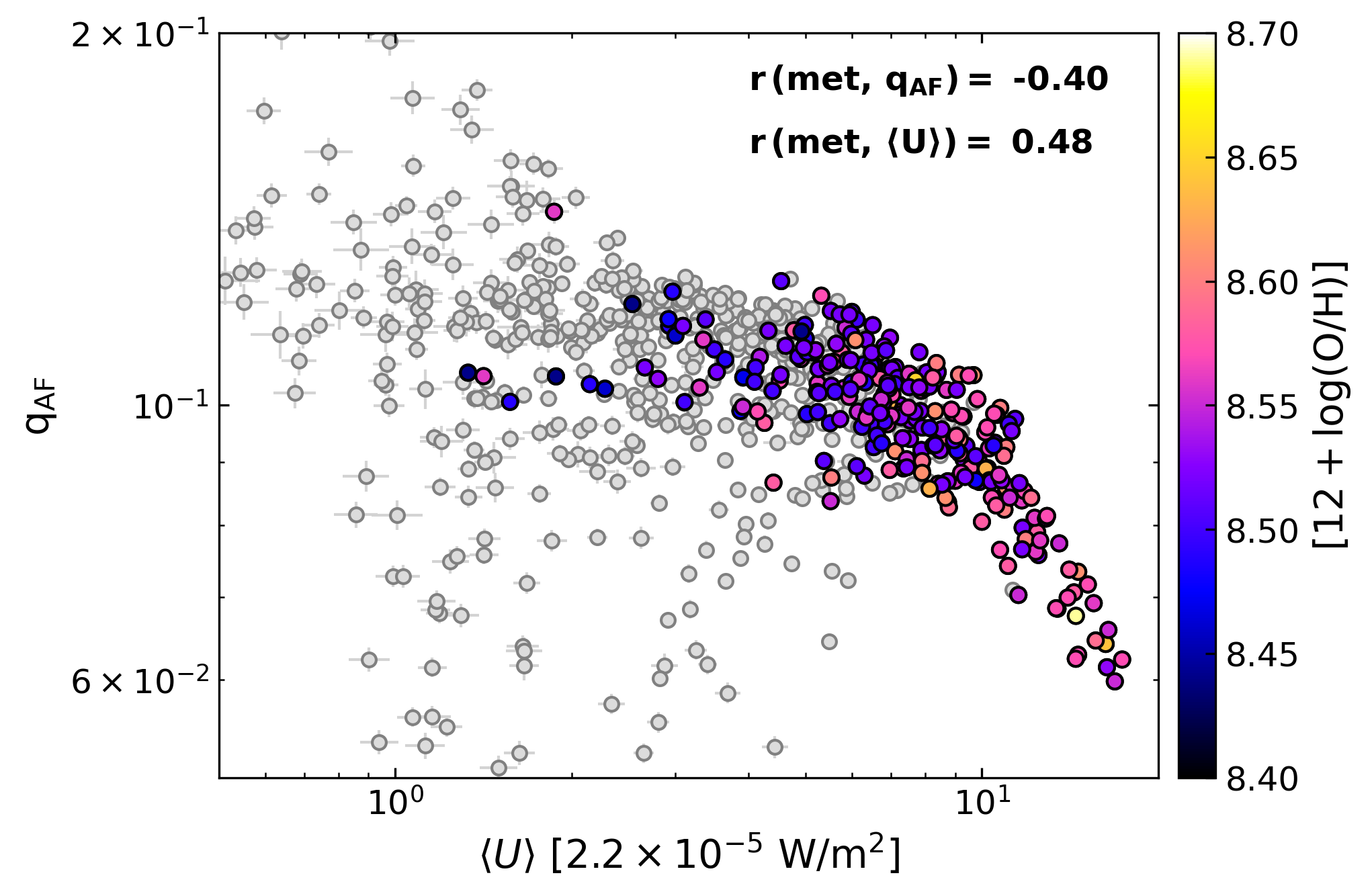}
\caption{Small grain fraction carrying MIR features ($q_{\rm AF}$) versus average interstellar radiation field intensity (${\langle}U{\rangle}$). Each filled circle represents a pixel of $8^{\prime\prime} \sim 560$ pc, while the angular resolution is $25^{\prime\prime} \sim 1.75$ kpc. Pearson correlation coefficients are listed in the top right.
\textit{Left panel}: pixels colour-coded by total gas surface density ($\Sigma_{\rm gas}^{\rm tot}$), a proxy for gas density. For reference, filled stars represent the disc, spiral arms and centre of M~99 and the red diamond the integrated values. 
Grey circles mark low S/N pixels excluded from analysis (Appendix \ref{App:masked_pixels}).
\textit{Right panel}: pixels colour-coded by gas-phase metallicity. Grey circles lie outside the metallicity (cf. Fig. \ref{fig-metallicity}). The green cross indicates the MW ($q_{\rm AF} = 0.17$).
}
\label{fig-qAF_Uav}
\end{figure*}

The fraction of small carbonaceous grains (Fig.~\ref{fig-dust_maps}, central panel) shows the opposite trend: $q_{\rm AF}$ is lowest in the centre and inner arms and highest in the outer disc of M~99, with typical statistical uncertainties of $\sim0.1$\%. Its spatial distribution anti-correlates with $\langle U \rangle$ (Fig.~\ref{fig-dust_maps}, bottom panel), and shows a clear deficiency in regions with enhanced star formation (see Fig.~\ref{fig-SFR}). The Pearson coefficient indicates a strong anti-correlation ($r = -0.733\pm0.009$; Fig. \ref{fig-qAF_Uav}, left panel) between q$_{\rm AF}$ and ${\langle}{U}{\rangle}$, while both correlate with total gas surface density in opposite directions. Small grains appear to be depleted in dense regions with strong radiation fields, whereas the diffuse outer regions of M~99 retain larger fractions of small grains.

Our results are consistent with previous studies of nearby galaxies \citep[][]{Chastenet2025ApJS..276....2C,Katsioli2023A&A...679A...7K, Katsioli2026MNRAS.tmp...78K}. 
Small carbonaceous grains are efficiently photo-destroyed by hard UV photons \citep[e.g.,][]{Luan1988BAAS...20..983L, Luan1990NASCP3084..108L, Cesarsky1996A&A...315L.309C, Madden2006A&A...446..877M, Galametz2013MNRAS.431.1596G, Galametz2016MNRAS.456.1767G, Galliano2018ARA&A..56..673G, Katsioli2023A&A...679A...7K}, and small grains regardless of composition are readily destroyed by SNII shock waves through kinetic sputtering \citep[e.g.][]{Draine1979ApJ...231..438D, Tielens1994ApJ...431..321T, Dwek1992ARA&A..30...11D, Borkowski1995ApJ...454..254B,Foster1993AAS...183.1404F, Hu2019MNRAS.487.3252H}. In the dense molecular clouds associated with star formation, small grains may also be efficiently depleted from the ISM via coagulation onto larger grains\footnote{Because their size ($\lesssim 100$ pc) is much smaller than our 1.75 kpc resolution, molecular clouds are blended with the surrounding irradiated medium and do not appear as a distinct low-$\langle U \rangle$ component.} \citep[e.g.][]{Chokshi1993ApJ...407..806C}.

The spatially resolved $q_{\rm AF}-{\langle}{U}{\rangle}$ trend aligns with our integrated results, although the integrated $q_{\rm AF}$ values are $\sim1.6$ times higher than the average pixel-based value. This difference likely reflects resolution effects, similar to those reported for other scale-dependent quantities such as the SFR \citep[e.g.,][]{Boquien2015A&A...578A...8B}.

Several studies have reported a deficit of aromatic feature carriers in the diffuse ISM of nearby galaxies at sub-solar metallicities \citep[e.g.,][]{Draine2007ApJ...657..810D, Smith2007ApJ...656..770S, Sandstrom2012ApJ...744...20S, Whitcomb2024ApJ...974...20W}. This has commonly been attributed to the destruction of small grains by energetic photons in environments that are both low in metallicity and diffuse \citep[e.g.,][]{Madden2006A&A...446..877M, Gordon2008ApJ...682..336G, Egorov2023ApJ...944L..16E}, where young stars are hotter \citep{Massey2005ApJ...627..477M} and dust attenuation is lower. To check whether this applies to M~99, we use the gas-phase metallicity measurements from \citet{DeVis2019A&A...623A...5D} and \citet{Kreckel2019ApJ...887...80K}. Fig.~\ref{fig-qAF_Uav} (right panel) shows that both $q_{\rm AF}$ and ${\langle}U{\rangle}$ exhibit only weak correlations with metallicity, with correlation coefficients of $r \sim -0.4$ for $q_{\rm AF}$ and $r \sim 0.5$ for ${\langle}U{\rangle}$. These results may partially reflect the limited coverage of our metallicity map (Fig.~\ref{fig-metallicity}), which only samples the central $\sim8$ kpc of M~99. The metallicity has typical uncertainties of $\sim0.1$ dex and exhibits a shallow radial gradient ($\Delta_r \sim 0.3$ dex). We conclude that radiation field intensity, rather than metallicity, is the main driver of small grain depletion in M~99, with metallicity possibly affecting only the diffuse outer ISM at low ${\langle}{U}{\rangle}$ and $q_{\rm AF}$. Extending the metallicity coverage would better constrain this relation.

\subsubsection{Spatial variations of the synchrotron spectral index}\label{Sect:alpha_sync}

The synchrotron spectral index ($\alpha_{\rm sync}$) traces the ageing and energy loss of cosmic ray electrons (CREs) accelerated by supernova shocks and propagating through the galactic magnetic field. Other processes, such as inverse Compton scattering, ionisation, and free–free losses, can also influence $\alpha_{\rm sync}$, particularly in dense or strongly irradiated regions \citep[see e.g.][]{longair2011high, Wills1997MNRAS.291..517W}.

\begin{figure}[!h]
\centering
\includegraphics[width=1\columnwidth]{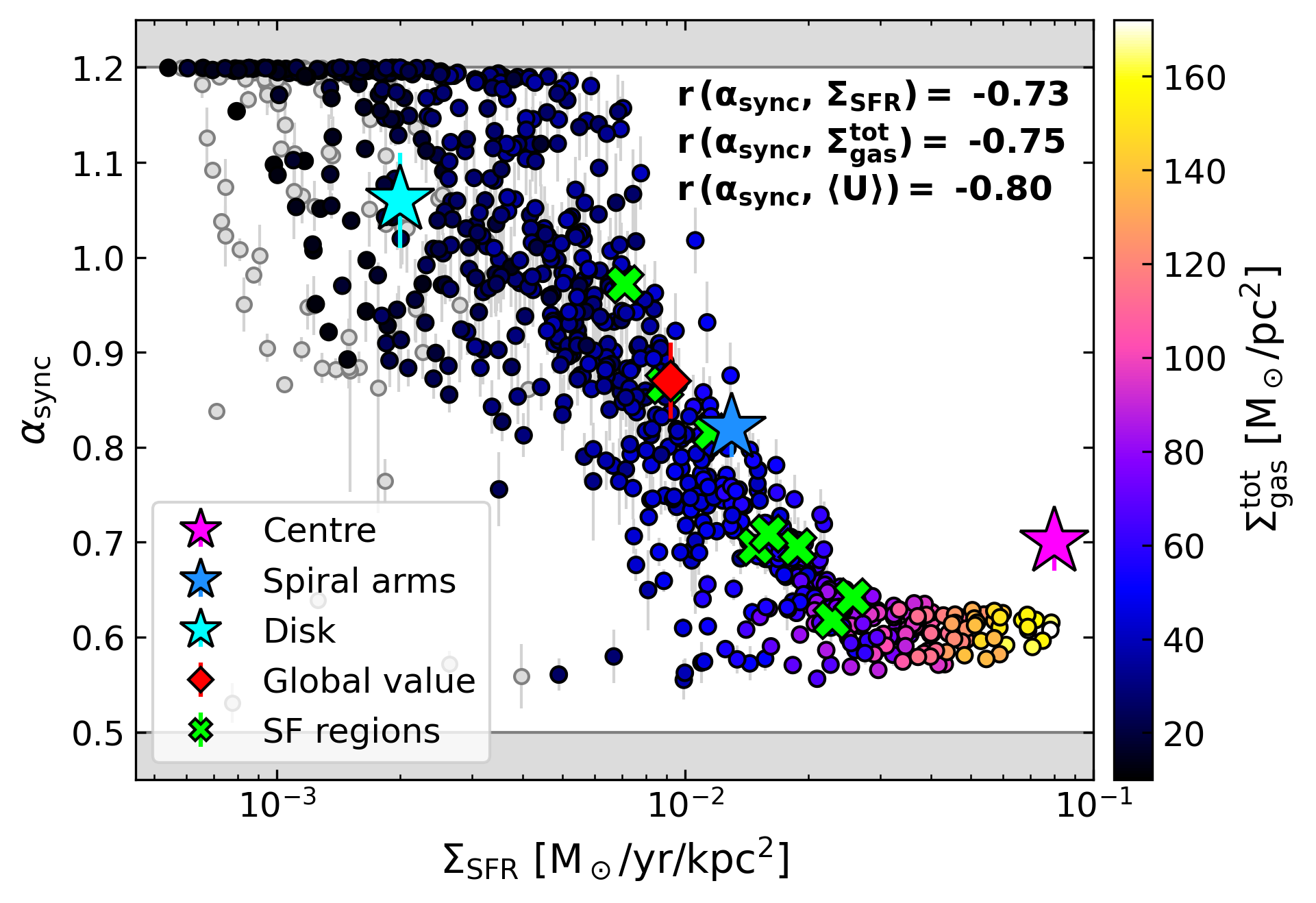}
\caption{Synchrotron spectral index ($\alpha_{\rm sync}$) vs. SFR surface density ($\Sigma_{\rm SFR}$), colour-coded by total gas surface density ($\Sigma_{\rm gas}^{\rm tot}$). The index $\alpha_{\rm sync}$ is constrained to the range [0.5, 1.2], as indicated by the grey shaded areas. Each circle is an $8^{\prime\prime}$ ($\sim 560$ pc) pixel, and the angular resolution is $25^{\prime\prime}$ ($\sim$1.75 kpc). Stars show integrated values for the disc (cyan), arms (blue), centre (magenta), and the red diamond, the global value. Green crosses mark eight SF regions from the sSFR map (Appendix~\ref{App:SF_regions}), measured within $25^{\prime\prime}$ apertures. Pearson coefficients are in the top right. Grey circles are low S/N pixels excluded from the analysis (Appendix~\ref{App:masked_pixels}). 
}
\label{fig-a_radio}
\end{figure}

We examined the variation of $\alpha_{\rm sync}$ relative to $\Sigma_{\rm gas}^{\rm tot}$, $\Sigma_{\rm SFR}$, and ${\langle}{U}{\rangle}$. Fig.~\ref{fig-a_radio} shows our pixel-based measurements of $\alpha_{\rm sync}$ versus $\Sigma_{\rm SFR}$, overlaid with the results of our integrated analysis (Sects.~\ref{Sect:globalSED} and \ref{sect:centre-arms-disc}). Eight regions of enhanced star formation, highlighted by green crosses, are selected from our sSFR map (defined as $\Sigma_{\rm SFR}/\Sigma_\star$; see Appendix~\ref{App:SF_regions}). These measurements integrate over $25^{\prime\prime}$ apertures and likely represent a mixture of emission from HII regions and the surrounding dense and diffuse ISM.

We find strong anti-correlations between $\alpha_{\rm sync}$ and $\Sigma_{\rm SFR}$ ($r = -0.730\pm0.003$), $\Sigma_{\rm gas}^{\rm tot}$ ($r = -0.750\pm0.004$), and ${\langle}{U}{\rangle}$ ($r = -0.803\pm0.004$), across the range $0.5 \lesssim \alpha_{\rm sync} \lesssim 1.2$, corresponding to the prior imposed on $\alpha_{\rm sync}$ during SED fitting (Table~\ref{tab:global_sed_paramvalue}). At low SFR ($\Sigma_{\rm SFR} \lesssim$ 0.01 M$_{\odot}\,{\rm yr}^{-1}\,{\rm kpc}^{-2}$), corresponding to $\Sigma_{\rm gas}^{\rm tot} \lesssim 60$, $\Sigma_{\rm SFR}\lesssim 0.01$ M$_{\odot}\,{\rm yr}^{-1}\,{\rm kpc}^{-2}$ and ${\langle}{U}{\rangle} \lesssim 7 {\langle}{U}{\rangle}_{\odot}$, $\alpha_{\rm sync}$ steepens to $\gtrsim 0.9$ (dark blue and black points in Fig.~\ref{fig-a_radio}). In these lower-density regions, we expect that the injection of fresh CREs is limited and the emission is dominated by older electrons that have diffused through the ISM without significant re-acceleration.

In the central $50^{\prime\prime}\,(\sim3.5$ kpc), where $\Sigma_{\rm SFR} \gtrsim$ 0.01 M$_{\odot}\,{\rm yr}^{-1}\,{\rm kpc}^{-2}$, $\alpha_{\rm sync}$  flattens to $\sim 0.6$, suggesting that CREs are more energetic in regions of high SFR \citep[e.g.][]{Condon1992ARA&A..30..575C, Thompson2006ApJ...645..186T}. The eight candidate star-forming regions (green crosses in Fig.~\ref{fig-a_radio}) exhibit synchrotron spectral indices of $\alpha_{\rm sync} \sim 0.6-0.9$. Most of these regions are located in areas with high $\Sigma_{\rm SFR}$ and $\Sigma_{\rm gas}^{\rm tot}$, where $\alpha_{\rm sync}$ typically lies in the range $\sim0.6-0.7$. These results are in agreement with previous spatially-resolved studies of radio emission in nearby galaxies \citep[e.g.][]{Murphy2006ApJ...651L.111M, Hughes2006MNRAS.370..363H, Basu2015MNRAS.449.3879B, Mulcahy2016A&A...592A.123M, Tabatabaei2017ApJ...836..185T, Tabatabaei2022MNRAS.517.2990T}, where flatter non-thermal spectral indices are interpreted as a combination of recent feedback-driven injection of young CREs and limited CRE diffusion/advection.

\subsubsection{Dust scaling relations}\label{Sect:dust-srel}
Dust is produced in the ejecta of supernovae and asymptotic giant branch (AGB) stars, but its abundance in galaxies is strongly shaped by ISM processes: grains can grow through metal accretion in dense environments or be destroyed by shocks and stellar radiation. As a result, the dust content of galaxies is expected to scale with their stellar mass, SFR, and metallicity.

In this Sect., we examine spatially resolved scaling relations for M~99, focusing on the dust-to-stellar mass ratio (DSR $=\Sigma_{\rm dust}/\Sigma_\star$) and dust-to-gas mass ratio (DGR $=\Sigma_{\rm dust}/\Sigma_{\rm gas}$) as functions of $\Sigma_\star$ and other ISM properties, including SFR, gas-phase metallicity, and molecular gas fraction, $f_{\rm mol}$ (see Appendix \ref{App:ancillary_maps}). The dust mass surface density ($\Sigma_{\rm dust}$) is derived from our \texttt{HerBIE} fits using both the \themis and \mbb models.\\

\textit{The dust-to-stellar mass ratio.} 
An anti-correlation between the DSR and stellar mass is well established in the literature, on integrated scales \citep{Cortese2012A&A...540A..52C, Clemens2013MNRAS.433..695C, Clark2015MNRAS.452..397C, DeVis2017MNRAS.464.4680D, Casasola2020A&A...633A.100C, DeLooze2020MNRAS.496.3668D} and at sub-kiloparsec resolution \citep[e.g.,][]{Viaene2014A&A...567A..71V}. This relation is also reproduced by theoretical models and hydrodynamical simulations \citep[e.g.,][]{Camps2016MNRAS.462.1057C, Calura2017MNRAS.465...54C, Lapi2020ApJ...897...81L}. The proposed explanation is that while both dust and stars are formed in star-forming regions, stellar mass accumulates continuously over time, whereas dust may be destroyed by shocks, radiation, and astration. 

\begin{figure}[!h]
\centering
\includegraphics[width=1\columnwidth]{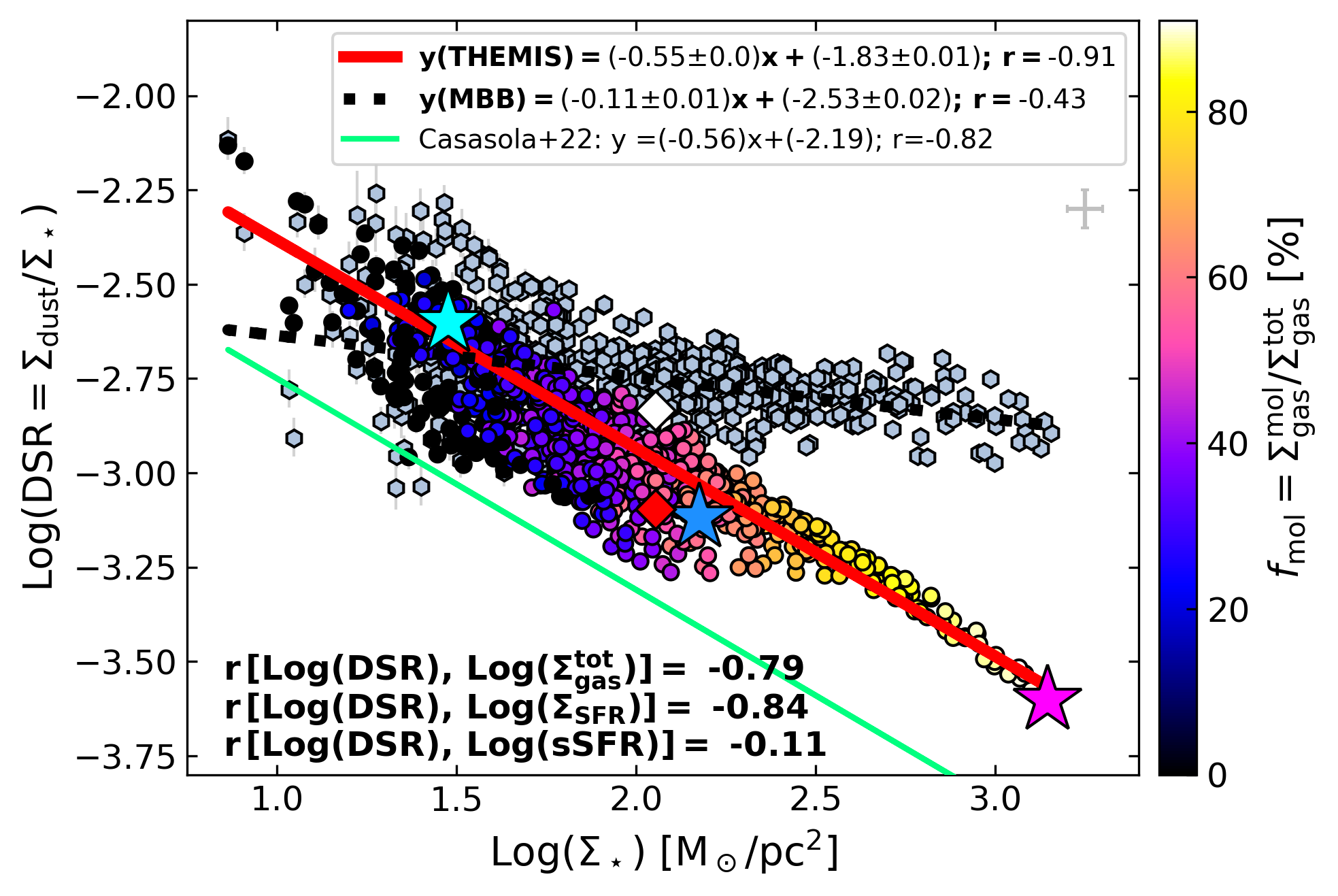}\\
\includegraphics[width=1\columnwidth]{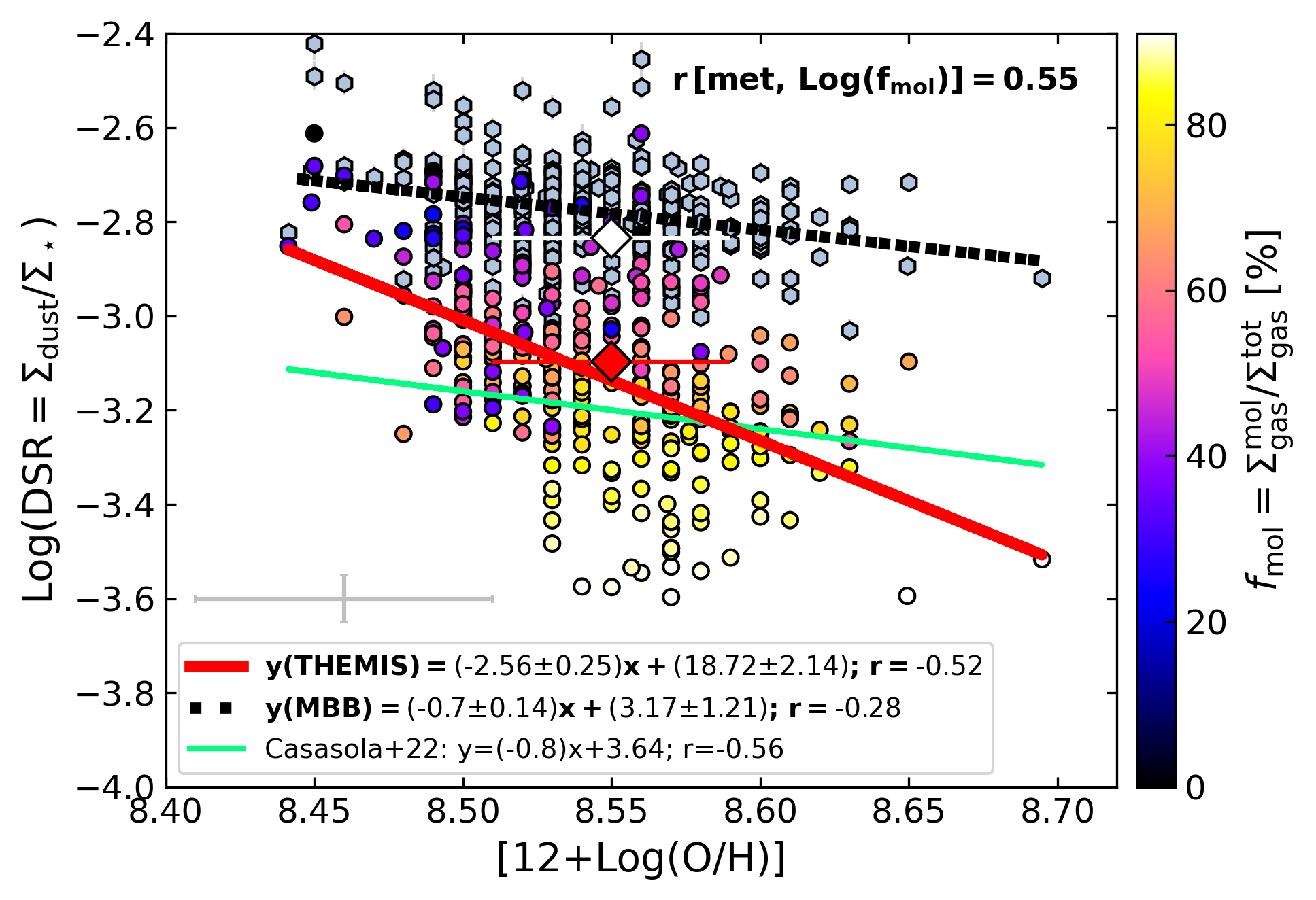}
\caption{Dust-to-stellar mass ratio (DSR) scaling relations with stellar mass surface density ($\Sigma_\star$, top) and gas metallicity (bottom) in M~99. Filled circles show the \themis dust masses, colour-coded by molecular gas fraction ($f_{\rm mol}$); grey hexagons correspond to single-T \mbb dust masses. Each symbol represents an $8^{\prime\prime} \sim 560$ pc pixel. The angular resolution is $25^{\prime\prime} \sim 1.75$ kpc. The bottom panel is restricted to the region covered by the metallicity map (Fig.~\ref{fig-metallicity}). Red solid and black dashed lines show best-fit relations for \themis and \mbb, respectively; green lines show correlations from \citet{Casasola2022A&A...668A.130C}. Red and white diamonds mark integrated values from \themis and \mbb fits, while filled stars in the top panel indicate disc (cyan), spiral arms (blue), and centre (magenta) averages. 
Calibration uncertainties are indicated by grey bars near the legend.
}
\label{fig-DSR}
\end{figure}

We find a strong anti-correlation between the DSR and $\Sigma_\star$ (Fig.\ref{fig-DSR}, top panel) using our \themis fit ($r = -0.909\pm0.004$, slope  $a \sim -0.5$) and a weaker trend for the \mbb ($r \sim -0.4$, $a \sim -0.11$). 
The flatter \mbb slope arises from higher inferred dust masses in dense, $\beta > 1.79$ regions. The DSR also correlates with $\Sigma_{\rm gas}^{\rm tot}$ and $\Sigma_{\rm SFR}$ for \themis ($r = -0.790\pm0.005$ and $r = -0.842\pm0.005$ respectively), but these correlations weaken using the \mbb model ($r = -0.289\pm0.009$ and $r =-0.40\pm0.01$). We do not detect any significant correlation with sSFR (\themis: $r \sim -0.1$; \mbb: $r \sim 0.05$). 

Our DSR-$\Sigma_\star$ relation is broadly consistent with the results of \citet[][green lines in Fig.~\ref{fig-DSR}]{Casasola2022A&A...668A.130C}, who reported a correlation with slope $a = -0.56 \pm 0.02$ and a Pearson coefficient of $r = -0.82$ for a sample of 18 large, face-on spiral galaxies from DustPedia (excluding M~99). Their best-fit relation is offset by approximately 0.2 dex toward lower DSR values compared to ours (Fig.~\ref{fig-DSR}), but still within the reported 1$\sigma$ scatter of 0.22 dex.
Relative to the \themis-based results, the DSR-$\Sigma_\star$ relation inferred from the \mbb modelling shows a more pronounced deviation from the trend of \citet{Casasola2022A&A...668A.130C} at high stellar surface densities ($\Sigma_\star > 10^{2}$ M$_\odot$ pc$^{-2}$), with offsets reaching up to 0.5 dex. However, a closer inspection of Fig.~5 in \citet{Casasola2022A&A...668A.130C} suggests a possible flattening of their relation at $\Sigma_\star > 10^{2}$~M$_\odot$~pc$^{-2}$ once NGC~3031 (M~81) is excluded. This brings their trend into closer agreement with our results.
Overall, the remaining discrepancies in slope likely reflect intrinsic differences between the ISM properties of M~99 and those of the DustPedia galaxies analysed by \citet{Casasola2022A&A...668A.130C}.

Differences in normalisation of the DSR-$\Sigma_\star$ relation may arise from variations in stellar mass calibrations (\citealt{Casasola2022A&A...668A.130C} follow \citealt{Querejeta2015ApJS..219....5Q}) or dust mass estimation methods \citep[e.g.,][see Sect.~5.1 of \citealt{Casasola2017A&A...605A..18C} for details on the DustPedia dust mass estimates]{Nersesian2019A&A...624A..80N, Relano2022MNRAS.515.5306R}. Another contributing factor may be differences in data resolution and spatial sampling (\citealt{Casasola2022A&A...668A.130C} adopted a physical scale of 3.4 kpc per pixel), since these determine the mixing within a resolution element and hence the data's sensitivity to local variations. The absence of a significant correlation between the DSR and sSFR suggests that the observed DSR-$\Sigma_\star$ anti-correlation is primarily driven by M~99's past star formation history (SFH), rather than ongoing star formation. 

The bottom panel of Fig.~\ref{fig-DSR} reveals a moderate anti-correlation between the DSR and gas-phase metallicity (\themis: $r \sim -0.5$; \mbb: $r \sim -0.3$). The metallicity itself correlates with the total gas surface density ($r \sim 0.6$), suggesting that dense, gas-rich regions in the centre of M~99 are more metal-enriched. Thus, the observed trend between DSR and gas-phase metallicity likely reflects a depletion of dust in evolved, metal-enriched, highly irradiated environments with high $\Sigma_\star$. 
We caution that the dust masses inferred from \themis may be underestimated in the densest regions (Sect.~\ref{Sect:globalSED}).\\

\begin{figure}[!h]
\centering
\includegraphics[width=1\columnwidth]{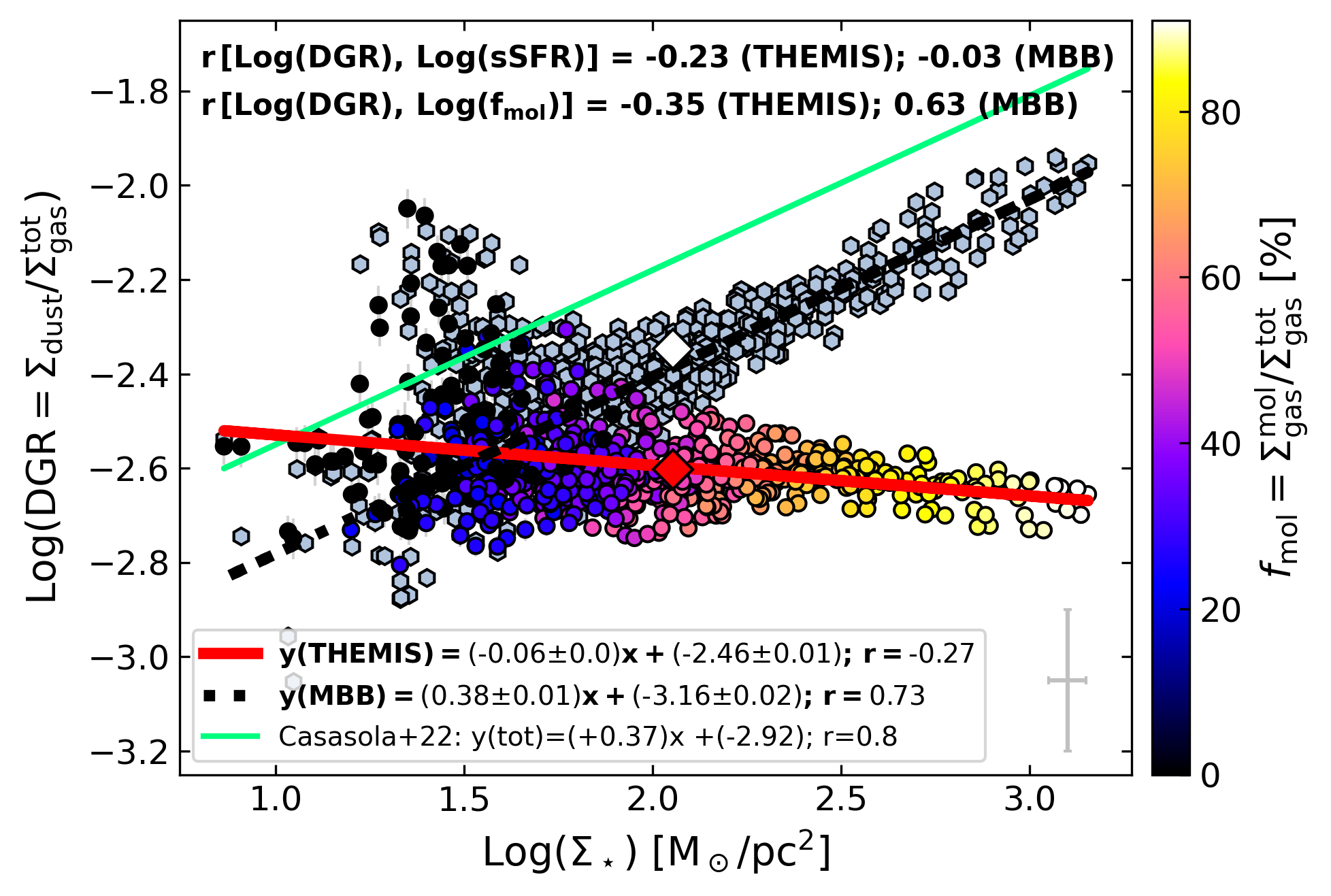}\\
\includegraphics[width=1\columnwidth]{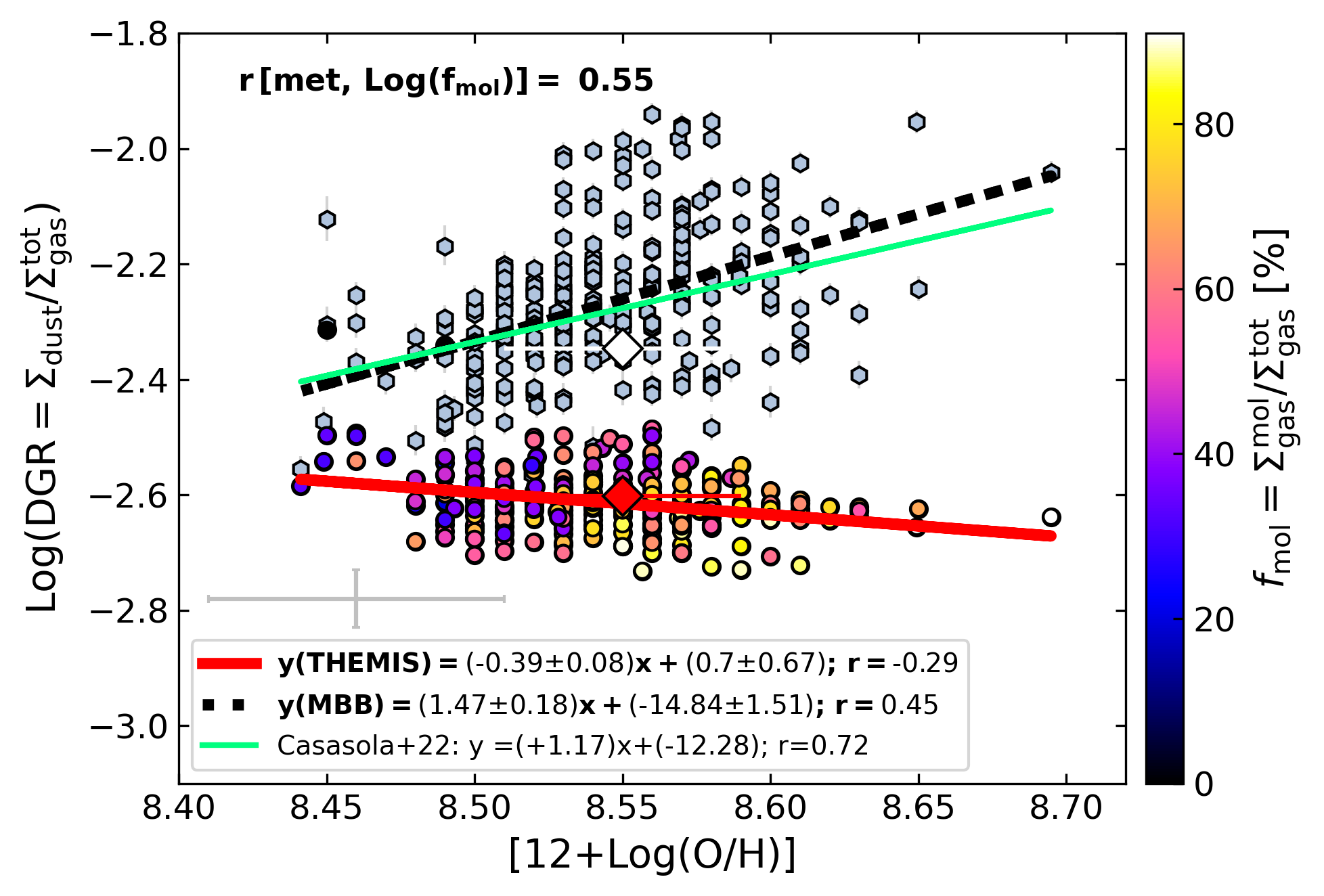}
\caption{Dust-to-(total) gas mass ratio (DGR) scaling relations with stellar mass surface density ($\Sigma_\star$, top) and gas metallicity (bottom) in M~99. Filled circles show \themis dust masses, colour-coded by molecular gas fraction ($f_{\rm mol}$); grey hexagons indicate single-T \mbb dust masses. Each point represents an $8^{\prime\prime} \sim 560$ pc pixel. The angular resolution is $25^{\prime\prime} \sim 1.75$ kpc. The bottom panel covers only regions with metallicity measurements (Fig.~\ref{fig-metallicity}). Red solid and black dashed lines show best-fit relations for \themis and \mbb, respectively; green lines show correlations from \citet{Casasola2022A&A...668A.130C}. Red and white diamonds mark integrated values for \themis and \mbb fits. Calibration uncertainties are indicated by grey bars near the legend.
}
\label{fig-DGR}
\end{figure}

\textit{The dust-to-gas mass ratio.}
In Fig.\ref{fig-DGR}, we present the pixel-by-pixel relation between the DGR and $\Sigma_\star$ (top panel) and gas-phase metallicity (bottom panel).  The DGR inferred using \themis is nearly flat with $\Sigma_\star$ ($a \sim -0.06$, $r = -0.27\pm$0.03), whereas the DGR inferred from an \mbb fit shows a positive trend ($a \sim 0.4$, $r = 0.73\pm0.03$). This discrepancy (up to 0.5 dex at high $\Sigma_\star$) is primarily due to the fixed $\beta$ adopted by the \themis model. \citet{Casasola2022A&A...668A.130C} found a strong positive correlation between DGR and $\Sigma_\star$ ($a \sim 0.37$, $r = 0.7$), in agreement with our \mbb fit, but with a $\sim0.2$ dex offset in normalisation. This offset may stem from differences in spatial resolution, gas and dust mass estimation methods, and the CO line transition used to trace molecular gas (\citealt{Casasola2020A&A...633A.100C} used CO(2--1) with fixed $r_{21}$).

Regions with higher $\Sigma_\star$ are generally expected to exhibit higher DGRs, consistent with a more metal-rich ISM resulting from prolonged, steady SFHs \citep[e.g.,][]{DeLooze2020MNRAS.496.3668D, Galliano2021A&A...649A..18G}. 
As expected for such a steady star formation scenario, we detect no correlation between the DGR and the sSFR. We find a tight positive correlation between the DGR and the fraction of molecular gas ($f_{\rm mol}$) using the \mbb model ($r=0.63\pm0.02$), while a weak anti-correlation is obtained for \themis ($r=-0.35\pm0.02$). A flat DGR with $f_{\rm mol}$ implies an almost constant fraction of dust in the neutral ISM, irrespective of whether the gas is predominantly atomic or molecular. A positive trend is more commonly reported in recent literature \citep{Casasola2022A&A...668A.130C}, especially for environments with relatively high $f_{\rm mol}$ ($\gtrsim 10\%$). This trend is often interpreted as evidence for dust grain growth in the ISM through accretion \citep[typical timescales are shorter in cold and dense environments;][]{Asano2013EP&S...65..213A, Vilchez2019MNRAS.483.4968V}.

The DGR is expected to increase with metallicity since more metals are available for grain growth. Using \mbb, we find a positive, superlinear trend ($a \sim 1.47$, $r \sim 0.45$). This is in good agreement with the relation reported by \citet{Casasola2022A&A...668A.130C} ($a \sim 1.17$, $r = 0.7$), and the full DustPedia sample of late-type galaxies \citep{DeVis2019A&A...623A...5D}, as well as with the predictions of several dust evolution models \citep[e.g.,][]{Feldmann2015MNRAS.449.3274F, DeVis2017MNRAS.471.1743D}. A superlinear slope implies a significant contribution from in-situ grain growth processes that become particularly important in high-metallicity environments. Our \themis model instead shows a weak negative correlation between the DGR and gas-phase metallicity, likely due to the underestimation of dust masses in dense, metal-rich regions.

\section{Summary}\label{Sect:conclusions}

We have analysed the mm emission of M~99 using new 1.15 and 2 mm continuum maps from the NIKA2/IRAM~30~m telescope (IMEGIN Guaranteed Time Large Programme) at angular resolutions of $12^{\prime\prime}$ and $18^{\prime\prime}$. We combined these data with extensive ancillary multi-wavelength photometry (from UV to radio) and CO and HI line maps tracing molecular and atomic gas. We modelled the IR-to-radio SED of M~99 on both integrated and spatially-resolved (1.75 kpc) scales using the hierarchical Bayesian fitting code \texttt{HerBIE} \citep{Galliano2018MNRAS.476.1445G}, decomposing dust, free-free, and synchrotron emission.
We described the dust emission with the \themis dust model \citep{Jones2017A&A...602A..46J} and a single-T \mbb. We summarise our main results and conclusions as follows.
\begin{itemize}
    \item We find substantial variations in the dust spectral index, $\beta$, with changes reaching up to $\sim0.6$ on integrated scales and up to $\sim0.9$ on resolved scales. $\beta$ is systematically lower in the diffuse ISM ($1.72 \pm 0.09$ in the disc) and increases in denser, star-forming regions ($1.92 \pm 0.08$ in spiral arms, $2.27 \pm 0.09$ in the centre). On kiloparsec scales, $\beta$ exhibits a bimodal distribution, with peaks near 1.7 and 2.1, consistent with dust grain reprocessing, likely driven by coagulation or changes in dust grain composition. Strong correlations between $\beta$, molecular gas surface density, and the interstellar radiation field support that the observed FIR slope steepening in M~99 results from dust evolution in dense or strongly irradiated environments. Accounting for variable $\beta$ increases dust mass estimates by a factor of 1.6 (mean value), compared to models assuming a fixed $\beta$, and up to a factor of $\sim4$ in the central regions of M~99.
    \item The fraction of small dust grains varies from a few percent in the galaxy centre to $\sim16\%$ in the disc, anti-correlated with the average interstellar radiation field intensity. This supports the idea that small grains, sensitive to energetic UV photons, are destroyed in intense radiation fields. Gas-phase metallicity plays only a marginal role in the central 8 kpc of M~99.
    \item The synchrotron spectral index varies significantly, from $\sim0.6-0.7$ in central star-forming regions to $\sim1.2$ in diffuse ISM, consistent with the ageing of CREs as they propagate far from the sites of star formation. 
    \item Correlations of DSR and DGR with stellar mass surface density, molecular gas fraction, and gas-phase metallicity are well reproduced only when allowing $\beta$ to vary.
    Models with fixed $\beta$ can underestimate the inferred DSR and DGR by up to 0.5 dex in star-forming regions with a high molecular gas fraction.
    \item These findings emphasise the need for physically motivated dust models, equipped with enough flexibility to match the different ISM conditions observed in galaxies and accurately retrieve the properties of dust grains.
\end{itemize}

\section*{Data availability}

Tables~\ref{tab:integrated_photometry}, \ref{tab:global_sed_paramvalue}, \ref{tab:env_photometry}, and \ref{tab:envs_sed_paramvalue}, along with the ancillary maps (Appendices~\ref{App:ancillary_maps}, \ref{App:masked_pixels}), the spatially resolved best-fit parameters (Figs.~\ref{fig-beta}, \ref{fig-qAF_Uav}), and their corresponding uncertainties in FITS format, are available in electronic form at the CDS via anonymous ftp to cdsarc.u-strasbg.fr (130.79.128.5) or via http://cdsweb.u-strasbg.fr/cgi-bin/qcat?J/A+A/.

\begin{acknowledgements}
    This work was funded by the P2IO LabEx (ANR-10-LABX-0038) in the framework "Investissements d'Avenir" (ANR-11-IDEX-0003-01) managed by the Agence Nationale de la Recherche (ANR, France). This work was supported by the Programme National Physique et Chimie du Milieu Interstellaire (PCMI) and the Programme National Cosmology et Galaxies (PNCG) of the CNRS/INSU with INC/INP co-funded by CEA and CNES.
    This work was partially funded by the Foundation Nanoscience Grenoble and the LabEx FOCUS ANR-11-LABX-0013. This work was supported by the French National Research Agency under the contracts "MKIDS", "NIKA" and ANR-15-CE31-0017 and in the framework of the "Investissements d’avenir” programme (ANR-15-IDEX-02). This work has benefited from the support of the European Research Council Advanced Grant ORISTARS under the European Union's Seventh Framework Programme (Grant Agreement no. 291294).
    This work is based in part on observations made with (i) the Spitzer Space Telescope, which was operated by the Jet Propulsion Laboratory, California Institute of Technology under a contract with NASA; (ii) Herschel, which was an ESA space observatory with science instruments provided by European-led Principal Investigator consortia and with important participation from NASA; (iii) the Galaxy Evolution Explorer (GALEX) mission; (iv) the Karl G. Jansky Very Large Array (VLA) from the National Radio Astronomy Observatory (NRAO) and the 100-m telescope of the MPIfR (Max-Planck-Institut für Radioastronomie) at Effelsberg. The NRAO is a facility of the U.S. National Science Foundation operated under cooperative agreement by Associated Universities, Inc.
    We would like to thank the IRAM staff for their support during the observation campaigns. The NIKA2 dilution cryostat has been designed and built at the Institut N\'eel. In particular, we acknowledge the crucial contribution of the Cryogenics Group, and in particular Gregory Garde, Henri Rodenas, Jean-Paul Leggeri, Philippe Camus.
    A.R. acknowledges financial support from the Italian Ministry of University and Research - Project Proposal CIR01$\_00010$. 
    E.A. acknowledges funding from the French Programme d’investissements d’avenir through the Enigmass Labex. 
    FG acknowledges support by the French National Research Agency under the contracts WIDENING (ANR-23-ESDIR-0004) and REDEEMING (ANR-24-CE31-2530). 
    M.M.E. acknowledges the support of the French Agence Nationale de la Recherche (ANR), under grant ANR-22-CE31-0010.
    M.B., A.N., and S.C.M. acknowledge support from the Flemish Fund for Scientific Research (FWO-Vlaanderen, research project G0C4723N). 
    L.P., M.B., and I.D.L. acknowledge funding from the Belgian Science Policy Office (BELSPO) through the PRODEX project ``JWST/MIRI Science exploitation'' (C4000142239).
    S.K. acknowledges support provided by the Hellenic Foundation for Research and Innovation (HFRI) under the 3rd Call for HFRI PhD Fellowships (Fellowship Number: 5357). 
    V.C. acknowledges funding from the INAF Mini Grant 2022 programme "Face-to-Face with the Local Universe: ISM’s Empowerment (LOCAL)" and the INAF Mini Grant 2024 programme "DustPedia meets Metal-THINGS: Dust-METAL".
\end{acknowledgements}

\bibliographystyle{aa.bst} 
\bibliography{bibliography.bib}

\begin{appendix}

\section{Millimetre observations of M~99 with NIKA2}\label{App:Scanam_nika_maps}
In this Appendix, we examine the mm morphology of M~99 and assess and quantify the effects of large-scale filtering.

\subsection{Millimetre morphology}

\begin{figure}[!h]
\centering
\includegraphics[width=.95\columnwidth]{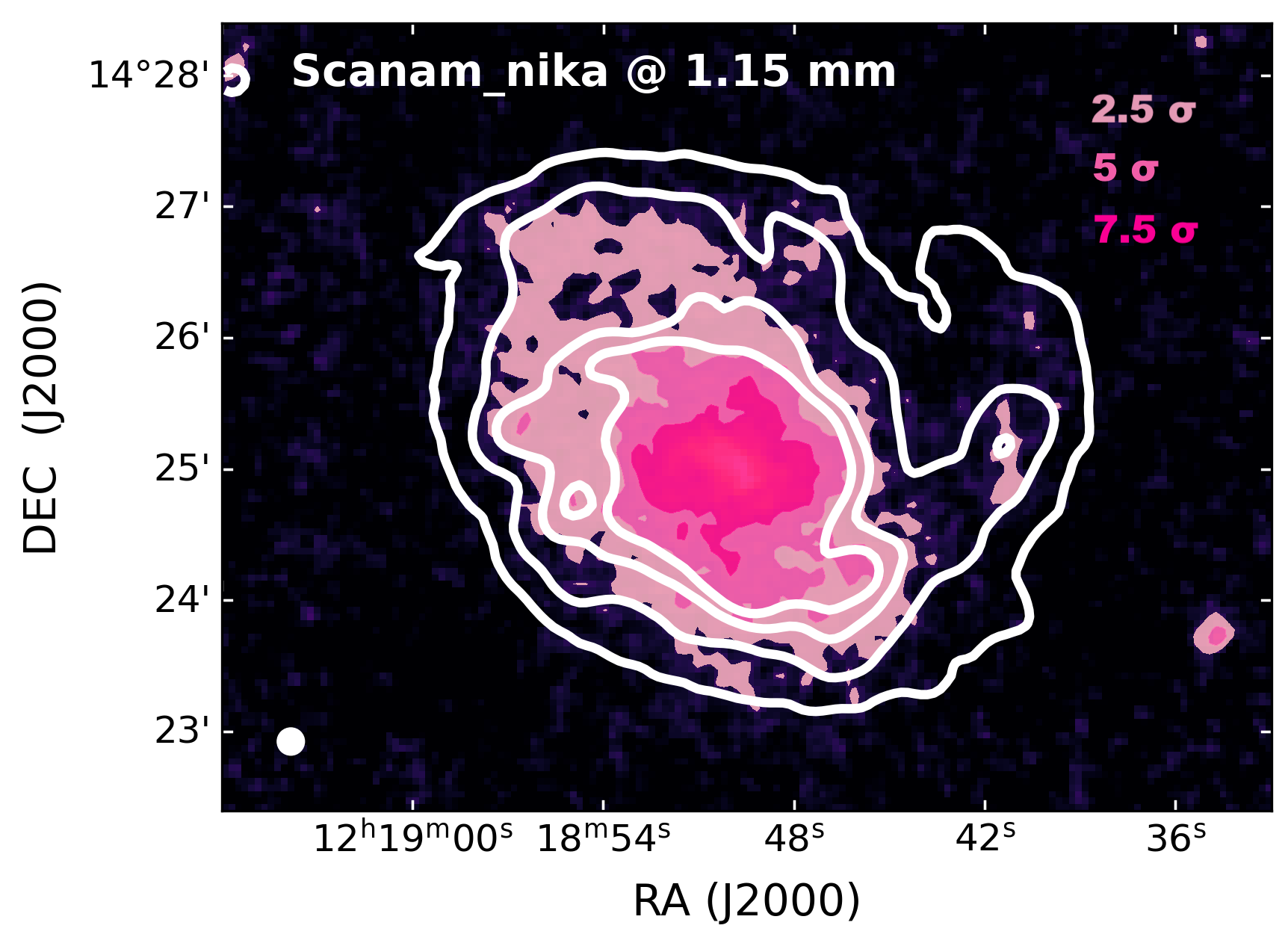}\\
\includegraphics[width=.95\columnwidth]{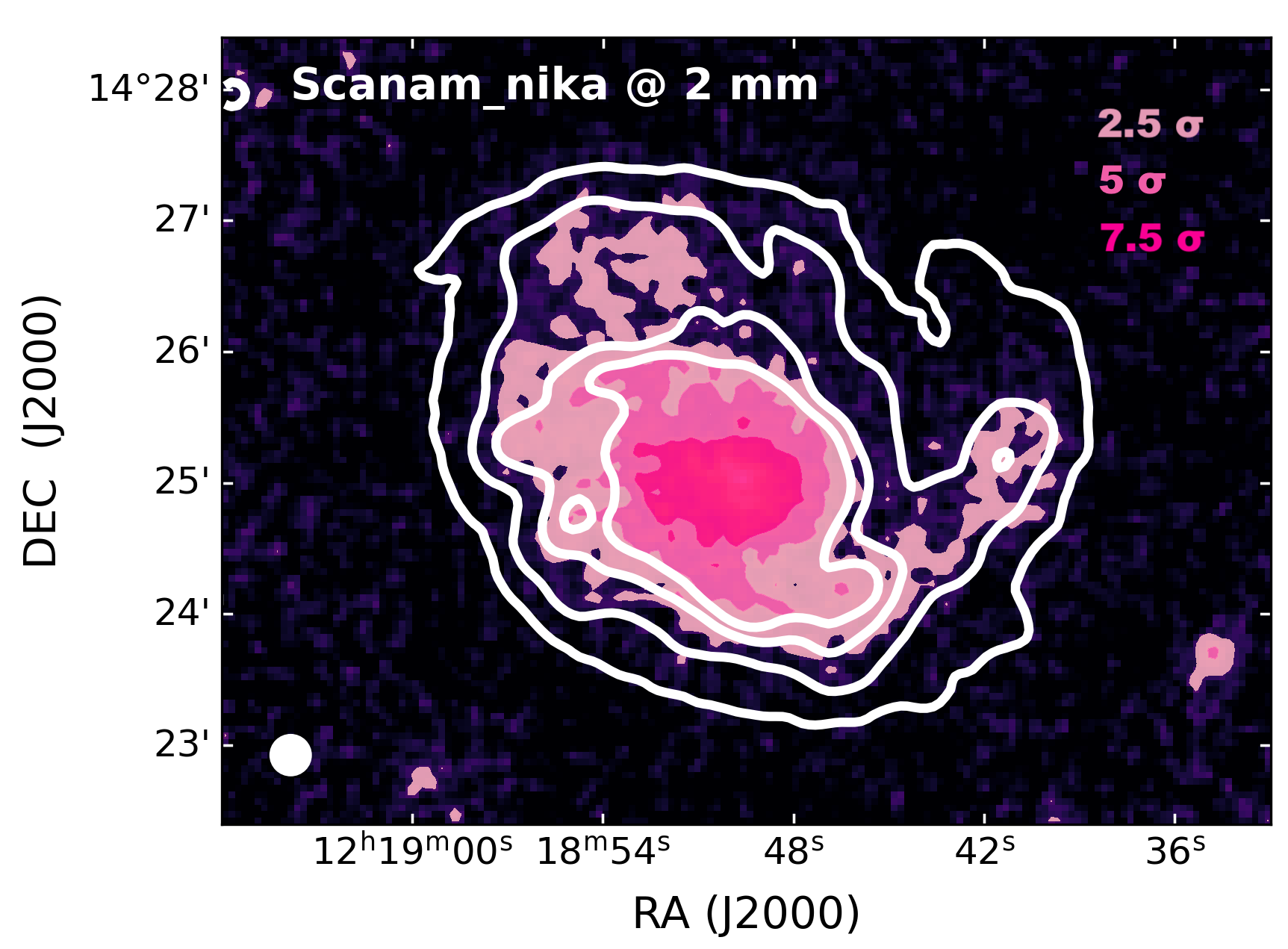}
\caption{Scanam\_nika filled contours of M~99 (shades of pink; levels of $[2.5,5,7.5]\times\sigma$) at 1.15 mm (top) and at 2 mm (bottom), overlaid with SPIRE~250~\upmicron contours (white solid lines; levels of $[5,15,35,55]\times\sigma$).}
\label{fig-SCANAM-morphology}       
\end{figure}

Fig. \ref{fig-SCANAM-morphology} compares the morphology observed with NIKA2 and SPIRE~250~\upmicron. Notably, the angular resolution of the SPIRE~250~\upmicron data ($18^{\prime\prime}$) matches that of NIKA2 at 2 mm. The NIKA2 $2.5\sigma$ contour (in light pink) closely follows the SPIRE~250~\upmicron $15\sigma$ contour, while higher NIKA2 significance levels ($5\sigma$ and $7.5\sigma$; darker pink) lie well within the $55\sigma$ SPIRE contour. 

Overall, the millimetre morphology traced by NIKA2 aligns well with the FIR morphology from SPIRE, within NIKA2’s detection limits, possibly reflecting an intrinsic compactness of M~99 at millimetre wavelengths.

\subsection{large-scale filtering}\label{App:large-scale-filtering}

As is the case for other ground-based millimetre observatories \citep[e.g., SCUBA2 on the JCMT;][]{Sadavoy2013ApJ...767..126S, Smith2021ApJS..257...52S, Pattle2023MNRAS.522.2339P}, NIKA2 observations of extended sources are affected by spatial filtering on large angular scales. This filtering leads to a partial loss of large-scale emission, arising from the combined effects of instrumental and atmospheric noise and the scanning strategy adopted for mapping extended regions.

The Scanam\_nika data-processing pipeline is specifically designed to mitigate this effect, as described in Appendix A.1 of \citet{Ejlali2025A&A...693A..88E}. Its core principle is to fully exploit the observational redundancy, with each position within the nominal sky coverage being sampled multiple times by different detectors under varying atmospheric conditions, while making minimal assumptions about the noise properties. The noise is decomposed into high- and low-frequency components, following the methodology detailed in Appendix A.2 of \citet{Ejlali2025A&A...693A..88E}. Further details will be presented in the forthcoming IMEGIN catalogue paper (Ejlali et al., in prep.).

\begin{figure}[h!]
\centering
\includegraphics[width=1.\columnwidth]{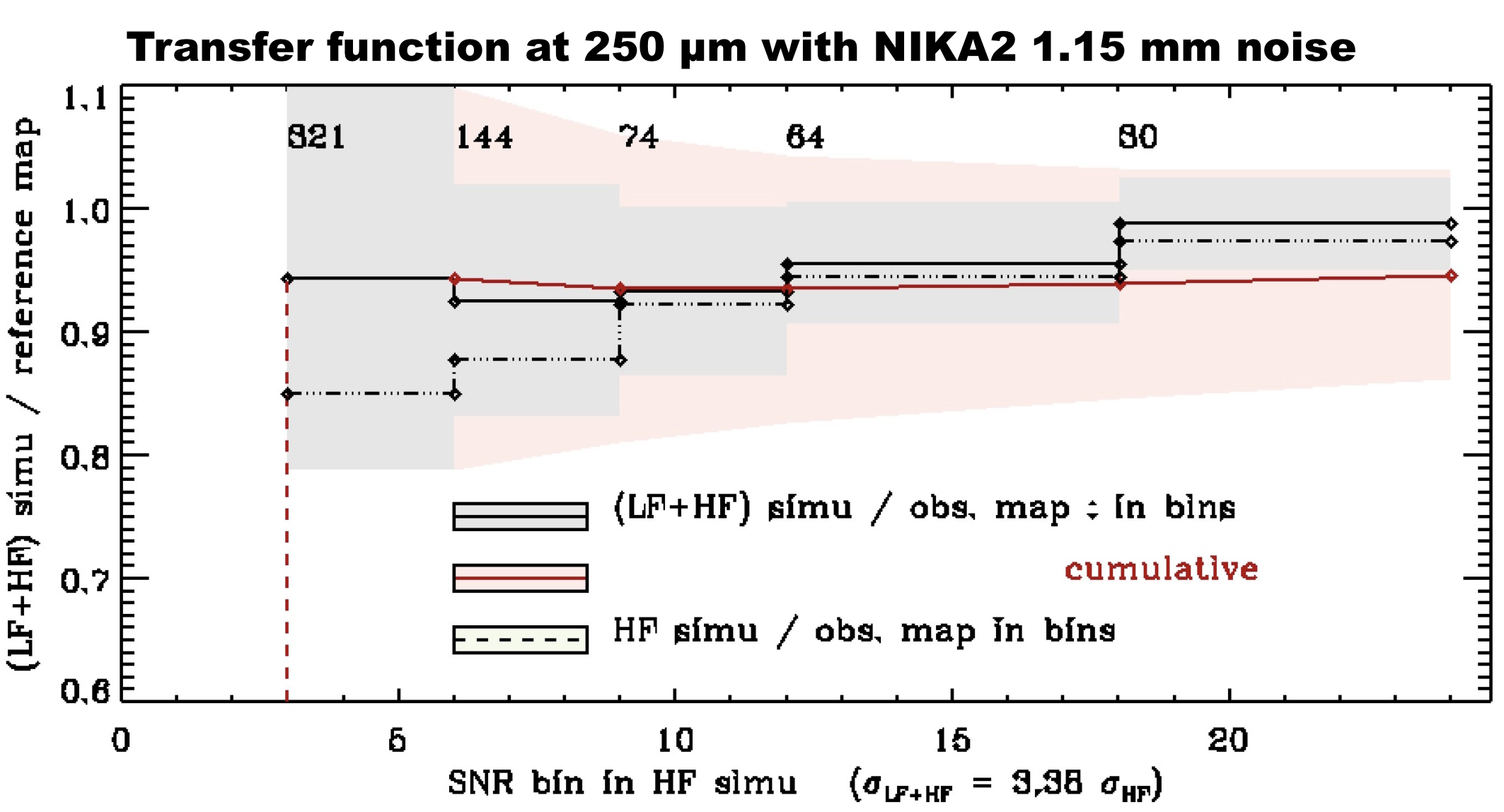}\\
\includegraphics[width=1.\columnwidth]{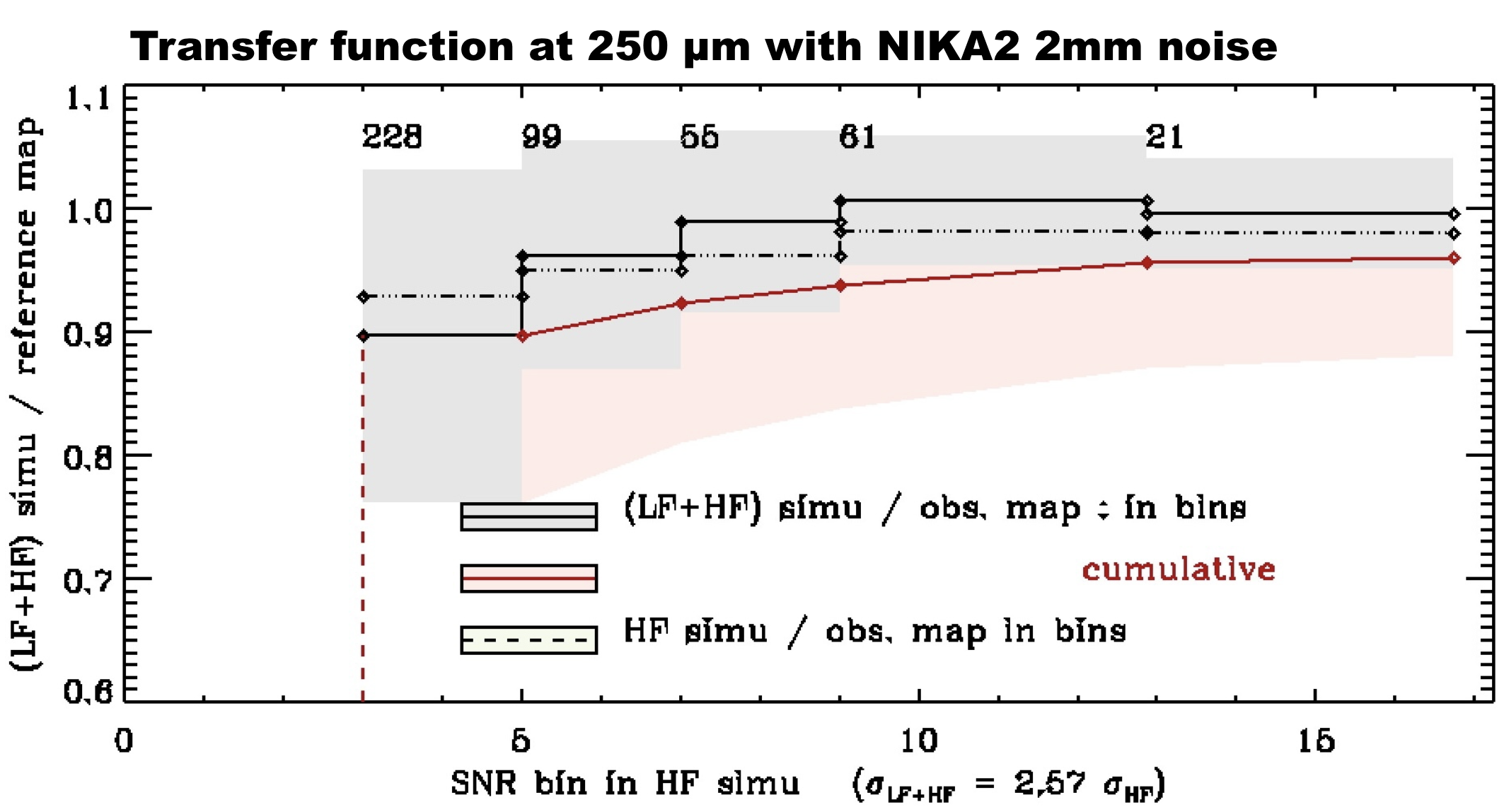}
\caption{Transfer functions from the SPIRE~250~\upmicron simulations binned as a function of S/N, with 1.15 mm (top panel) and 2 mm (bottom panel) total noise (black solid lines; grey shaded areas show the associated uncertainties), i.e. the sum of high frequency (HF) and low frequency noise (LF). The noise standard deviation $\sigma_{\rm HF}$ is from a simulation including only high-frequency noise (dashed line). The red curve and red shaded area show the cumulative fraction of the flux that is recovered and the associated uncertainties.
}
\label{fig-FILTERFUNCTIONS}
\end{figure}

We compute the transfer functions of the NIKA2 maps of M~99 using the SPIRE~250~\upmicron image, i.e. our morphological reference. We simulate the SPIRE~250 map through the Scanam\_nika pipeline after injecting the NIKA2 total noise at 1.15 mm and 2 mm, following the procedure described in \citet[][Appendix A.3]{Ejlali2025A&A...693A..88E}. Comparing the simulated output maps with the original input maps directly yields the spatially resolved transfer functions. The transfer functions are presented in bins of S/N ratio (Fig.~\ref{fig-FILTERFUNCTIONS}), with the low-S/N tail corresponding to the most diffuse emission at large scales. At 1.15 mm, the transfer function reaches approximately 95\% flux recovery up to the $3\sigma$ bin. At 2 mm, the recovered flux is about 90\% in the $3\sigma$ bin, increases to nearly 100\% at $7\sigma$, and remains flat at higher S/N ratios.
\begin{figure}[!h]
\centering
\includegraphics[width=1\columnwidth]{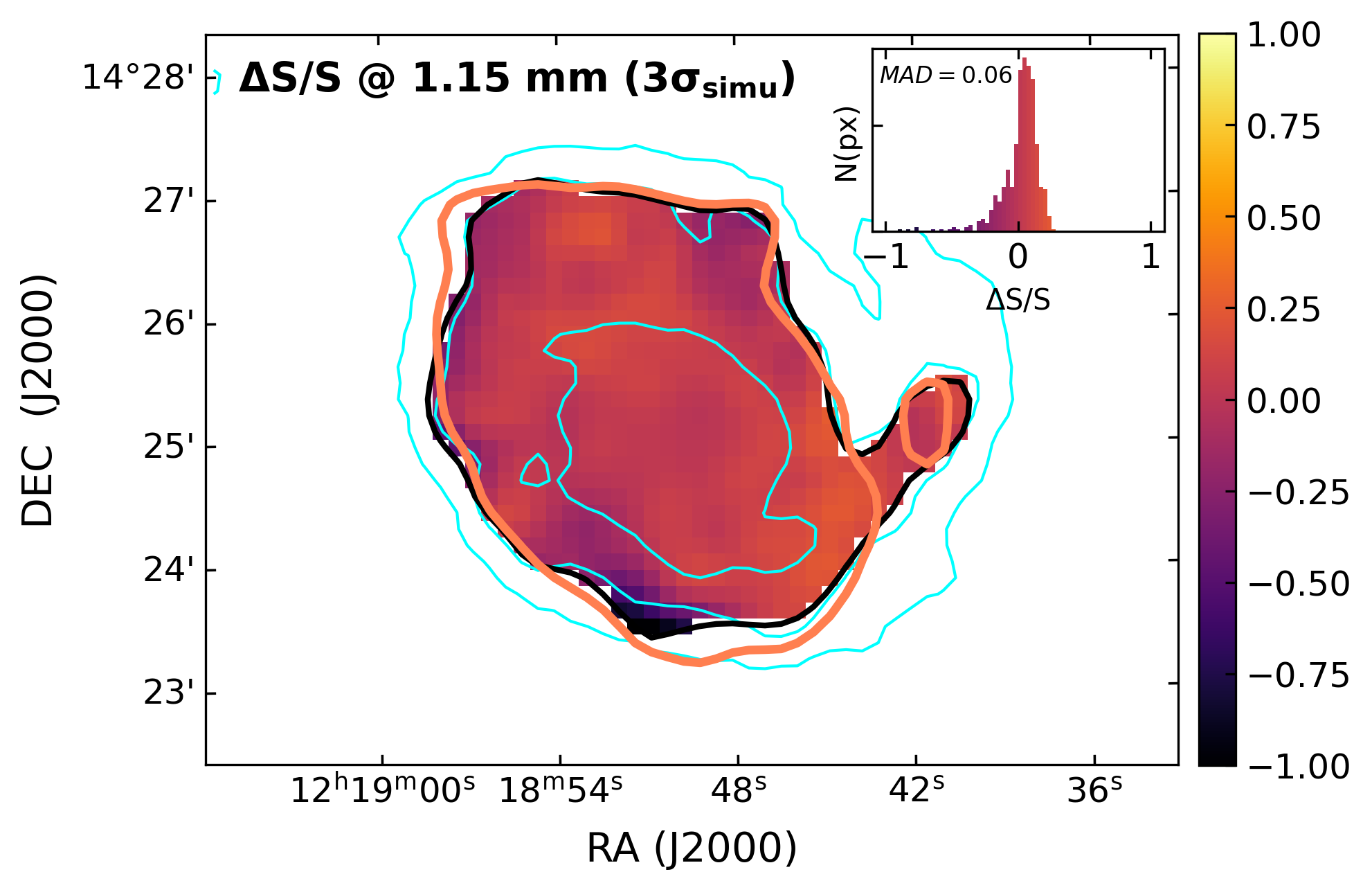}\\
\includegraphics[width=1\columnwidth]{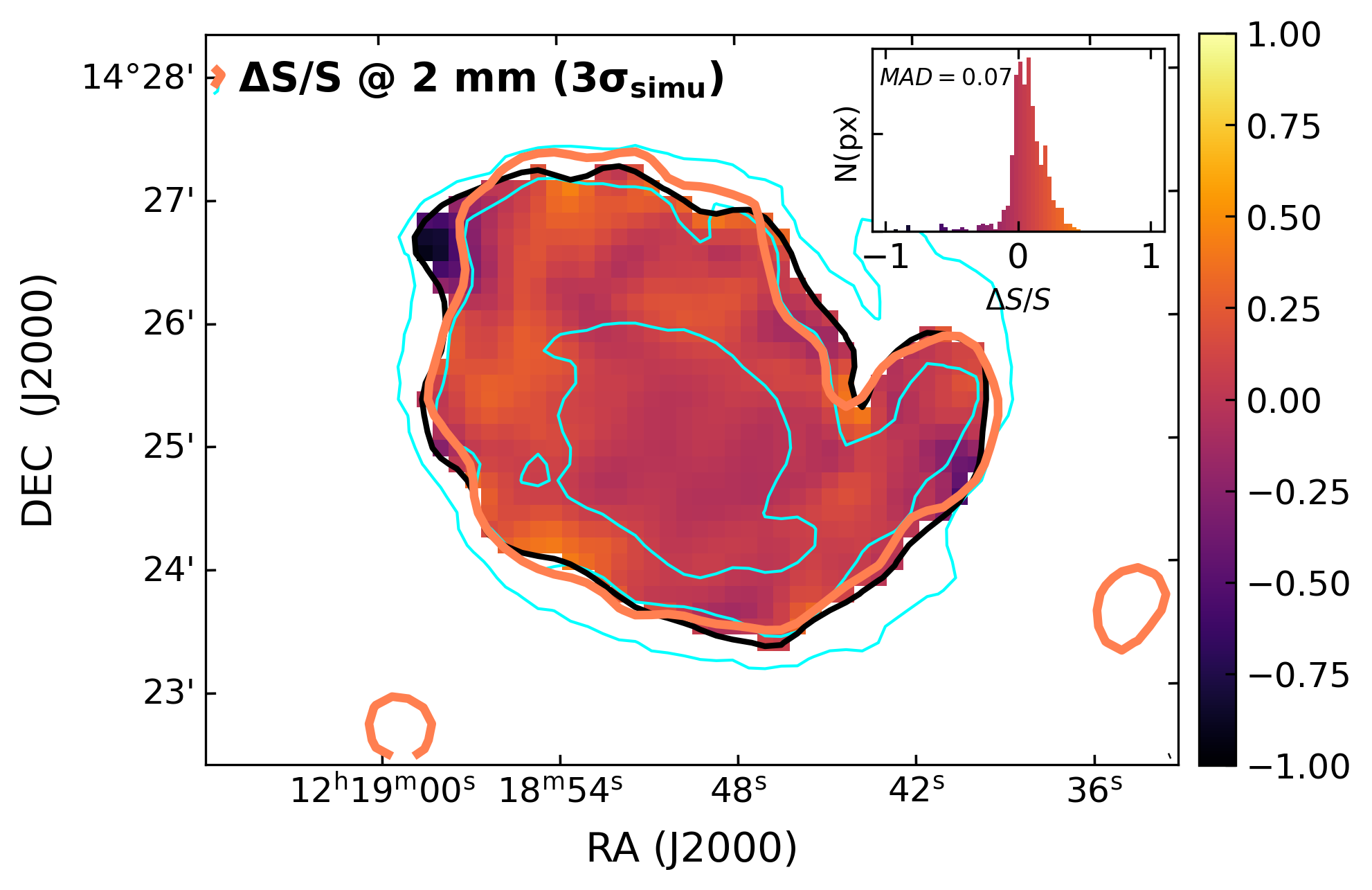}
\caption{Relative differences between the original and simulated SPIRE~250~\upmicron maps, using 1.15 mm noise and mask (top) and 2 mm noise and mask (bottom). All images are degraded and resampled to the SPIRE 350~\upmicron resolution. The $3\sigma$ level from the simulated maps ($3\sigma_{\rm simu}$) is shown in black, with corresponding NIKA2 contours in orange. Cyan contours show the original SPIRE~250~\upmicron emission (as in Fig.~\ref{fig-SCANAM-morphology}). Figure insets display the distribution of relative differences ($\Delta S$), with MAD values of 0.06 (1.15 mm) and 0.07 (2 mm), both approximated to 0.1.}
\label{fig-relative-diff-filtering-pixel}
\end{figure}

Given the relatively minor filtering effects within the $3\sigma$ threshold, we incorporate this contribution into the overall uncertainty budget. This additional uncertainty is treated as a systematic error and combined in quadrature with the statistical uncertainties. The relative uncertainty introduced by large-scale filtering in the NIKA2 maps is estimated from the flux difference between the original and simulated data. On an integrated scale (within the ellipse used for integrated photometry; Fig. \ref{fig-SPIRE250_photometry}), these relative uncertainties amount to $\sim5\%$ at 1.15 mm and 15\% at 2 mm. When considering distinct morphological components, the relative uncertainties at 1.15 mm are 5\% for the spiral arms and disc and 1\% for the centre. At 2 mm, the uncertainties are 7\% for the spiral arms, 16\% for the disc, and 2\% for the centre.

The pixel-scale uncertainty is quantified using the following procedure. We construct the map (Fig.~\ref{fig-relative-diff-filtering-pixel}) of the relative difference between the original and the simulated SPIRE~250~\upmicron image, i.e. $(S_{\rm orig}-S_{\rm simu})/S_{\rm orig}$, within the $3\sigma$ level measured on the simulated map (i.e., $3\sigma_{\rm simu}$; cf. Fig. \ref{fig-FILTERFUNCTIONS}). Then, we compute the distribution of the relative difference (figure insets). The spread of this distribution, which we quantify using the median absolute deviation (MAD), provides an estimate of the flux recovery accuracy within $3\sigma_{\rm simu}$.
The MAD is approximately 10\% at both 1.15 mm and 2 mm. The NIKA2 flux densities corresponding to the $3\sigma$ level of the simulated maps are 0.24 mJy at 1.15 mm and 0.022 mJy at 2 mm (orange solid contours in Fig.~\ref{fig-relative-diff-filtering-pixel}). These values imply flat noise levels of 0.024 mJy at 1.15 mm and 0.0022 mJy at 2 mm.
This flat noise component is added in quadrature, on a pixel-by-pixel basis, to the NIKA2 RMS maps.

\subsection{Comparison with Planck observations}\label{App:mm-maps}

To further validate the NIKA2 maps produced with the Scanam\_nika software, we compare the integrated NIKA2 fluxes with the corresponding values derived from Planck, as described below.

We model and subtract the sky emission from Planck images, dominated by the CMB, using \texttt{HIP} (see Appendix \ref{App:hip}), as displayed in Fig. \ref{fig-Planck_BKG_1mm}.

Since M~99 is unresolved and the coarse pixel size causes the galaxy's emission to spread into adjacent pixels, we measure the integrated flux within a circular aperture, with a radius matching the Planck beam ($r = 300^{\prime\prime}$; yellow solid line in the top panels of Figs.~\ref{fig-Planck_BKG_1mm}).
The resulting integrated flux of M~99 at 1.38 mm, measured on the sky-subtracted Planck map, 
is $0.37\pm0.07$ Jy (versus the $0.52\pm0.06$ Jy measured on the original map).   
At 2 mm, the integrated flux we measure on the original Planck map is $-0.07\pm0.04$ Jy. 
After correcting for the background emission, we measure an integrated flux of $0.03\pm0.05$ Jy, which is consistent with 0. Thus, Planck does not detect M~99 at 2 mm.

\begin{figure*}[h!]
\centering
\includegraphics[width=0.48\hsize]{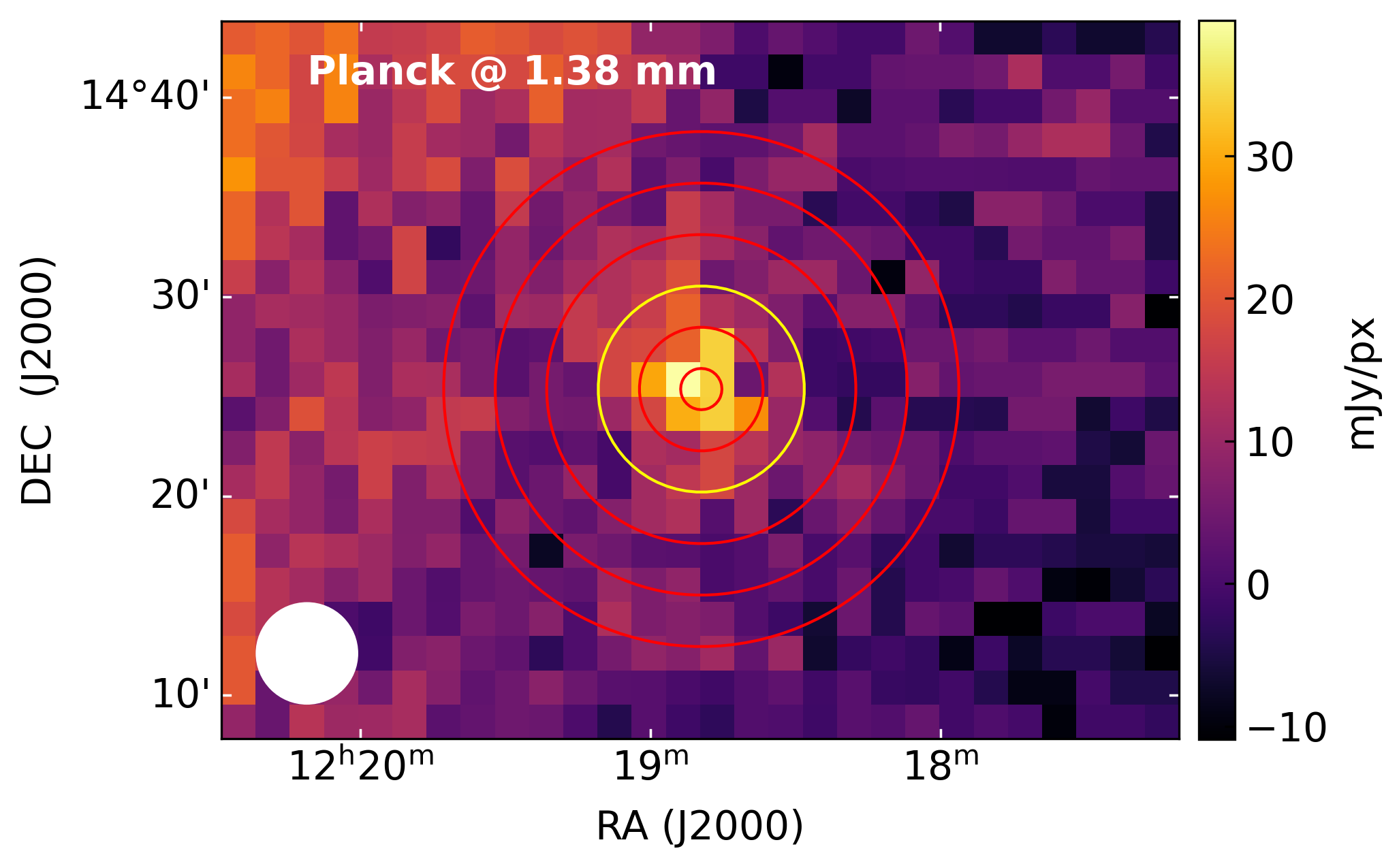}
\includegraphics[width=0.475\hsize]{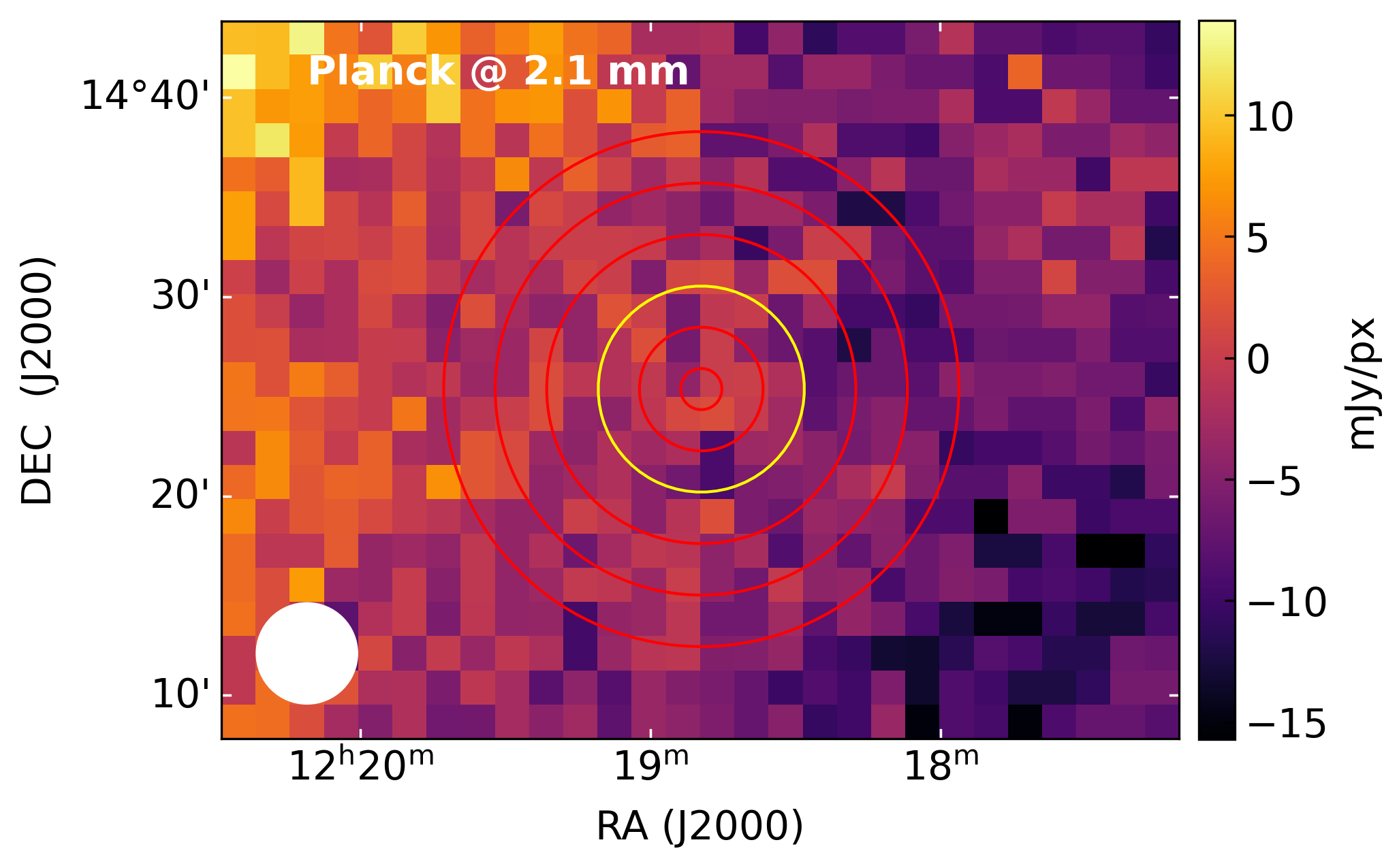}\\
\includegraphics[width=0.482\hsize]{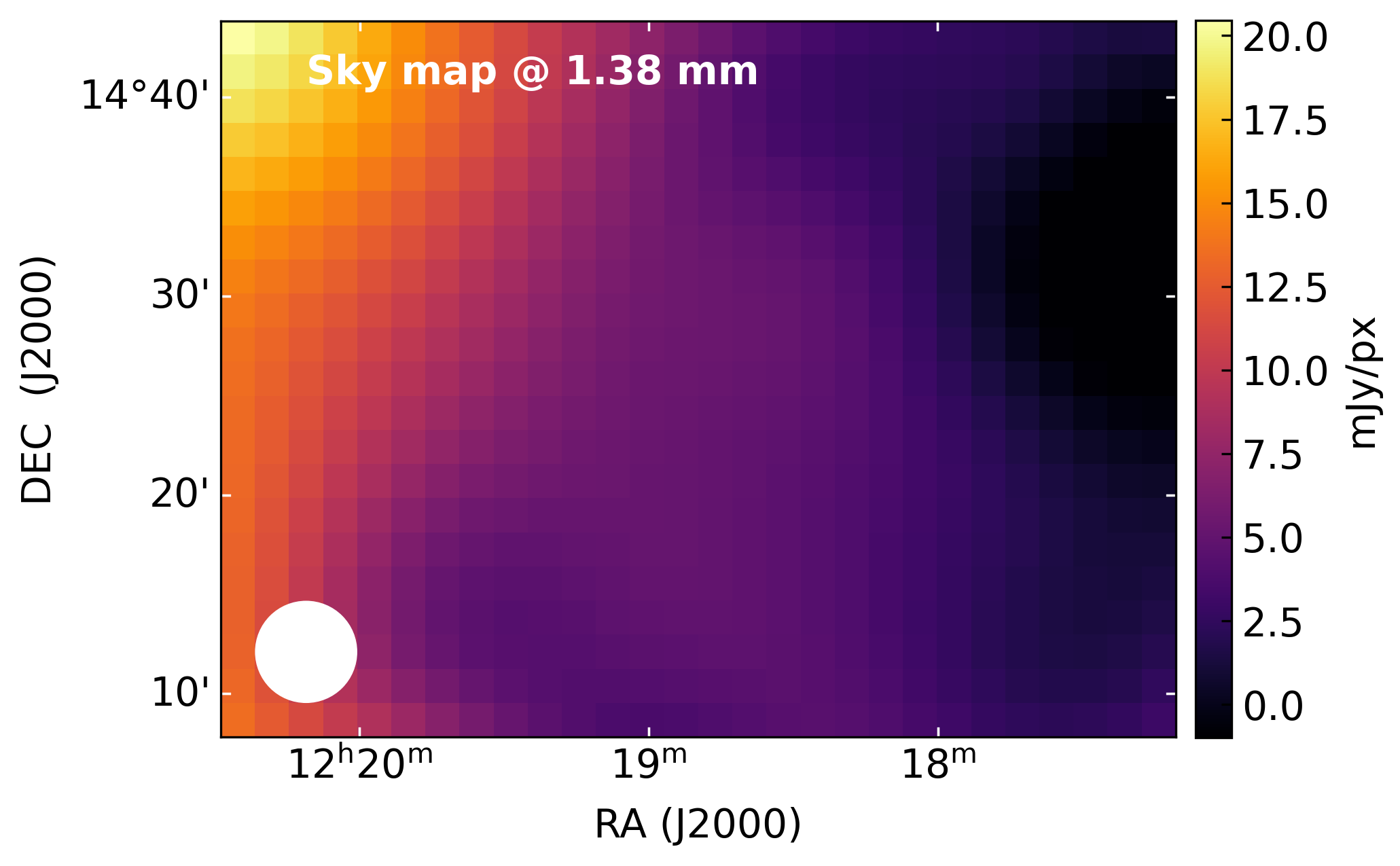}
\includegraphics[width=0.475\hsize]{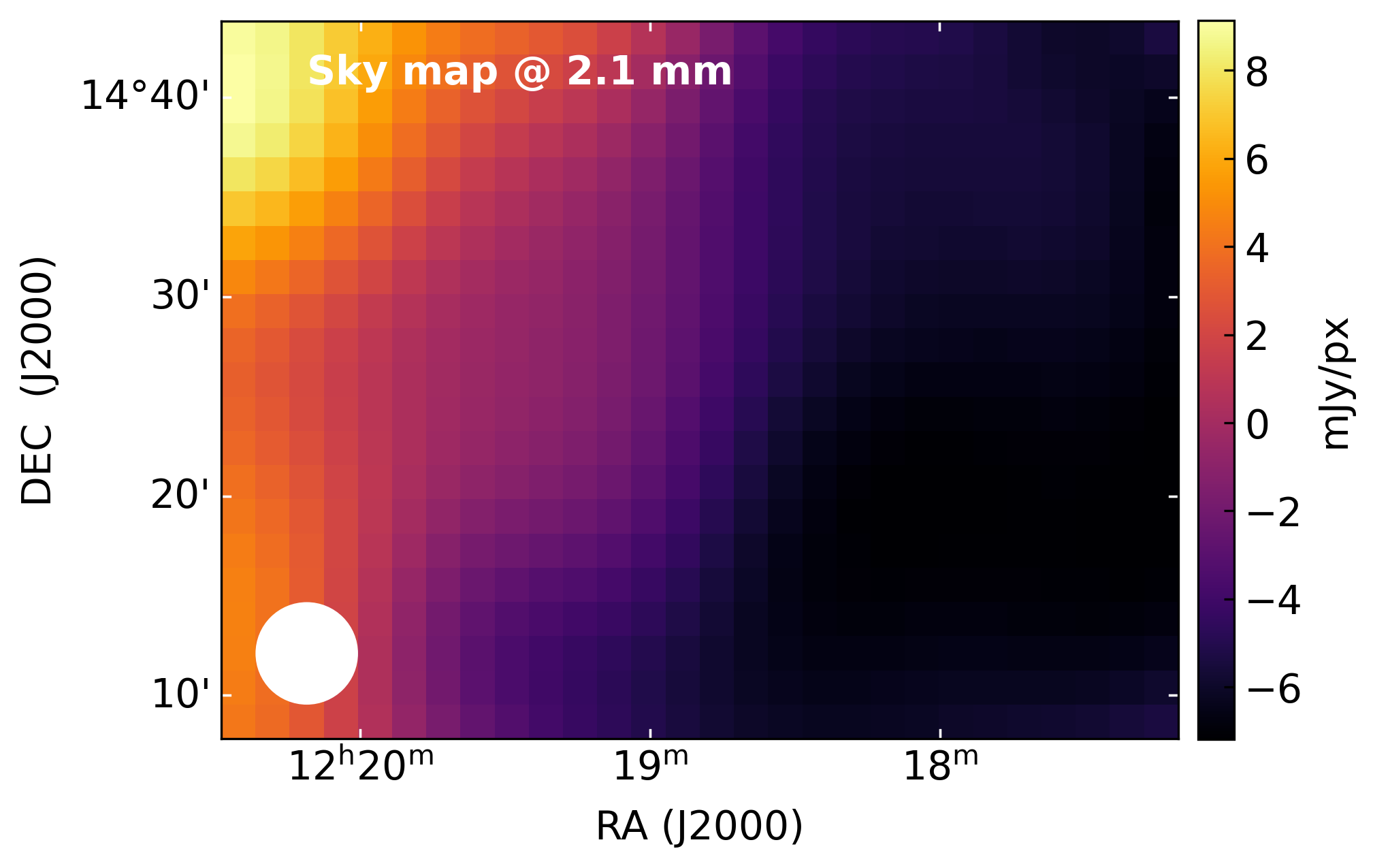}\\
\includegraphics[width=0.438\hsize]{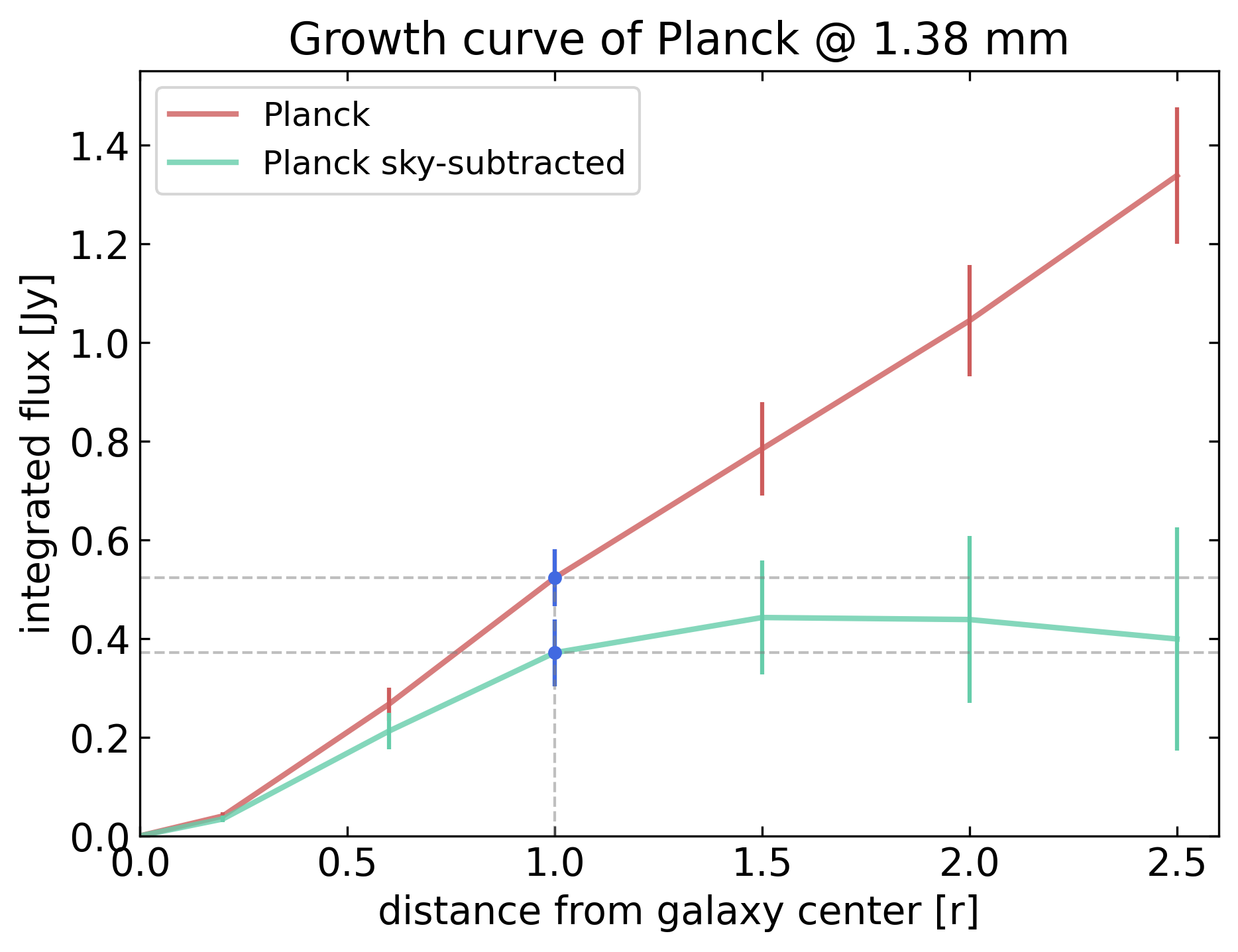}
\includegraphics[width=0.45\hsize]{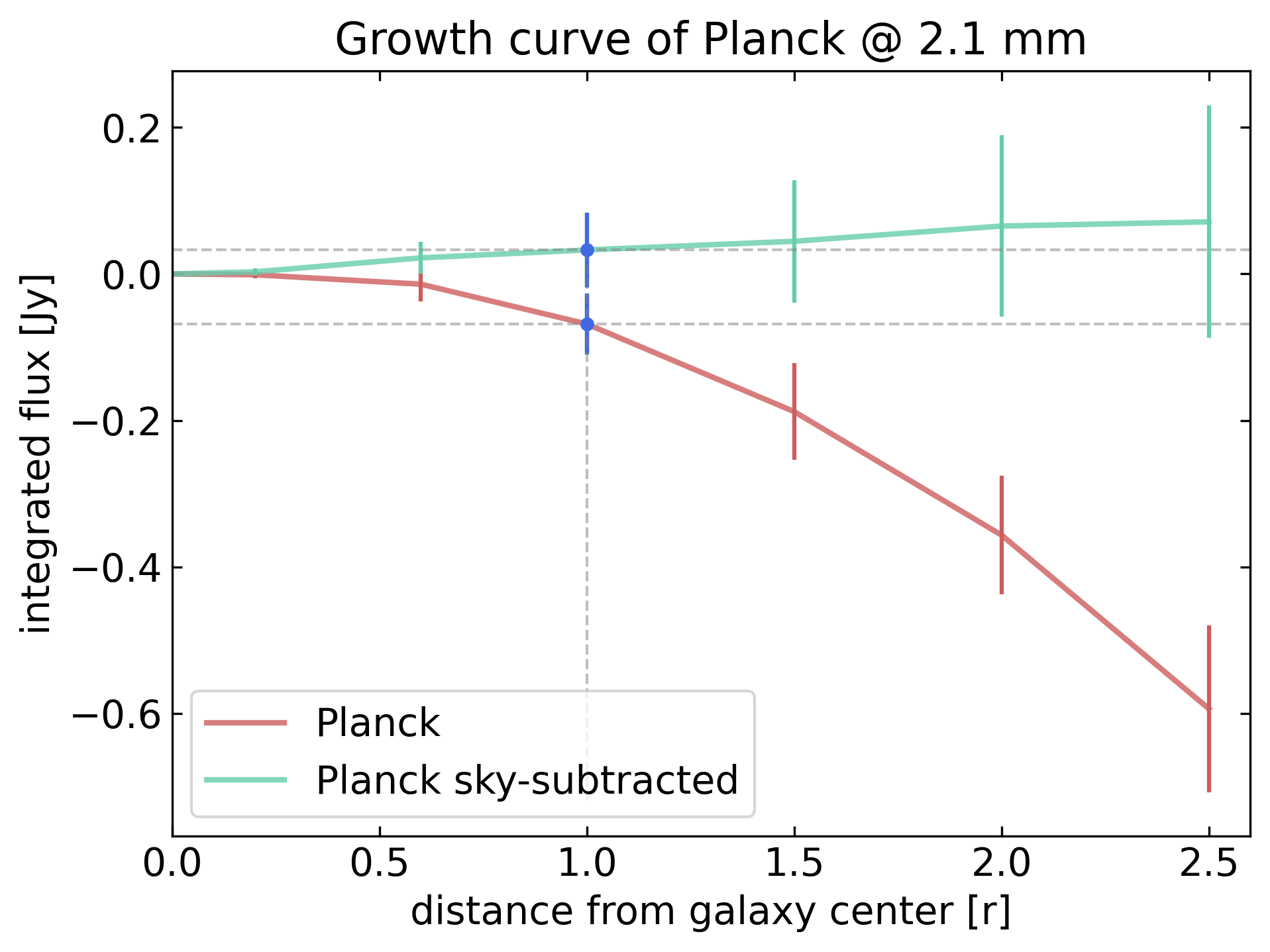}
\caption{Original Planck maps of M~99 at 1.38 mm and 2.1 mm (top panels), overlaid with the circles used for computing the growth curve. The modelled sky maps are shown in the central panels. 
Bottom panels show the growth curves computed on the original Planck maps (red solid lines) and on the sky-subtracted maps (green solid lines). The blue-filled circles indicate the value of the integrated flux, corresponding to the yellow circles in the top panels ($r=300^{\prime\prime}$).}
\label{fig-Planck_BKG_1mm}
\end{figure*}

Since Planck and NIKA2 have different filter shapes and central wavelengths, we need to perform a ``colour correction'', which implies the assumption of a reference model.
Dust emission at mm wavelengths is reasonably well described by the Rayleigh-Jeans approximation for $T_{\rm dust}\gtrsim20$ K; thus a sensible reference model is:
\begin{equation}
    S_{RJ}(\lambda) \propto \lambda^{-(2+\beta)},
\end{equation}
where $\beta\sim 2$ and $S_{RJ}$ is the observed flux in the  Rayleigh-Jeans tail.
We use \texttt{HerBIE}'s synthetic photometry module \citep{Galliano2018MNRAS.476.1445G} for integrating, in the range 0.3 mm $<\lambda< 3$ mm, NIKA2 and Planck bandpasses with central wavelengths 1.15 mm and 1.38 mm. The following expression gives the ``colour correction'':
\begin{equation}
    C = S_{\rm RJ}(\lambda_{\rm NIKA2})/S_{\rm RJ}(\lambda_{\rm Planck}) \sim 1.7,
\end{equation}
which implies
\begin{equation}\label{Eq:coulorcorrected-HFI4flux}
  S_{\rm C} = S(\lambda_{\rm Planck})\times C \sim 0.37\times 1.7 \sim 0.63 \,{\rm Jy}.
\end{equation}
$S_{\rm C}$ is the colour-corrected Planck flux at $\lambda=1.15$ mm, i.e. the total flux we expect to measure on the NIKA2 1.15 mm map of M~99, given the flux we measure on the Planck 1.38 mm map.
This value is consistent with the Scanam\_nika integrated flux, i.e. $0.65\pm0.04$ Jy.

\section{Ancillary maps}\label{App:ancillary_maps}

This Appendix outlines the methodology used to derive the ancillary maps.
\subsection{Stellar mass surface density}\label{Sect:stellarsurfdensity}
We derive the stellar mass surface density map of M~99 using the calibration by \citet[][Eq. C1]{Leroy2008AJ....136.2782L}, under the assumption of a \cite{Kroupa2001MNRAS.322..231K} Initial Mass Function (IMF). This calibration is based on Spitzer/IRAC 3.6~\upmicron surface brightness and reads:
\begin{equation}\label{Eq:StellarMass}
    \frac{\Sigma_{*}}{{\rm M}_\odot \,{\rm pc}^{-2}} = 280 \,\frac{I_{3.6}}{{\rm MJy}\, {\rm sr}^{-1}} \, \cos{i},
\end{equation}
where $i$ is the galaxy inclination and $I_{3.6}$ is the Spitzer/IRAC surface brightness at 3.6~\upmicron.
 
The calibration relies on an empirical conversion from 3.6~\upmicron emission to K-band intensity, assuming a standard mass-to-light ratio of 0.5 M$_\odot$/L$_\odot$ typical of star-forming galaxies. This assumption dominates the calibration uncertainty, with $\sim0.1$ dex scatter due to variations in SFH, metallicity, and IMF \citep{Leroy2008AJ....136.2782L}.

\begin{figure}[!h]
\centering
\includegraphics[width=.98\columnwidth]{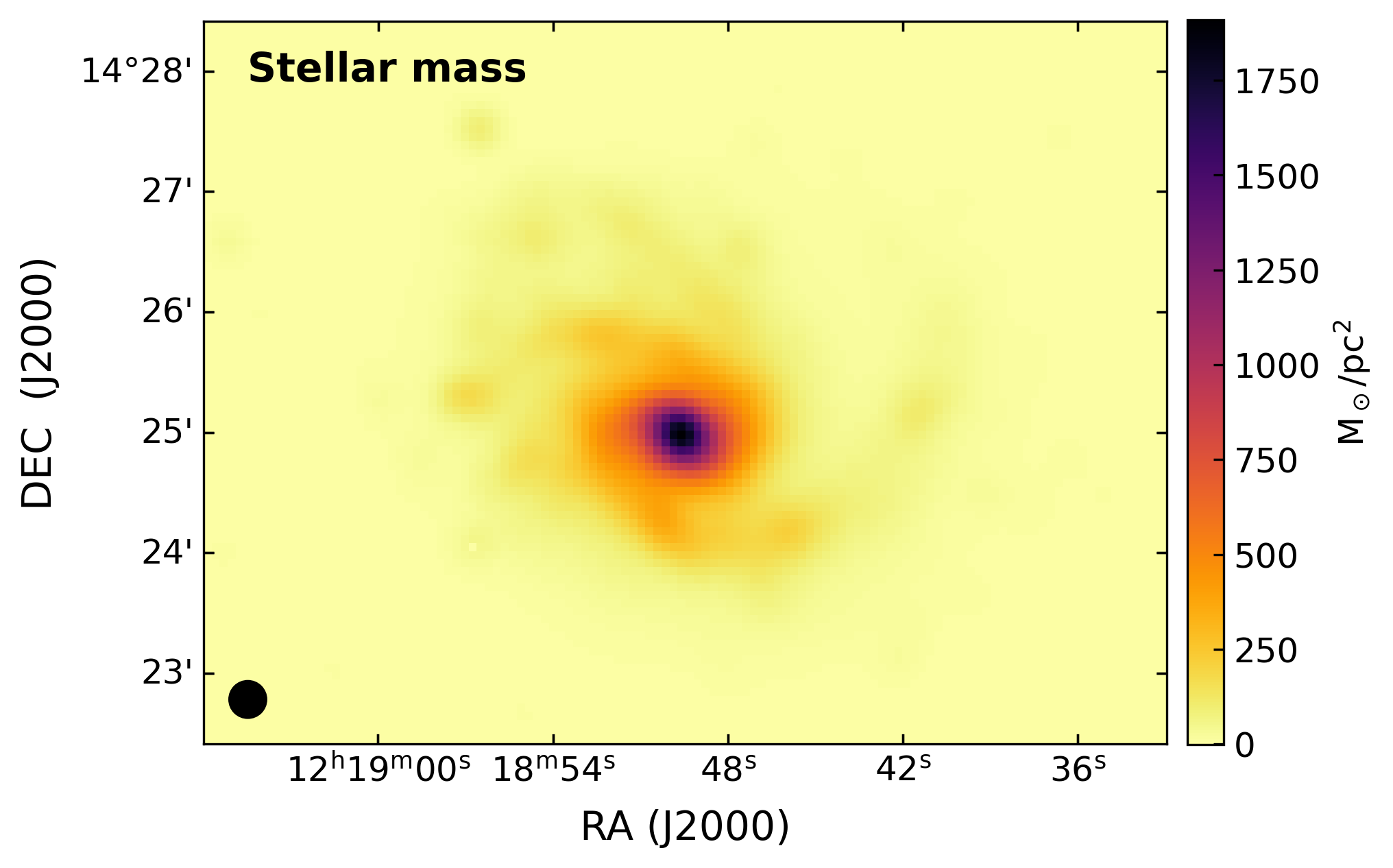}
\caption{Stellar mass surface density of M~99. The map is convolved to match NIKA2 angular resolution at 2 mm and pixel size.}
\label{fig-stellarmass}
\end{figure}

In Fig. \ref{fig-stellarmass} we show the resulting stellar mass surface density map of M~99. The associated uncertainties are estimated via the MC approach with $10^3$ iterations.
From our map we derive a total stellar mass of $(4 \pm 1) \times 10^{10}$ M$_\odot$ for M~99, consistent with literature values \citep[e.g.][]{Chemin2016A&A...588A..48C, Leroy2021ApJS..257...43L}.

\subsection{Star formation rate surface density}\label{Sect:SFRsurfdensity}
We derive the SFR surface density of M~99 using the calibration by \citet[][Eq. D1]{Leroy2008AJ....136.2782L}, with 1$\sigma$ scatter of 0.22 dex:
\begin{equation}\label{Eq:SFR calibration}
    \frac{\Sigma_{\rm SFR}}{{\rm M}_\odot {\rm yr}^{-1}{\rm kpc}^{-2}} = 0.081\, \frac{I_{\rm FUV, \,SF}}{{\rm MJy}\, {\rm sr^{-1}}} + 0.0032\,\frac{I_{\rm 24\,\mu m,\,SF}}{{\rm MJy}\, {\rm sr^{-1}}} \times \cos{i},
\end{equation}
where $I_{\rm FUV}$ and $I_{24,\mu\mathrm{m}}$ are the GALEX FUV and Spitzer/MIPS 24~\upmicron surface brightnesses, respectively (Table \ref{tab:ancillary_SEDfit}), and $i$ is the inclination angle in radians (Table \ref{tab:NGC4254prop}).

\begin{figure}[!h]
\centering
\includegraphics[width=.98\columnwidth]{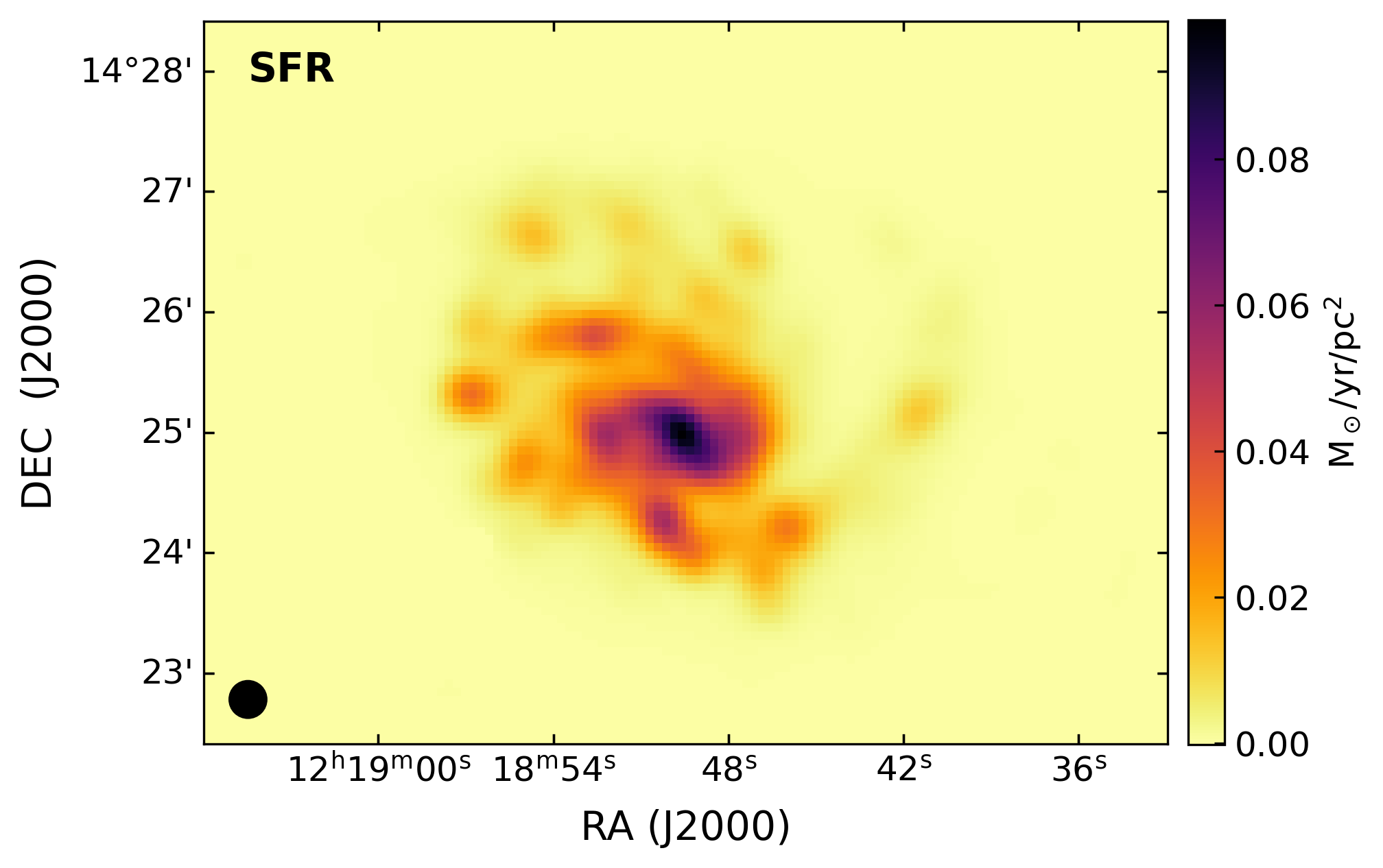}
\caption{SFR density of M~99. The map is shown at $18^{\prime\prime}$ angular resolution and $3^{\prime\prime}$ pixel size, matching NIKA2 at 2 mm.}
\label{fig-SFR}
\end{figure}

We correct the GALEX FUV and Spitzer/MIPS 24~\upmicron maps for emission from evolved stars using the Spitzer/IRAC 3.6~\upmicron map (Table \ref{tab:ancillary_SEDfit}), following the method of \citet[][]{Pattle2023MNRAS.522.2339P}:
\begin{equation}
  \begin{array}{cc}
     I_{\rm FUV, \,SF} = I_{\rm FUV} - 10^{-3}\times I_{\rm 3.6\, \mu m} \quad [{\rm MJy\,sr^{-1}}]\\
     I_{\rm 24 \,\mu m, \,SF} = I_{\rm 24\, \mu m} - 0.1\times I_{\rm 3.6\, \mu m} \quad [{\rm MJy\,sr^{-1}}].\\
\end{array}  
\end{equation} 
Figure~\ref{fig-SFR} shows the resulting SFR surface density map. Uncertainties from the intensity maps and the intrinsic scatter of the calibrations are propagated via a Monte Carlo method with $N_{\rm MC} = 10^3$ iterations.

From our map we measure a total SFR of 3 M$_\odot$ yr$^{-1}$, with a 3$\sigma$ confidence interval of [0.66, 13.7] M$_\odot$ yr$^{-1}$. This value agrees with estimates from the literature \citep[e.g.][]{Tabatabaei2017ApJ...836..185T, Leroy2021ApJS..257...43L}, despite differences in the adopted calibrations \citep[e.g.][]{Boselli2018A&A...615A.114B}. 

\subsection{Atomic and molecular gas masses}\label{Sect:gassurfdensity}

We derive the atomic gas mass surface density ($\Sigma_{\rm HI}$) from the 21 cm line intensity map ($I_{21\,\mathrm{cm}}$) of M~99 (Table \ref{tab:ancillary_spectrallines}), by \citet{Chung2009AJ....138.1741C}. We assume optically thin HI emission, following standard practice \citep[e.g.,][]{Schruba2011AJ....142...37S, Casasola2017A&A...605A..18C}:
\begin{equation}\label{Eq:HI}
    \frac{\Sigma_{\rm HI}}{{\rm M}_\odot \,{\rm pc}^{-2}} = 0.02\,\frac{I_{21\,cm}}{\rm K\,km\,s^{-1}}  \times \cos{i}
\end{equation}
where $i$ is the inclination angle of the galaxy. The associated uncertainty ranges between approximately 0.1 and 0.15 dex \citep{Aniano2020ApJ...889..150A, Casasola2020A&A...633A.100C, Casasola2022A&A...668A.130C}. 

\begin{figure}[!h]
\centering
\includegraphics[width=.98\columnwidth]{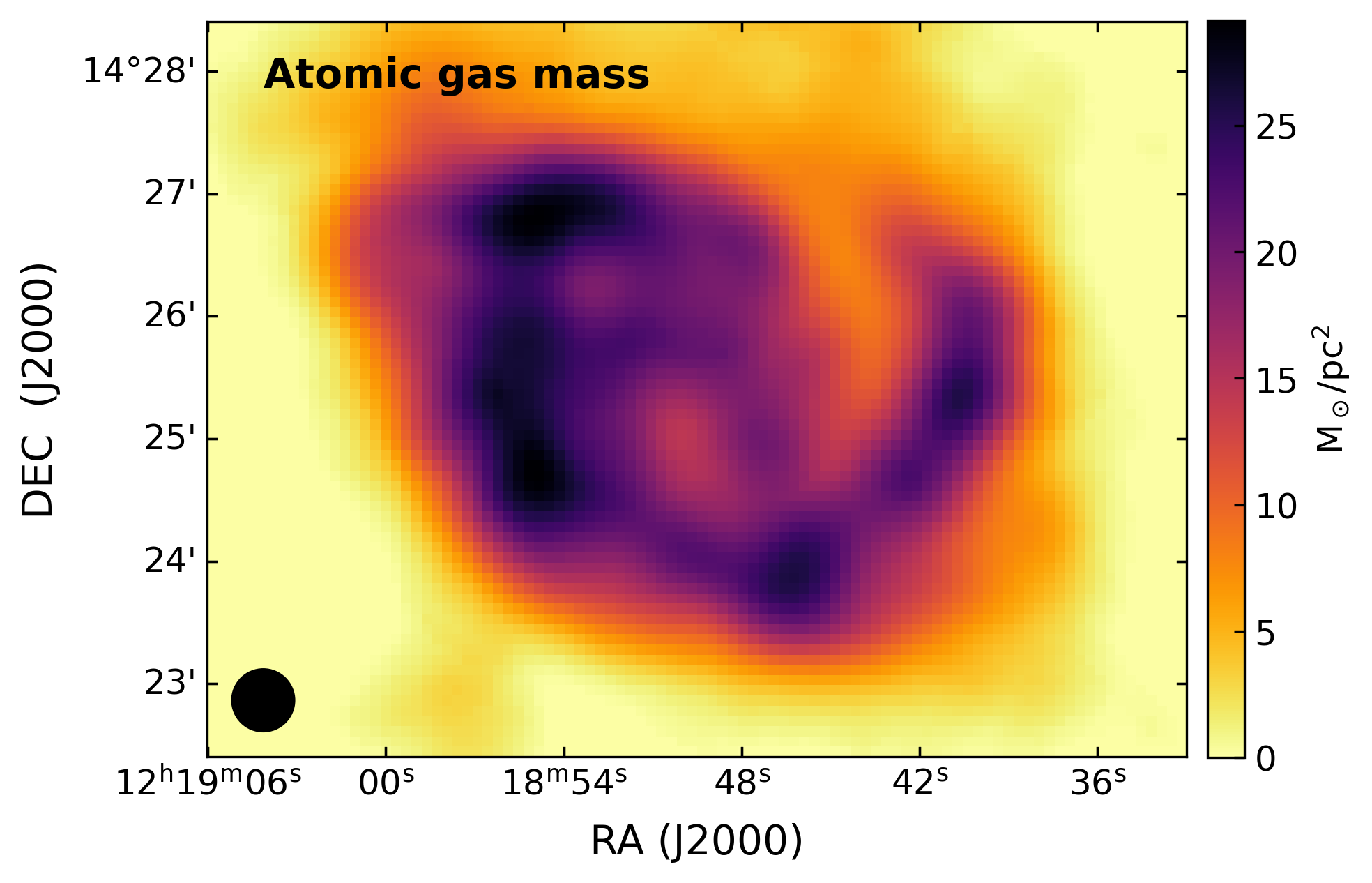}
\caption{Atomic gas mass surface density of M~99. Spatial resolution and pixel size are the same as the original HI map, i.e. $30^{\prime\prime}$ and $5^{\prime\prime}$, respectively.}
\label{fig-HI}
\end{figure}

Figure~\ref{fig-HI} presents the resulting HI mass surface density map at the original angular resolution of $30^{\prime\prime}$. From our map we find a total atomic hydrogen mass of $M_{\rm HI} = (7 \pm 1) \times 10^{9}$ M$_\odot$, in agreement with the literature \citep[e.g.,][]{Chemin2016A&A...588A..48C, Haynes2018ApJ...861...49H}.

We derive the molecular gas mass surface density ($\Sigma_{\rm gas}^{\rm mol}$) using the CO(1--0) line intensity map by the EMPIRE programme \citep[][Table \ref{tab:ancillary_spectrallines}]{Jimenez-Donaire2019ApJ...880..127J}. We adopt a constant CO-to-H$_2$ conversion factor\footnote{The CO-to-H$_2$ conversion factor has been suggested to be strongly dependent on the gas-phase metallicity \citep[e.g.,][]{Genzel2012ApJ...746...69G, Schruba2011AJ....142...37S, Amorin2016A&A...588A..23A, Accurso2017MNRAS.470.4750A}. Many works in the literature showed that this dependence, if taken into account, can lead to systematically larger conversion factors, up to a factor of 3 \citep[e.g.,][]{Casasola2017A&A...605A..18C, Salvestrini2025A&A...699A.346S}. However, this systematics is consistent, within $2\sigma$, with the uncertainty driven by the variety of gas-phase metallicity calibrations \citep[i.e., $\sim0.2$ dex;][]{Amorin2016A&A...588A..23A}.}, i.e. X$_{\rm CO} = 2.0 \times 10^{20}$ cm$^{-2}$ (K km s$^{-1})^{-1}$, following \cite{Bolatto2013ARA&A..51..207B}. Specifically, X$_{\rm CO}=$ $N$(H$_2$)/$I_{\rm CO}$, where $N$(H$_2$) is the molecular gas column density in cm$^{-2}$, and $I_{\rm CO}$ is the CO line intensity in K km s$^{-1}$. This value corresponds to $\alpha_{\rm CO} = 3.2$ M$_\odot$ pc$^{-2}$ (K km s$^{-1})^{-1}$ \citep{Narayanan2012MNRAS.421.3127N}, where $\alpha_{\rm CO} =M$(H$_2$)/$L$(CO), with $M$(H$_2$) being the H$_2$ mass in M$_\odot$ and $L$(CO) the luminosity of the CO line in K km s$^{-1}$ pc$^2$. 
Ultimately, we use the calibration by \citet[][see also \cite{Casasola2017A&A...605A..18C}]{Schruba2011AJ....142...37S}:
\begin{equation}\label{Eq:H2}
   \frac{\Sigma_{\rm H_2}}{{\rm M}_\odot \,{\rm pc}^{-2}} = 4.17\, \frac{I_{\rm CO(1-0)}}{\rm K\,km\,s^{-1}}\times \cos{i} 
\end{equation}
where $i$ is the inclination angle of the galaxy (Table \ref{tab:NGC4254prop}). The associated scatter is $\sim 0.3$ dex, mostly due to the uncertainties on the constant CO-to-H$_2$ conversion factor \citep{Bolatto2013ARA&A..51..207B}. 

\begin{figure}[!h]
\centering
\includegraphics[width=.98\columnwidth]{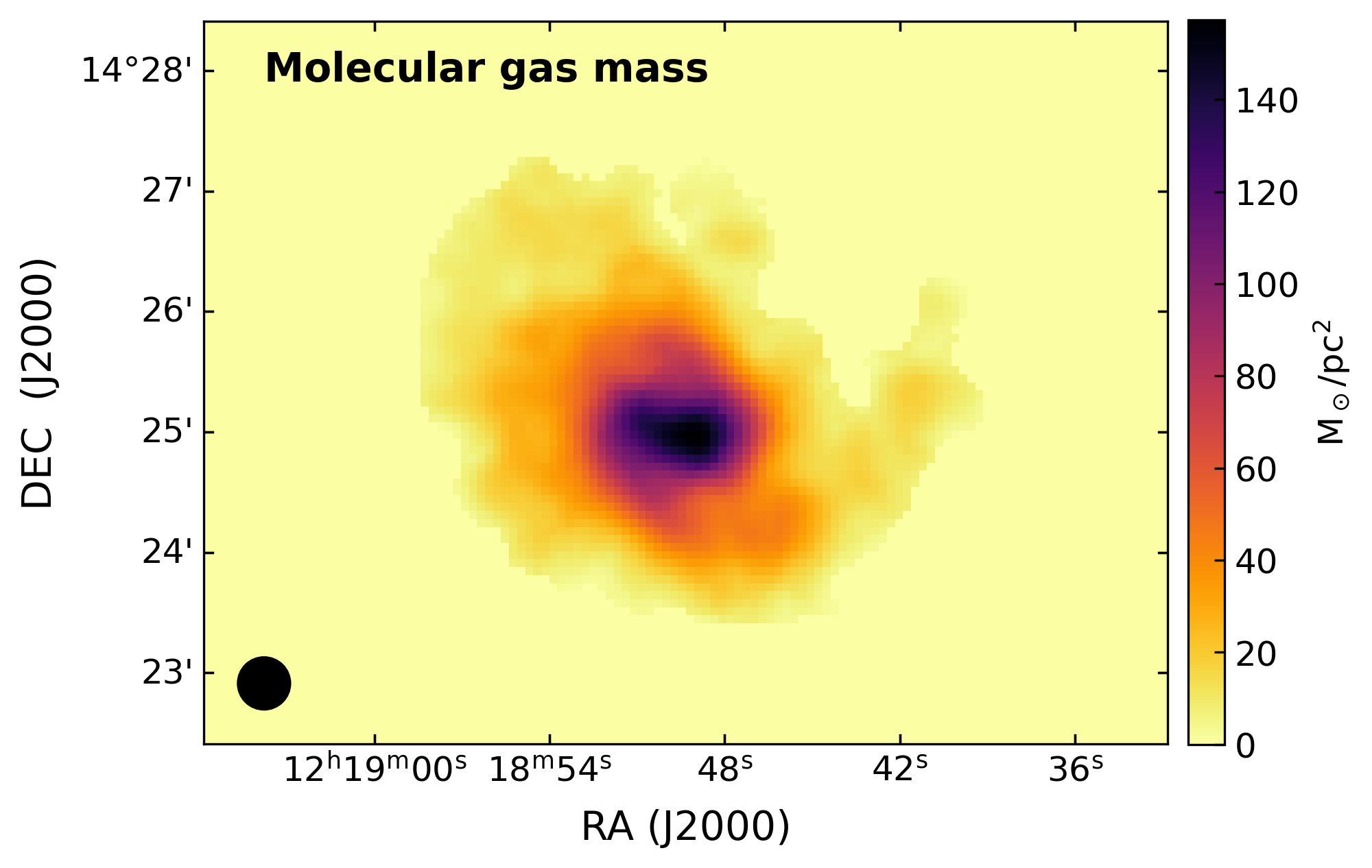}
\caption{Molecular gas surface density map of M~99. Spatial resolution and pixel size are the same as the original CO(1--0) map, i.e. $25.6^{\prime\prime}$ and $4^{\prime\prime}$, respectively.}
\label{fig-CO}
\end{figure}

In Fig.\ref{fig-CO} we show the surface density map of the molecular gas in M~99 at native angular resolution of $25.6^{\prime\prime}$. We measure a total molecular gas mass of $M_{\rm H_2} = (6\pm2)\times 10^9$ M$_\odot$, consistent with the literature \citep[e.g., ][]{Chemin2016A&A...588A..48C, Casasola2020A&A...633A.100C}. 

In addition, we compute the molecular gas surface density of M~99 at an improved angular resolution of $18^{\prime\prime}$, matching that of the NIKA2 2 mm map, using the CO(2--1) intensity map from \citet[][Table~\ref{tab:ancillary_spectrallines}]{Jimenez-Donaire2019ApJ...880..127J}. The CO(2--1)/CO(1--0) ratio, $r_{21}$, is derived pixel-by-pixel at the native CO(1--0) resolution ($25^{\prime\prime}$), yielding a mean value of $\langle r_{21} \rangle = 0.7 \pm 0.2$ with no significant spatial variation, consistent with MW-like galaxies \citep[e.g.,][]{Keenan2025ApJ...979..228K}. The CO(2--1) map is then convolved to $25^{\prime\prime}$ to match CO(1--0), and Eq.~\eqref{Eq:H2} is applied to derive the molecular gas surface density.

Both calibrations, Eqs.~\eqref{Eq:HI} and \eqref{Eq:H2}, include a factor of 1.36 to account for helium and heavier elements. We propagate uncertainties through the MC approach with $N_{\rm MC} = 10^3$ iterations, incorporating the intrinsic scatter of the calibrations.

\subsection{Metallicity}\label{Sect:metallicity}
We construct the gas-phase metallicity map of M~99 by leveraging recent PHANGS-MUSE measurements \citep{Kreckel2019ApJ...887...80K} combined with the compilation of approximately 1900 HII regions from the DustPedia collaboration \citep{DeVis2019A&A...623A...5D}. Metallicities are calibrated using the S calibration of \citet{PilyuginGrebel2016MNRAS.457.3678P}.

\begin{figure}[!h]
\centering
\includegraphics[width=.98\columnwidth]{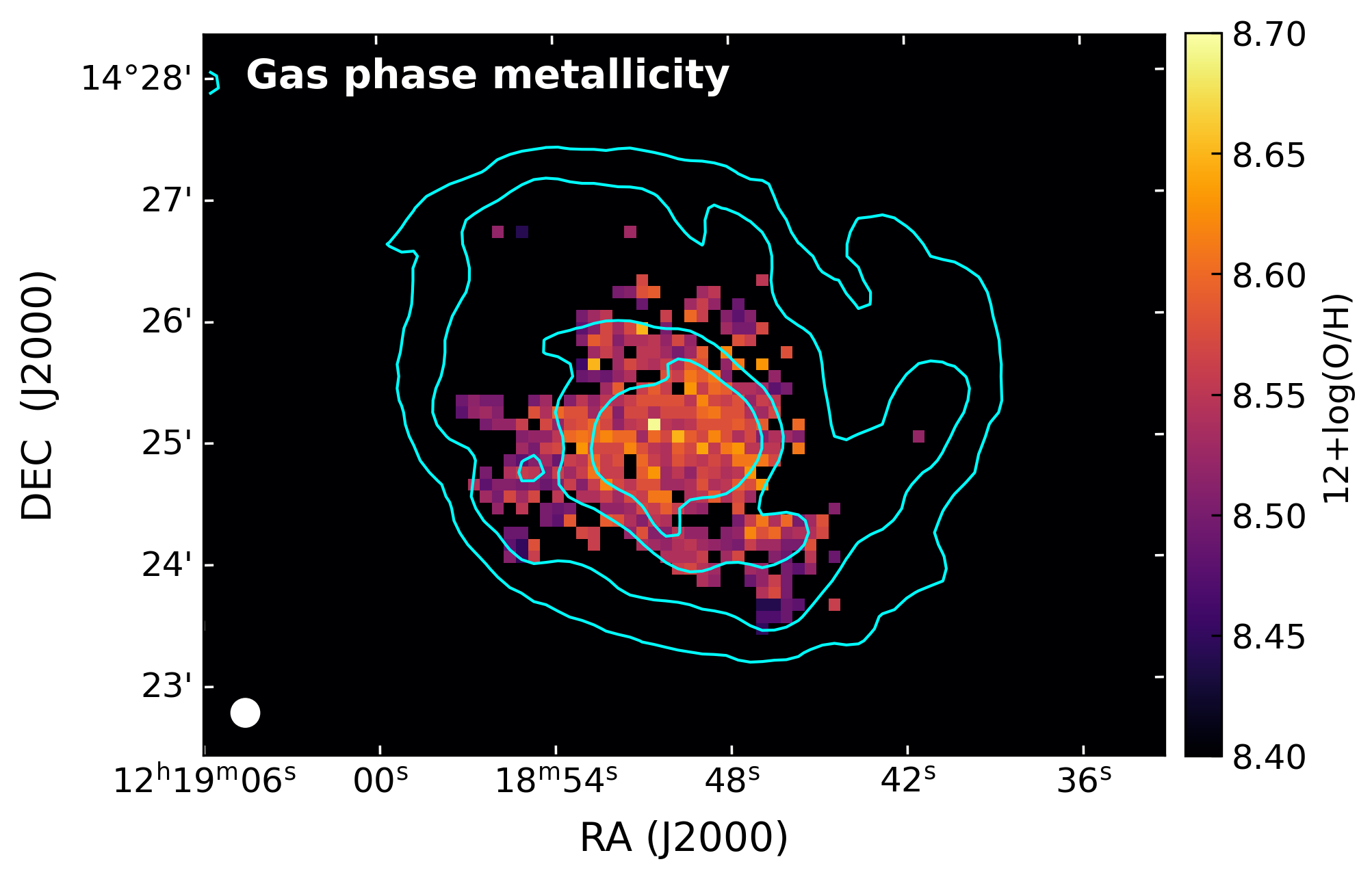}
\caption{Metallicity map of M~99, overlaid with SPIRE~250~\upmicron contours (cyan solid lines; levels of $[5,15,55,100]\times\sigma$). Spatial resolution and pixel size are the same as the SPIRE~250~\upmicron map.}
\label{fig-metallicity}
\end{figure}

Metallicity measurements are typically obtained within $1^{\prime\prime}$ apertures, well below our working angular resolution ($25^{\prime\prime}$) and pixel size $6^{\prime\prime}$. Using the SPIRE~250~\upmicron map as a reference, we averaged all metallicity measurements that fall within each $6^{\prime\prime}$ pixel. Fig.~\ref{fig-metallicity} shows the resulting metallicity map of M~99. The associated uncertainties mostly range between 0.01 and 0.05 in units of $12 + \log(\mathrm{O/H})$, corresponding to $\sim0.1$ dex.

\section{Homogenisation for IMEGIN Photometry post-processing pipeline (\texttt{HIP})}\label{App:hip}

The Homogenisation for IMEGIN Photometry post-processing pipeline (\texttt{HIP}) performs multiple steps to remove astrophysical and artificial contamination from the images, homogenise the dataset, and propagate uncertainties. \texttt{HIP} offers flexible control over the selection and sequencing of these steps, as well as the option to enable or disable uncertainty propagation. In the following sections, we detail each step we executed for the processing of M~99, following the order applied.
\begin{figure}[!h]
\centering
\includegraphics[width=1\columnwidth]{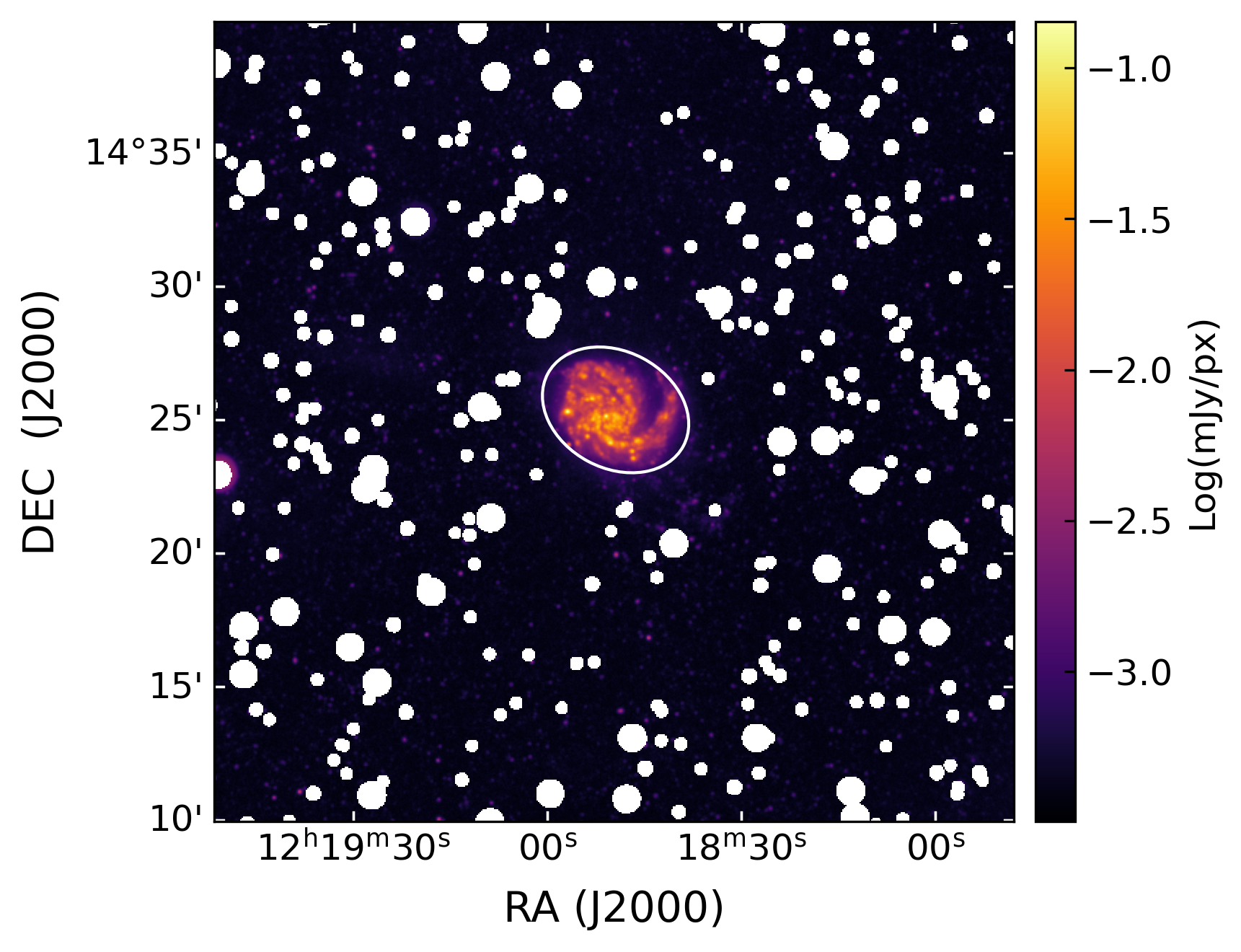}
\caption{GALEX NUV map of M~99 (in log scale and at native angular resolution, i.e. $5.3^{\prime\prime}$) where the emission coming from the brightest foreground stars is masked. The position and the geometry of M~99 are highlighted with a white ellipse, defined in Appendix \ref{App:photometry}.}
\label{fig-PS}
\end{figure}

\subsection{Foreground stars}\label{Sect:foreground_stars}

\texttt{HIP} identifies foreground stars in the vicinity of the target galaxy by querying the point source catalogue of \citet{Cutri2003yCat.2246....0C}. The stars are selected to have $\rm{J} < 40$ mag. \texttt{HIP} masks each star using a circular region with a default radius twice the map’s spatial resolution (FWHM), accounting for the point source emission spreading into adjacent pixels. Additionally, it applies a magnitude-dependent scaling factor to address pixel saturation and bleeding effects\footnote{For J magnitudes $<13.5$, $<14$, $<15.5$, $<16$, $<18$, $<40$ we use, in the order, the following factors: 4.6, 3.0, 2.1, 1.4, 1.15, 0.7.}. 

We ran this step for wavelengths $\lambda < 24$~\upmicron, where stellar emission dominates. 
Fig.~\ref{fig-PS} illustrates the GALEX NUV map of M~99 after masking the brightest foreground stars with \texttt{HIP}.

\subsection{Sky subtraction}\label{Sect:skysub}

\texttt{HIP} models the large-scale sky structures using the \texttt{Background2D} class from \texttt{Photutils}\footnote{\url{https://photutils.readthedocs.io/en/stable/}} \citep{larrybradley202410967176}. 
\texttt{HIP} divides the input map into a square mesh, with each cell size defined as an integer multiple of the map’s spatial resolution, in order to capture local variations of the sky. The default multiplier is 10 and the total number of cells follows accordingly. For each cell, it calculates the mean or median sky emission (depending on the number of pixels), after applying a 3$\sigma$ threshold to reject outliers. Then, it performs a two-dimensional third-order spline interpolation between these values. To avoid bias, \texttt{HIP} masks the target source during modelling, then recovers the large-scale sky emission at the galaxy’s position by interpolating the emission in adjacent cells.
This step efficiently models and removes contamination from the near-infrared sky brightness, Galactic cirrus affecting the mid-IR to (sub)mm sky emission, the Cosmic Infrared Background (CIB), the CMB, and instrumental gradients. 

\begin{figure}[!h]
\centering
\includegraphics[width=0.99\columnwidth]{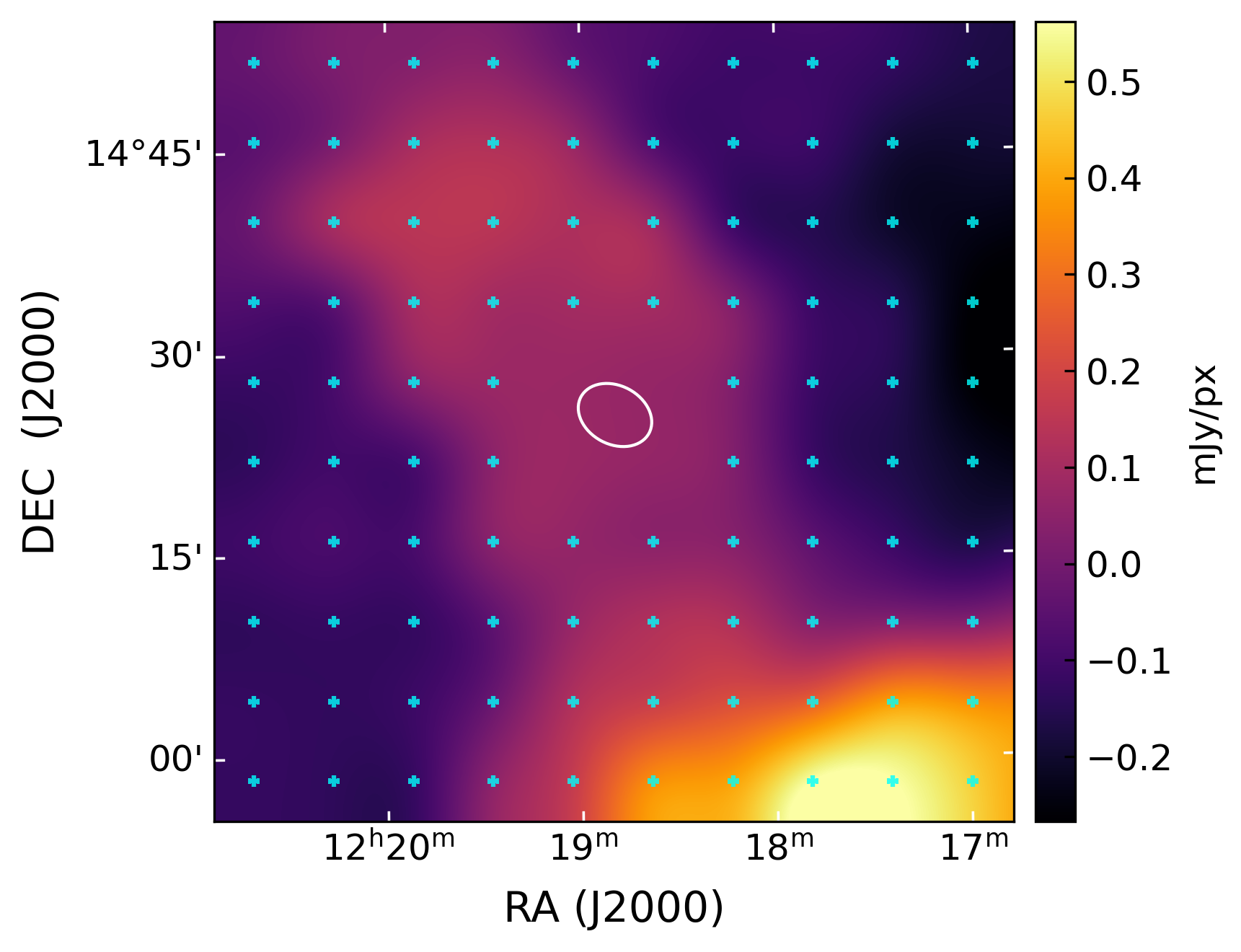}
\caption{Modelled sky emission (3$\sigma$ threshold) in the SPIRE~250~\upmicron map of M~99 at native angular resolution, i.e. $18^{\prime\prime}$. Cyan crosses mark the centres of the mesh cells. The white ellipse is the same as Fig. \ref{fig-PS}.}
\label{fig-bkg}
\end{figure}

We disabled the sky-flattening function when processing the NIKA2 images of M~99, as this correction is already applied during the data reduction with Scanam\_nika.
Figure~\ref{fig-bkg} illustrates the modelled large-scale sky contribution to the SPIRE~250~\upmicron map of M~99.

\subsection{Degrading to the same angular resolution}\label{Sect:degrading}

\texttt{HIP} homogenises the angular resolution of multi-wavelength images by convolving them with appropriate kernels. It primarily uses the publicly available kernels from \citet{Aniano2011PASP..123.1218A}\footnote{\url{https://www.astro.princeton.edu/~draine/Kernels.html}}, though Gaussian kernels can also be applied. Convolution is performed using a Python function by H. Salas, adapted from the original IDL routine\footnote{\url{https://www.astro.princeton.edu/~draine/Kernels/convolve_image.pro}} and available on GitHub\footnote{\url{https://github.com/hsalas/convolution}}. Prior to convolution, image units are converted from Jy~px$^{-1}$ to $L_\odot\,{\rm Hz}^{-1}\,{\rm pc}^{-2}$ to ensure flux conservation.

Fig.~\ref{fig-GALEXFUV_homo} shows the original GALEX FUV map of M~99 (top panel), at its native angular resolution of $4.3^{\prime\prime}$, and the same map after we degrade it to the SPIRE 500~\upmicron resolution ($36^{\prime\prime}$, central panel).

\subsection{Reprojection and resampling to the same grid}\label{Sect:reprojection} %

The final step in the homogenisation process reprojects all images to a common reference frame and resamples them to a uniform pixel size using the \texttt{reproject\_inter} Python function\footnote{\url{https://reproject.readthedocs.io/en/stable/api/reproject.reproject\_interp.html}}. 
After resampling, images are converted back to units of Jy~px$^{-1}$. 

\begin{figure}[!h]
\centering
\includegraphics[width=.97\columnwidth]{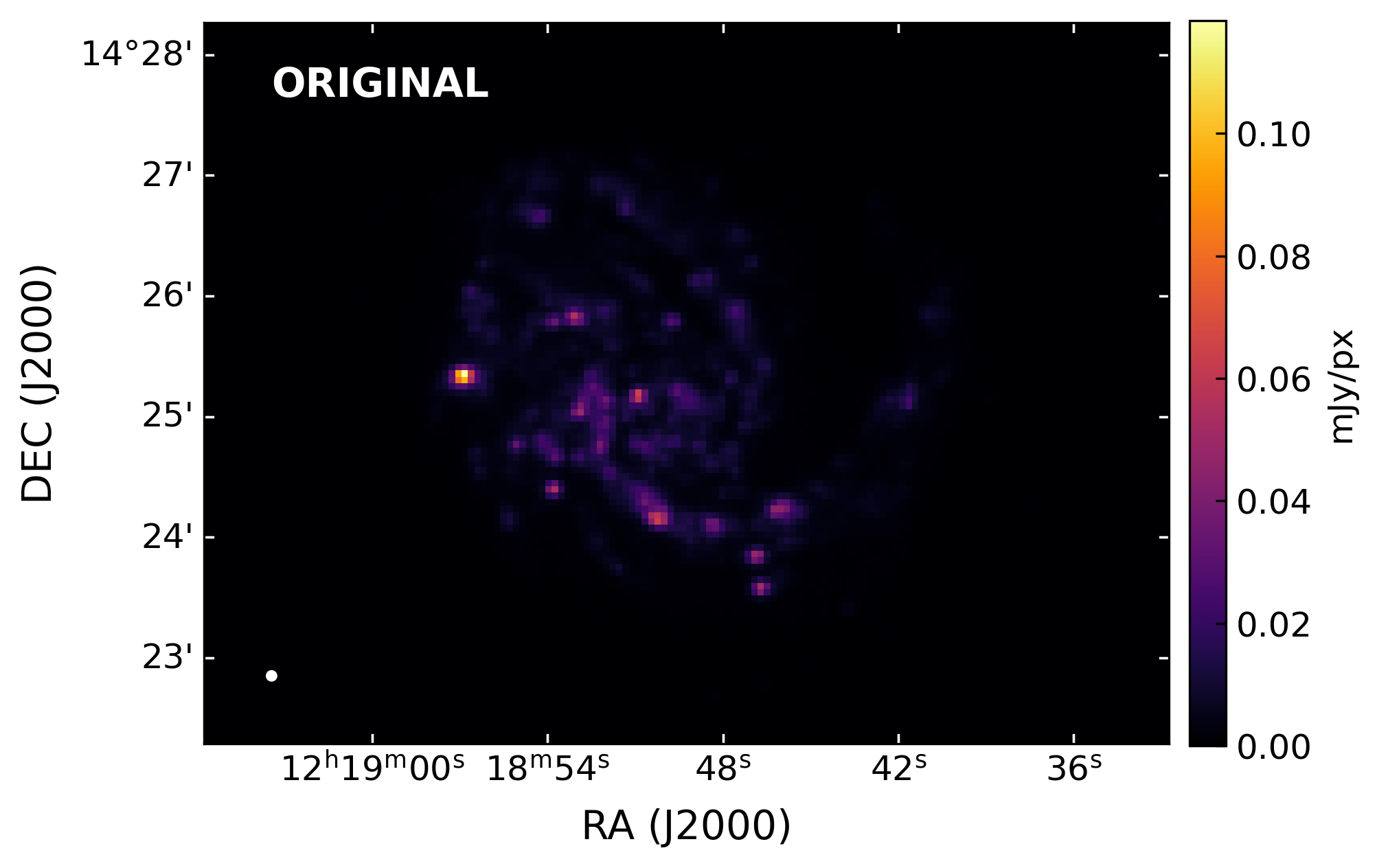}
\includegraphics[width=.98\columnwidth]{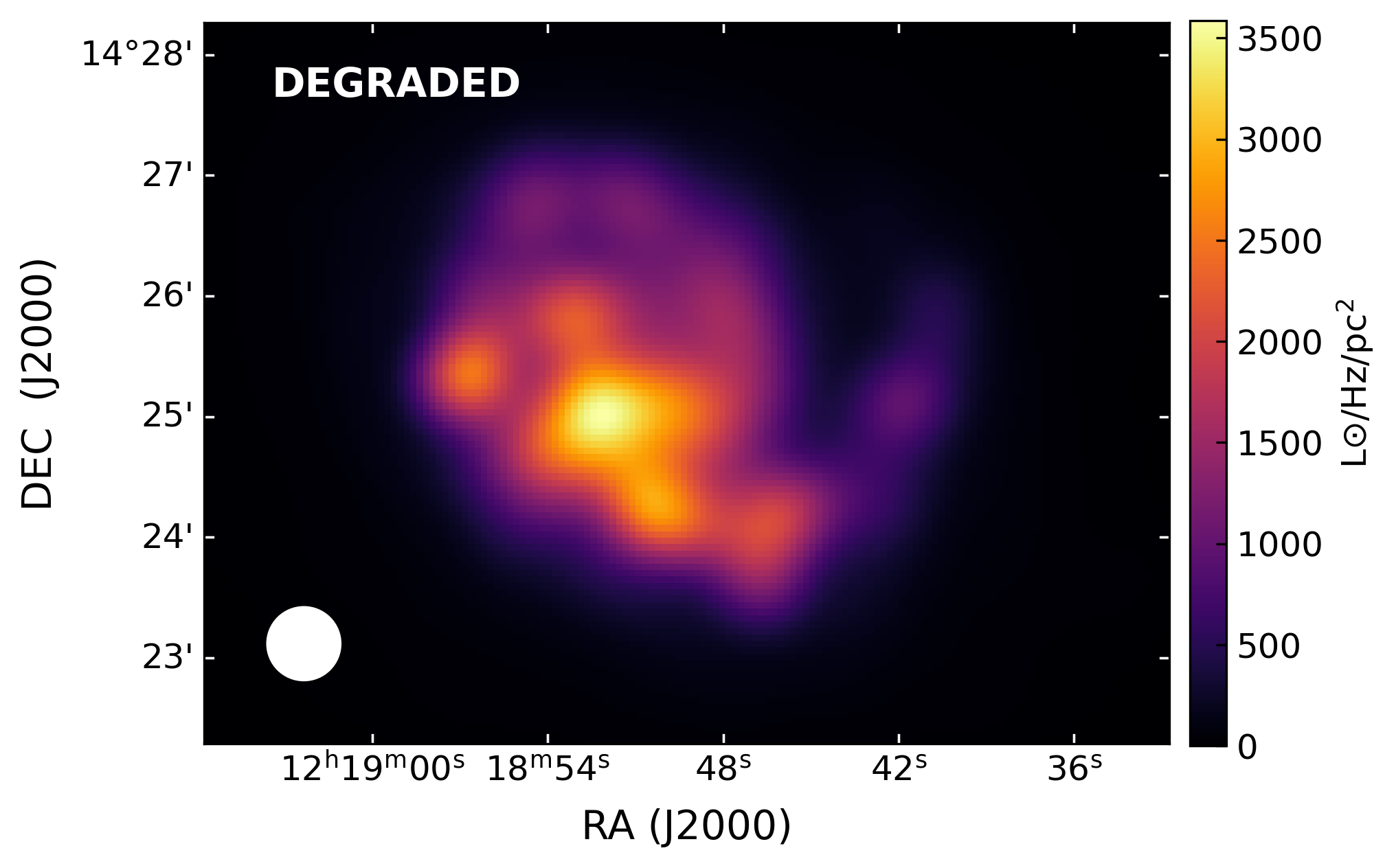}
\includegraphics[width=.98\columnwidth]{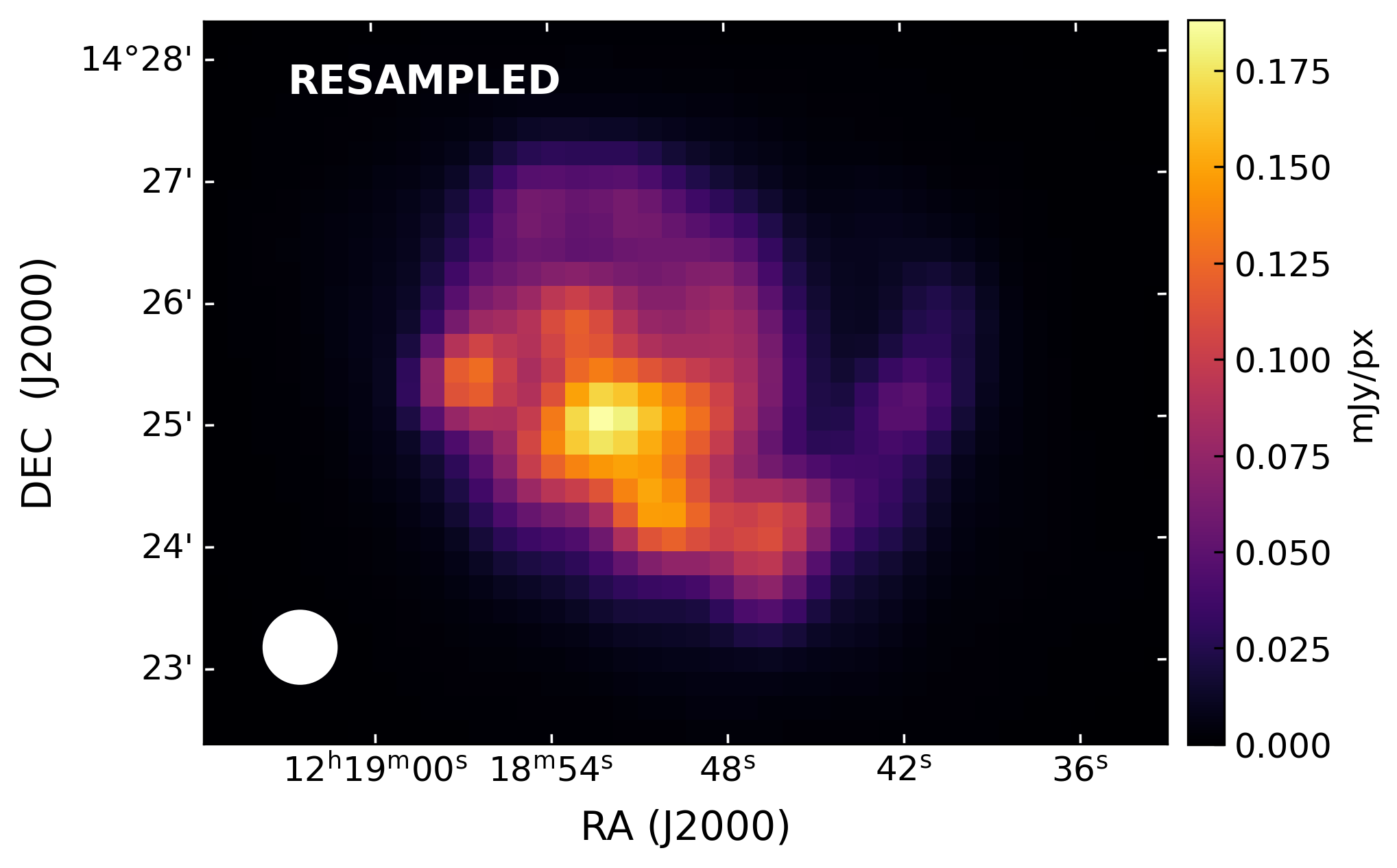}
\caption{GALEX FUV map of M~99. Top panel: original map in mJy/px, with an angular resolution of $4.3^{\prime\prime}$ and pixel size of $\sim3^{\prime\prime}$. Central panel: degraded map to the SPIRE 500~\upmicron angular resolution (i.e., $36^{\prime\prime}$) in surface brightness units ($L_\odot\,$Hz$^{-1}\,$pc$^{-2}$). Bottom panel: re-projected and re-sampled map to the same orientation and pixel size as SPIRE 500~\upmicron (i.e., px $\sim12^{\prime\prime}$) in units of mJy px$^{-1}$.}
\label{fig-GALEXFUV_homo}
\end{figure}

The bottom panel of Fig.~\ref{fig-GALEXFUV_homo} shows the GALEX FUV map of M~99 after convolution, reprojection, and resampling to the SPIRE 500~\upmicron grid.

\subsection{Aperture photometry}\label{App:photometry}

\texttt{HIP} performs aperture photometry on the target galaxy and estimates the integrated flux uncertainty via a Monte Carlo method (Sect.~\ref{Sect:unc_prop}).

\begin{figure}[!h]
\centering
\includegraphics[width=1\columnwidth]{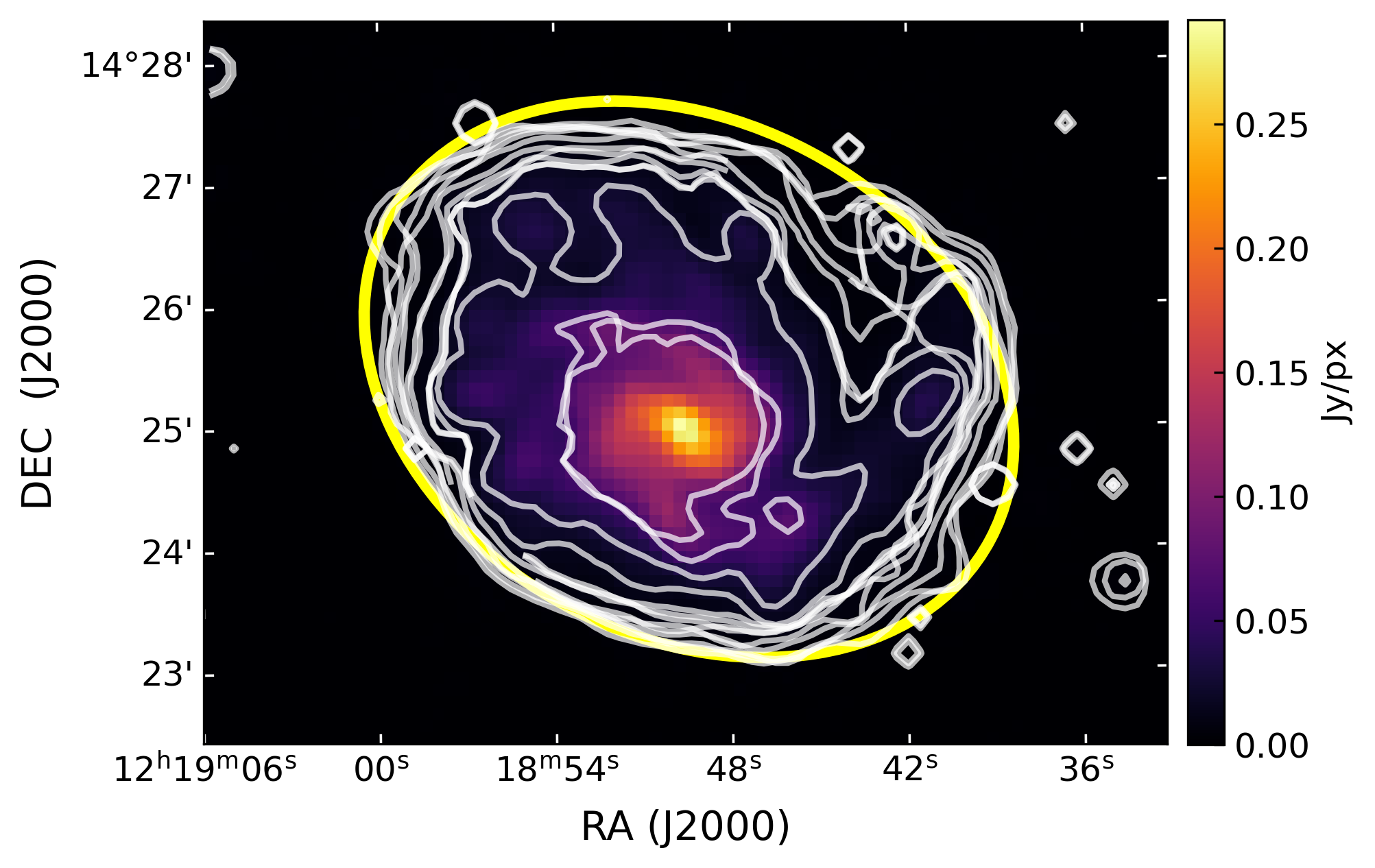}
\caption{SPIRE~250 cutout of M~99, overlaid with the ellipse we use for computing the aperture photometry of the galaxy (yellow solid line). For reference, we show the multi-wavelength $5\sigma$ contours (white solid lines) of M~99 of all the bands used for SED fitting, listed in Tabs. \ref{tab:ancillary_SEDfit} and \ref{tab:integrated_photometry}, at native angular resolution and pixel size.}
\label{fig-SPIRE250_photometry}
\end{figure}

Due to M~99’s intrinsic asymmetry, most notably its extended western spiral arm, we adopt and modify the elliptical aperture from \citet{Clark2018A&A...609A..37C} based on visual inspection of our multi-wavelength data. The resulting aperture, shown in Fig.~\ref{fig-SPIRE250_photometry}, better aligns with the $5\sigma$ contours (white solid lines) and excludes low S/N pixels. Restricting the analysis to pixels within this aperture also reduces the computational cost of the spatially-resolved SED fitting.
For the Planck/HFI4 band, we measure integrated flux within a circular aperture of radius 300$^{\prime\prime}$ to account for the large beam and coarse pixel scale, which spread the galaxy flux over a broader area (Appendix~\ref{App:mm-maps}).

\begin{figure}[!h]
\centering
\includegraphics[width=0.95
\columnwidth]{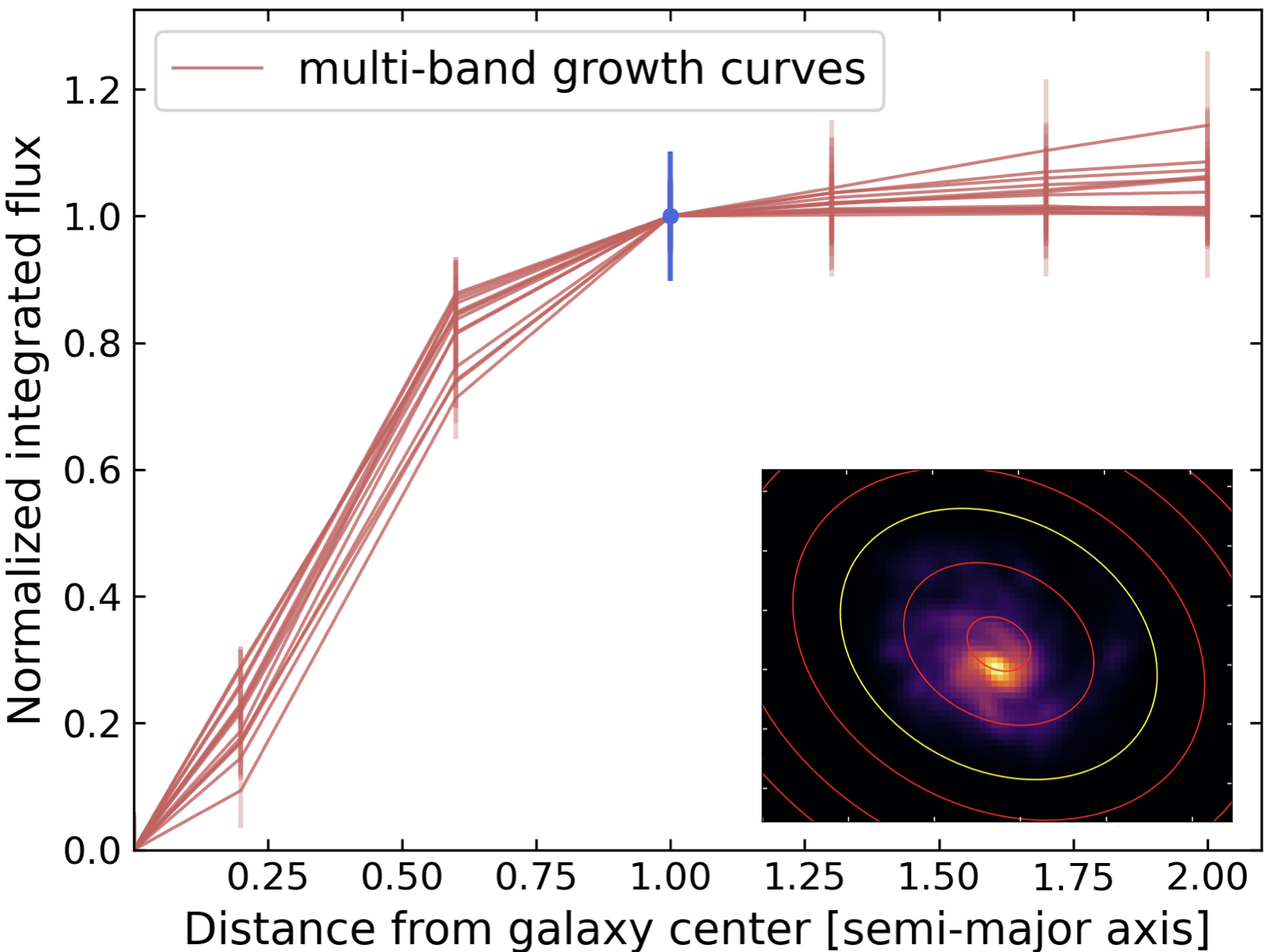}
\caption{Multi-band growth curves built by \texttt{HIP} while computing the aperture photometry of M~99. Each measurement (red bars) is computed within larger and larger co-focal ellipses (shown in the inset, overlaid on the SPIRE~250~\upmicron cutout), and then normalised to the corresponding integrated flux (blue dot). The latter is obtained integrating within the yellow ellipse in the inset (cf. Table \ref{tab:NGC4254prop}). Vertical bars show the error (both RMS and flux calibration uncertainty) on the integrated flux. NIKA2 maps are not included here, since the sky subtraction is already performed during the data reduction with Scanam\_nika. Planck/HFI4 growth curve is shown in Fig. \ref{fig-Planck_BKG_1mm}.}
\label{fig-MULTI-BAND_GC}
\end{figure}

\texttt{HIP} additionally computes and plots growth curves (i.e., integrated flux as a function of radius) out to $r=2a$, where $a$ is the semi-major axis of the elliptical aperture. Fluxes are summed within co-focal ellipses, serving as diagnostics for sky subtraction quality: the curve should rise within $r < a$, dominated by galaxy emission, and flatten beyond as noise dominates. Visual inspection confirmed expected growth curve behaviour across bands. Figure~\ref{fig-MULTI-BAND_GC} shows the normalised multi-band growth curves, with the adopted aperture indicated in the inset.

\subsection{CO(2--1) contamination in the mm continuum}\label{Sect:CO_sub}

The CO(2--1) molecular emission line at 230.5 GHz ($\lambda = 1.3$ mm) lies within the passbands of both NIKA2 and Planck/HFI4, which have central wavelengths of 1.15 mm and 1.38 mm, respectively, and thus contributes to the measured continuum emission.

\begin{figure}[!h]
\centering
\includegraphics[width=1\columnwidth]{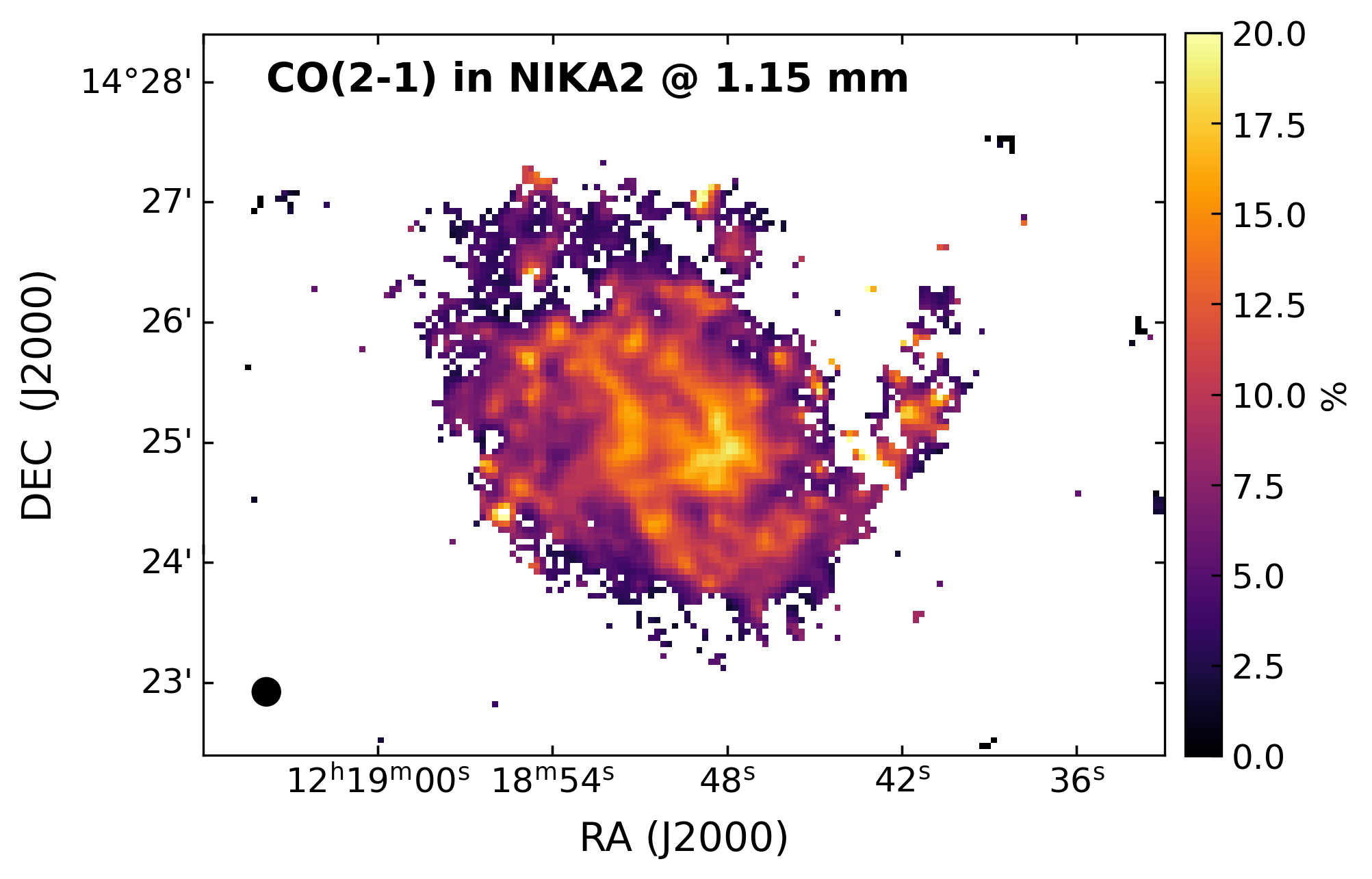}
\caption{Percentage of flux in the NIKA2 1.15 mm map due to CO(2--1) emission line. The beam size ($\sim13^{\prime\prime}$) is shown at the bottom left.}
\label{fig-COfrac}
\end{figure}

\texttt{HIP} includes a built-in routine that estimates and subtracts the CO(2--1) line contamination from the millimetre continuum, both on pixel-by-pixel and integrated scales. This routine implements the method described by \citet{Drabek2012MNRAS.426...23D}. First, it ensures that the CO(2--1) intensity map and the continuum image share the same angular resolution, pixel size, and orientation (performing regridding if necessary). Then, it computes the conversion factor $C$, which quantifies the fraction of the line flux contributing to the continuum, following Eq.~(8) in \citet{Drabek2012MNRAS.426...23D}:
\begin{equation}\label{Eq:COcont}
    \frac{C}{\mathrm{mJy\, beam}^{-1} \,\mathrm{per}\,\mathrm{K\, km\,s}^{-1} } 
    = \frac{2k\nu^3}{c^3} \frac{g_\nu(\mathrm{line})}{\int g_\nu\,{\mathrm d}\nu} \, \Omega_b  
\end{equation}
where $k$ is Boltzmann’s constant; $\nu$ is the line frequency (in GHz); $c$ is the speed of light in km s$^{-1}$; $g_\nu$(line) is the filter passband at the frequency of the molecular line; $\int g_\nu\,d\nu$ is the integrated filter passband across the full range of filter frequencies; $\Omega_b$ is the main beam size. 

For M~99, we use the CO(2--1) intensity map from the HERACLES survey \citep[][Table~\ref{tab:ancillary_spectrallines}]{Leroy2009} and compute a contamination factor $C = 0.0476$ for NIKA2 and $C = 50.3$ for Planck/HFI4. After subtracting the CO(2--1) contribution, the integrated flux in the NIKA2 1.15~mm map is $0.59 \pm 0.04$~Jy, with CO(2--1) accounting for $\sim9$\% of the observed flux ($0.65 \pm 0.04$~Jy before correction). For Planck/HFI4, the CO-subtracted flux is $0.35 \pm 0.07$~Jy, where CO(2--1) contributes about 5\% of the total observed $0.37 \pm 0.07$~Jy.

Figure~\ref{fig-COfrac} shows the spatial distribution of the CO contamination fraction in the NIKA2 1.15~mm continuum (pixels $>3\sigma$), ranging from a few percent up to 20\%.

\subsection{Uncertainty treatment}

\texttt{HIP} implements a rigorous treatment of uncertainties and ensures their propagation throughout all steps of the pipeline.

\subsubsection{Statistical uncertainty}\label{Sect:stat_unc}
 
Statistical uncertainty is traditionally estimated from the RMS of flux values measured in regions well separated from the target source. More robust measurements are the 5$\sigma$-trimmed standard deviation or the Median Absolute Deviation (MAD), defined as:
\begin{equation}
    \rm{MAD}(\rm{X}) = k\times \rm{med(\,|X-med(X)|\,)}
\end{equation}
where $k=1.4826$ for Gaussian distributions. In \texttt{HIP}, the statistical uncertainty is quantified using the MAD computed over the entire image after masking the science source and foreground stars. This approach minimises the impact of random fluctuations from other astrophysical signals or bad pixels.

For images with associated error maps\footnote{Error maps quantify the instrumental noise.} (for M~99: Herschel, Spitzer/IRAC 5.2~\upmicron, 8~\upmicron, Spitzer/MIPS 24~\upmicron maps), \texttt{HIP} computes the median of the error map (${\rm Me}_{\rm Err}$) and compares it with the MAD. If ${\rm MAD}>{\rm Me_{\rm Err}}$, ${\rm Me_{\rm Err}}$ is subtracted in quadrature from the MAD to obtain a flat noise estimate representing residual random fluctuations. This flat noise is then added in quadrature to the original error map pixel-by-pixel, producing the final statistical uncertainty map. Since statistical errors are typically independent or weakly correlated, summing and subtracting in quadrature is justified.

For maps without error maps, the statistical uncertainty is assumed uniform and set equal to the MAD. For NIKA2 maps, the statistical uncertainty is taken as the sky RMS estimated by the Scanam\_nika reduction software.

\subsubsection{Systematic uncertainty}\label{Sect:sys_unc}

\texttt{HIP} accounts for the absolute calibration uncertainty, which represents the bias in the flux measurement introduced by the instrument’s calibration. Table~\ref{tab:ancillary_SEDfit} lists the calibration uncertainties for each instrument used in this work. Note that the calibration uncertainty is independent of the noise quantified in Sect.~\ref{Sect:stat_unc}.

\subsubsection{Uncertainty propagation}\label{Sect:unc_prop}

Homogenising multi-wavelength images in angular resolution and pixel size reduces statistical uncertainties by averaging over multiple pixels but introduces correlations between adjacent pixels. As a result, standard error propagation methods assuming independent pixels are no longer valid.

To account for this, \texttt{HIP} implements uncertainty propagation via the MC approach. For each image, it draws random noise values per pixel from a normal distribution centred at zero with a standard deviation equal to the pixel’s uncertainty. These noise values are added to the original map, which is then processed through the full homogenisation pipeline. This procedure is repeated over $N_{\rm MC} > 100$ iterations, generating a cube of perturbed and homogenised maps. The statistical uncertainty map is then computed as the pixel-wise standard deviation across the MC realisations.
The total uncertainty map is obtained by adding in quadrature the statistical uncertainty and the calibration uncertainty (Sect.~\ref{Sect:sys_unc}).

The same MC approach is used to estimate uncertainties on aperture photometry: the integrated flux is measured on each perturbed map, and the standard deviation across the $N_{\rm MC}$ fluxes provides the statistical uncertainty. Calibration uncertainty is again added in quadrature to derive the total uncertainty.

\section{Validation of our modelling with machine learning predictions}\label{App:val_Deb}

In this Appendix, we validate our modelling by comparing it with the dust maps from \citet{Paradis2024A&A...691A.241P}, available in HEALPix format via the CADE website\footnote{\url{https://cade.irap.omp.eu/dokuwiki/doku.php?id=nnpredictions}}. These maps were generated using machine learning techniques to predict emission across various Galactic environments and extragalactic sources in the two Planck/HFI bands centred at 850~\upmicron and 1.38 mm. The predicted maps have an angular resolution of approximately $37^{\prime\prime}$, comparable to that of SPIRE 500~\upmicron, and include contributions from synchrotron and free-free emission, which remain below a few percent. Emission from M~99 was predicted based on the KINGFISH survey \citep{Kennicutt2011PASP..123.1347K} and converted to WCS FITS format using the Drizzweb interface\footnote{\url{http://drizzweb.irap.omp.eu/}}.

We performed aperture photometry on the predictions from \citet{Paradis2024A&A...691A.241P}, obtaining integrated flux densities of $S_{850} = 1.55 \pm 0.06$ Jy and $S_{1.38} = 0.40 \pm 0.03$ Jy, after modelling and subtracting the Galactic cirrus contribution. These values agree, within uncertainties, with the fluxes derived from our integrated-scale modelling using \texttt{HerBIE}: $S_{850} = 1.76 \pm 0.18$ Jy and $S_{1.38} = 0.40 \pm 0.04$ Jy (\themis), and $S_{850} = 1.55 \pm 0.16$ Jy and $S_{1.38} = 0.31 \pm 0.05$ Jy (\mbb). Our values, however, suggest a slightly steeper millimetre slope.

To enable a direct, spatially resolved comparison, we used our pixel-by-pixel SED decomposition to generate separate dust and radio emission maps of M~99 at 850~\upmicron and 1.38 mm, degraded to the angular resolution of the SPIRE 500~\upmicron data. At a pixel scale of $15^{\prime\prime}$ ($\sim1.1$ kpc), matching that of the \citet{Paradis2024A&A...691A.241P} predictions, the relative differences in flux density between our maps and theirs remain within 10\%, in line with typical uncertainties at this resolution. This agreement holds for both the \themis and \mbb fits.

\section{Pixel masking at low S/N}\label{App:masked_pixels}

\begin{figure}[!h]
\centering
\includegraphics[width=.82\columnwidth]{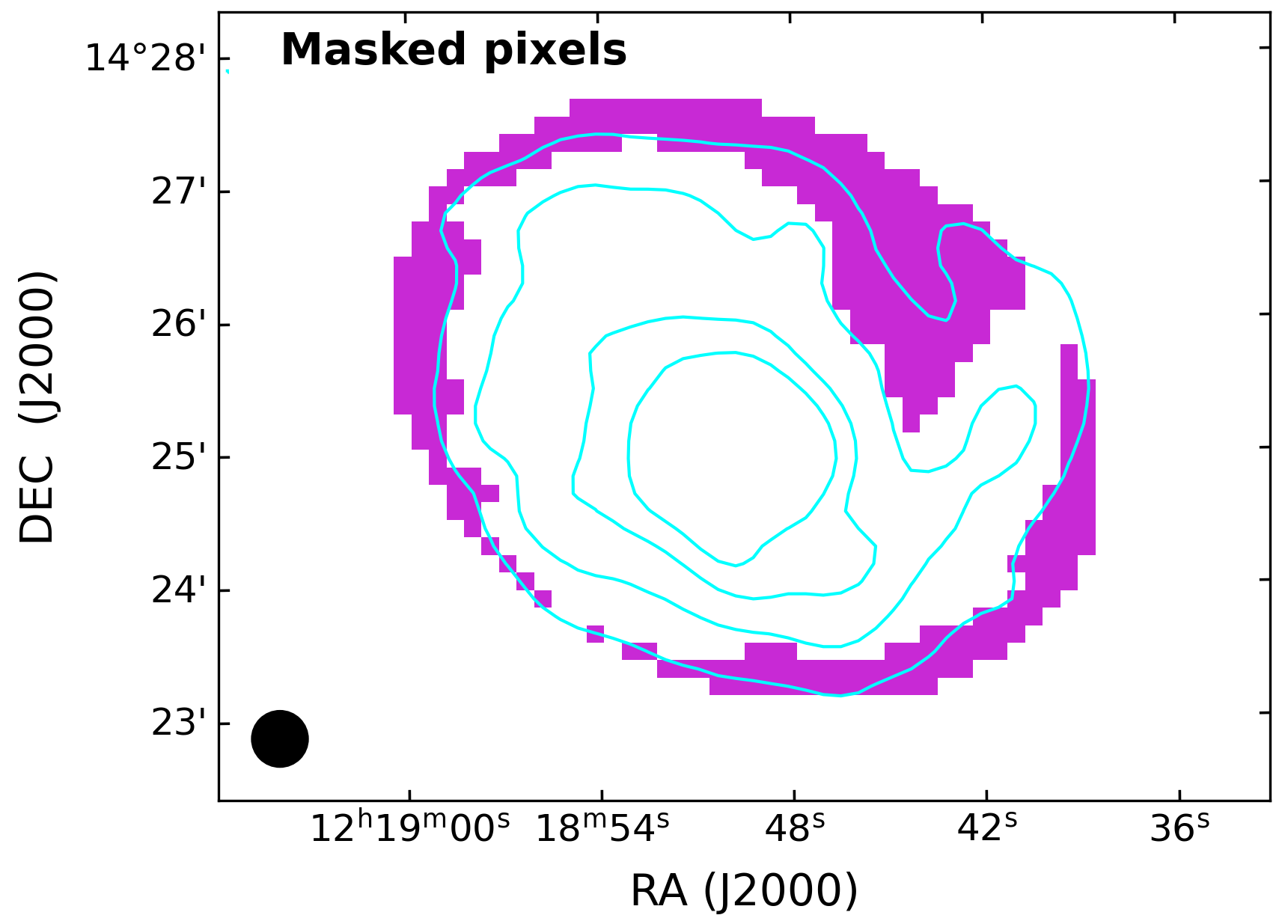}
\caption{
Magenta regions indicate areas with S/N $<3$ in the NIKA2 maps, within the ellipse defining M~99's geometry (Table~\ref{tab:NGC4254prop}). These low-S/N regions often show anomalous or extreme model parameter values. Cyan contours show SPIRE 350~\upmicron emission at [5, 15, 35, 55]$\times\sigma$. The same low-S/N pixels are marked in grey in Figs.~\ref{fig-qAF_Uav} and~\ref{fig-a_radio} for reference.}
\label{fig-pixel_mask}
\end{figure}

In Fig.~\ref{fig-pixel_mask}, we highlight in magenta pixels with low S/N, typically located in regions above the $5\sigma$ level in the SPIRE 350~\upmicron map but below $3\sigma$ in the NIKA2 data (see Figs.~\ref{fig-NIKA2_SCANAM} and~\ref{fig-SCANAM-morphology}). These pixels often yield unphysical or poorly constrained fitted parameters and are therefore excluded from all quantitative analyses. 

\section{Candidate star-forming regions in M~99}\label{App:SF_regions}

In this appendix, we present eight candidate star-forming (SF) regions in M~99, visually selected on the sSFR map as the brightest, isolated, and unresolved sources, consistent with H~II regions being smaller than our working resolution of $25^{\prime\prime}$.

\begin{figure}[!h]
\centering
\includegraphics[width=1\columnwidth]{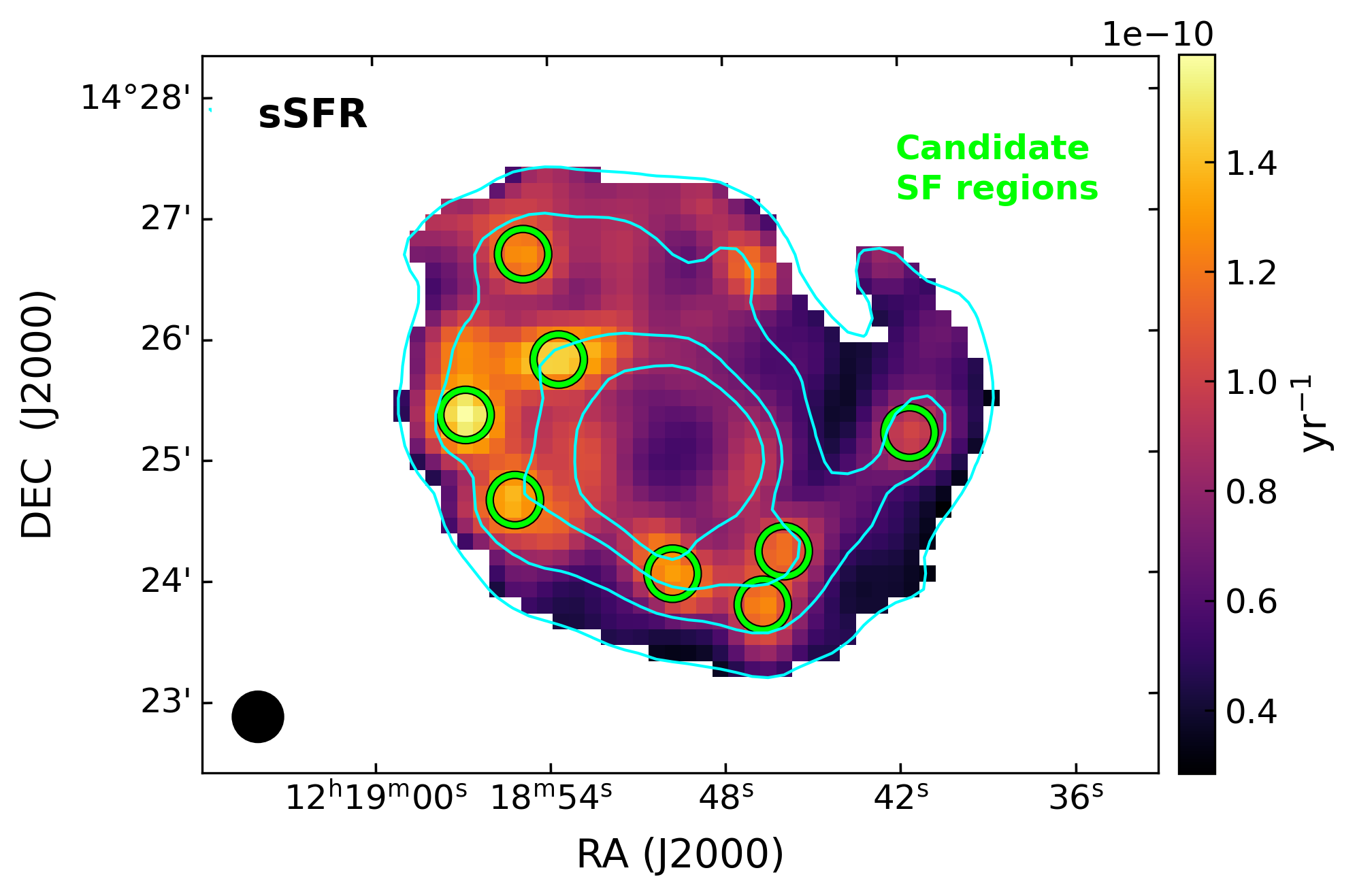}
\caption{Candidate SF regions (green circles) overlaid on the sSFR map of M~99 at $25^{\prime\prime}$ resolution. For reference, cyan contours represent SPIRE 350~\upmicron emission at $[5, 15, 35, 55] \times \sigma$ levels.}
\label{fig-SFregions}
\end{figure}

Figure~\ref{fig-SFregions} shows the selected SF regions overlaid on the sSFR map. For quantitative analysis, we measured integrated properties within circular apertures of radius $12.5^{\prime\prime}$. The sSFR map was derived by dividing the SFR surface density (Fig.~\ref{fig-SFR}) by the stellar mass surface density (Fig.~\ref{fig-stellarmass}).

\end{appendix}

\end{document}